\RequirePackage{fix-cm}
\documentclass[smallextended]{svjour3}       

\usepackage{amsmath}
\usepackage{graphicx}
\usepackage{setspace}
\usepackage[subrefformat=parens,labelformat=parens]{subfig}
\usepackage{float}

\usepackage{epstopdf}
\usepackage{xcolor}

\smartqed  
\usepackage{graphicx}
\newcommand{\beq}{\begin{equation}}
\newcommand{\eeq}{\end{equation}}

\newcommand{\stochDiff}{\tilde{D}}
\newcommand{\stochChemo}{\tilde{\chi}}

\newcommand{\pospart}[1]{\left(#1\right)^+}

\newcommand{\nene}[1]{<#1>}
\newcommand{\mean}[1]{\left<#1\right>}

\begin{document}

\title{Mesoscopic and continuum modelling of angiogenesis
}


\author{F. Spill \and P. Guerrero \and T. Alarcon\and P.K. Maini\and H.M. Byrne
}


\institute{F. Spill, H.M. Byrne, \at
              OCCAM, Mathematical Institute, University of Oxford, Oxford OX2 6GG, UK\\
	      \email{spill,byrne@maths.ox.ac.uk}           
              \and
              P. Guerrero,  T. Alarcon \at
              Centre de Recerca Matematica, Campus de Bellaterra, Edifici C, 08193 Bellaterra, Barcelona, Spain \\ 
	      \email{talarcon@crm.cat}
	      \and
	      P. Guerrero \at
              Department of Mathematics, University College London, Gower Street, London WC1E 6BT, UK \\
              \email{pguerrero@ucl.ac.uk}
              \and
              T. Alarcon \at
              Departament de Matem\`atiques, Universitat Aton\`oma de Barcelona, 08193 Bellaterra (Barcelona), Spain.
	      \and
	      P.K. Maini \at
              Wolfson Centre for Mathematical Biology, Mathematical Institute, University of Oxford, Oxford OX2 6GG, UK\\
              \email{maini@maths.ox.ac.uk} 
	      \and
	      H.M. Byrne, \at
              Department of Computer Science, University of Oxford, Oxford OX1 3QD, UK
}

\date{Received: date / Accepted: date}

\maketitle

\begin{abstract}
 Angiogenesis is the formation of new blood vessels from pre-existing ones in response to chemical signals secreted by, for example, a wound or a tumour. In this paper, we propose a mesoscopic lattice-based model of angiogenesis, in which processes that include proliferation and cell movement are considered as stochastic events. By studying the dependence of the model on the lattice spacing and the number of cells involved, we are able to derive the deterministic continuum limit of our equations and compare it to similar existing models of angiogenesis. We further identify conditions under which the use of continuum models is justified, and others for which stochastic or discrete effects dominate. We also compare different stochastic models for the movement of endothelial tip cells which have the same macroscopic, deterministic behaviour, but lead to markedly different behaviour in terms of production of new vessel cells. 
\end{abstract}

\section{Introduction}
Continuum models are widespread in biology, where their use, as in other sciences, is typically justified when the length scale of the problem of interest is considerably larger than the length scale of the underlying microscopic elements of the model, and when the averaged microscopic elements form well defined, continuous functions on the macroscale. For example, when studying the invasion of a tumour, which contains billions of cells, one may justify modelling the tumour in terms of a macroscopic, continuously varying cell density function (see the reviews \cite{moreira2002cellular,araujo2004history,roose2007mathematical,byrne2010dissecting}). Typically, cell proliferation and death rates depend on nutrients such as oxygen, as well as signalling molecules, 
and these are typically modelled as continuous concentrations, evolving according to partial differential equations (PDEs) of reaction-diffusion type. 

A hallmark of cancer is its ability to stimulate angiogenesis, which is the formation of new blood vessels from pre-existing ones \cite{folkman1995angiogenesis,carmeliet2000angiogenesis,hanahan2000hallmarks,hanahan2011hallmarks}. If the tumour cannot incorporate existing vasculature, as well as growing new vessels, the size of the tumour would be limited by the diffusion range of nutrients. The diffusion range of oxygen is about $100 \mu m$ to several $mm$ \cite{folkman1990evidence}, restricting the avascular tumour size to be of the order of a few millimetres. Due to its importance in tumour growth, targeting angiogenesis is an active area of cancer research. The initial aim was to prevent angiogenesis, and hence reduce the delivery of nutrients and thus stop the growth of the tumour \cite{folkman1971tumor}. Even though a number of anti-angiogenic molecules have been identified, treatment with only these molecules does not necessarily improve tumour prognosis, and may
even lead to a worse prognosis by selecting for more aggressive phenotypes \cite{norden2009antiangiogenic,ebos2011antiangiogenic}. Therapeutic effects have, however, been observed when anti-angiogenic compounds are combined with other treatments, such as chemotherapy. In such situations, the angiogenic inhibitors act to transiently normalise the notoriously leaky tumour vasculature and thereby to improve the delivery of blood-borne drugs to the tumour \cite{jain2005normalization,goel2011normalization,potente2011basic,carmeliet2011molecular}.

In order to understand angiogenesis and its interaction with drugs, huge efforts have been undertaken by the biological and medical community in recent decades (see the reviews \cite{risau1997mechanisms,carmeliet2000mechanisms,potente2011basic}). Angiogenesis in initiated typically by hypoxic cells, which secrete a range of angiogenic factors (AFs) such as vascular endothelial growth factor (VEGF) \cite{carmeliet2011molecular}. These AFs diffuse through the tissue and stimulate endothelial cells to become migratory tip cells. These tip cells secrete proteases which break down the basement membrane enabling tip cells to migrate via chemotaxis up spatial gradients of AFs. Stalk cells located behind the tip cells proliferate. Once tip cells encounter 
other tip cells or a vessel, loops can form via a process called anastomosis. The stalk cells can then form lumen through which blood may flow. The tip and stalk cells then mature, which is itself a complex process, involving other vessel cells such as pericytes and smooth vessel cells. 

\begin{figure}[h!]
\subfloat[Spatial tips profile evolving in time]
{
\includegraphics[width=0.48\linewidth]{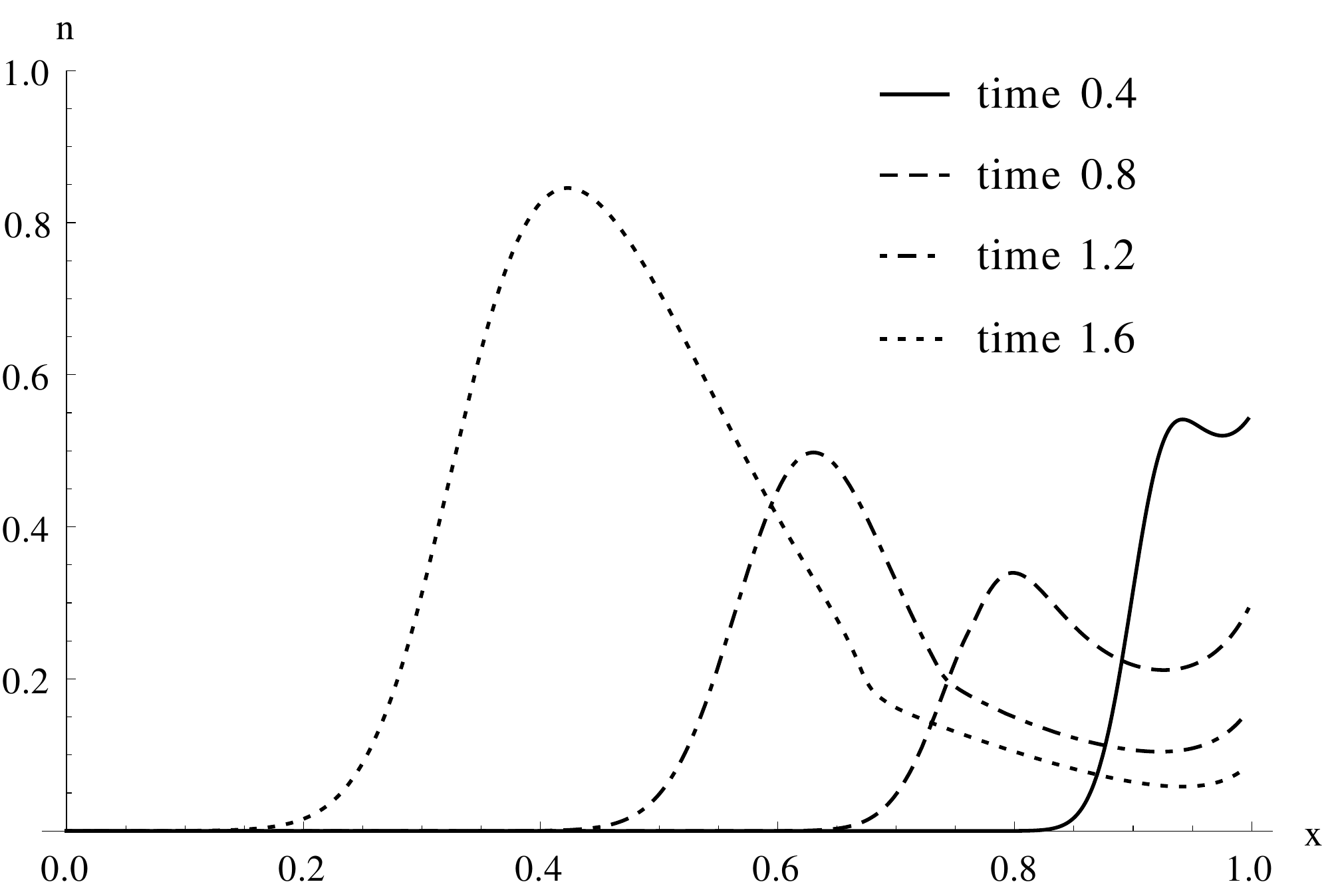}
\label{fig:PDEFullModelTips}
}
\subfloat[Spatial vessel profile evolving in time]
{
\includegraphics[width=0.48\linewidth]{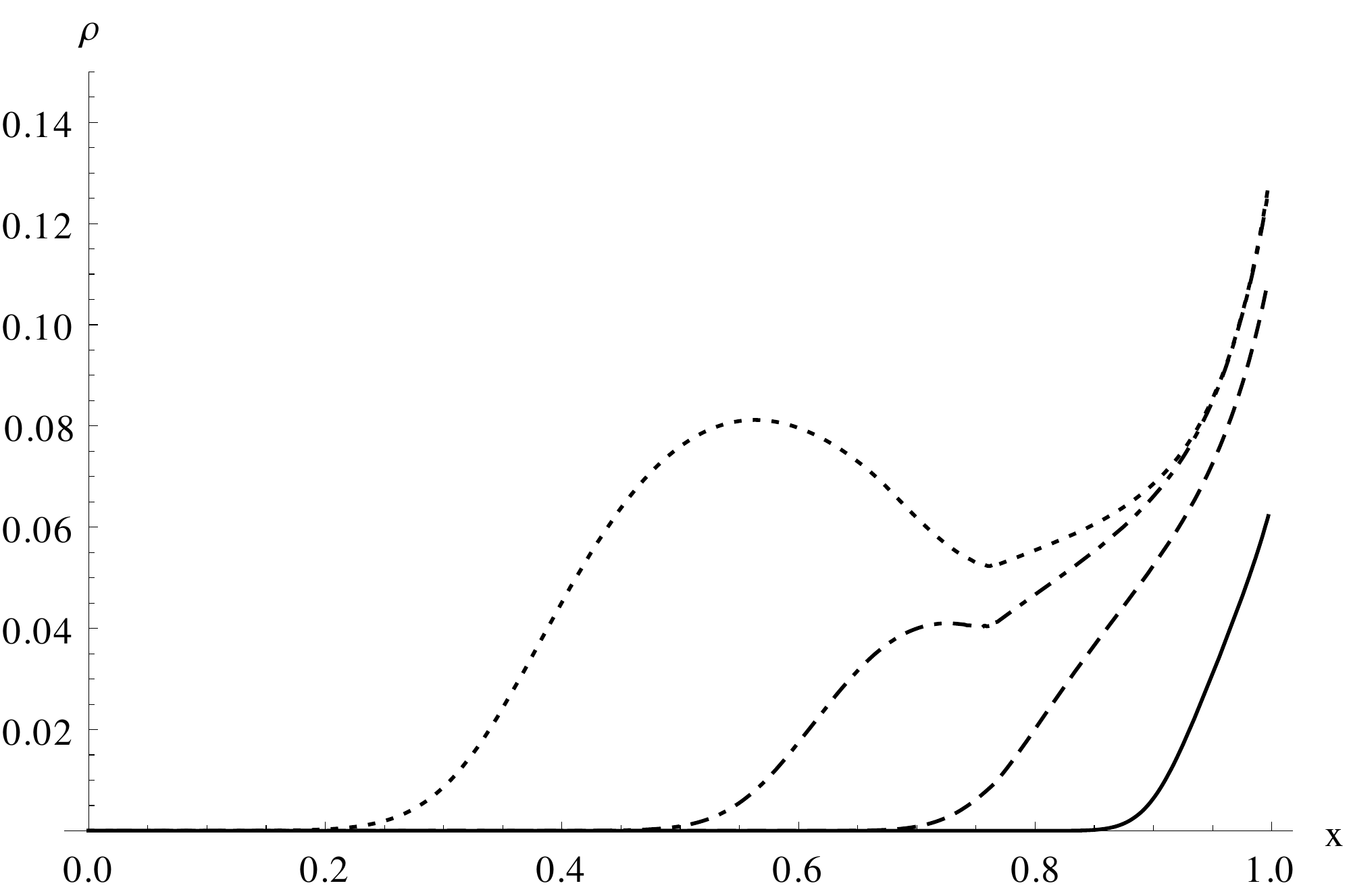}
\label{fig:PDEFullModelVessels}
}
\caption{\label{fig:PDEFullModel} A tumour acts as a source for an AF on the left boundary $(x=0)$, leading to tip migration from a parent vessel on the right boundary $(x=1)$ to the left and proliferation of tip and vessel cells. These profiles were obtained by solving the partial differential equations \eqref{eq:PDEFullModel} numerically using the parameter values stated in section \ref{sec:simulationFullModel} and based on \cite{byrne1995mathematical,mantzaris2004mathematical}.}
\end{figure}

To assist in understanding the complexities of angiogenesis, and to predict the growth of the vasculature and the impact of changes of external conditions, such as the growth of a tumour or the application of drugs, a large number of mathematical models of angiogenesis have been developed (see the reviews \cite{chaplain2000mathematical,mantzaris2004mathematical,chaplain2006mathematical,peirce2008computational}). Early models describe the evolution of tip cell densities, proliferating stalk or vessel cells and concentrations of AFs by systems of coupled PDEs \cite{balding1985mathematical,chaplain1993model,byrne1995mathematical}, and were motivated by similar models describing the growth of fungal networks \cite{edelstein1982propagation}. The tip cells evolve via a reaction-advection-diffusion equation, the advection term modelling the chemotactic migration of the tip cells up
the gradient of the AF. The evolution of the stalk or vessel cell densities is driven by a term proportional to the flux of tip cells, a phenomenon termed the "snail-trail". The typical behaviour of such a snail-trail model, \cite{byrne1995mathematical}, is shown in Figure \ref{fig:PDEFullModel}, where angiogenesis in a corneal assay was modelled. In these assays, a tumour is implanted into the cornea of an animal such as a rabbit or a mouse. Due to the transparency of the cornea, the growing blood vessels can hence be easily observed. The tumour in this model is considered as a source for an AF on the left boundary, and a parent vessel acts as a source for tip and vessel cells at the right boundary. Figure \subref*{fig:PDEFullModelTips} shows tips migrating with increasing density towards the tumour. Vascularisation occurs behind the evolving tips, as shown in Figure \subref*{fig:PDEFullModelVessels}. 

Whereas PDE models treat populations of cells as continua, individual based models distinguish single cells. In \cite{stokes1991analysis}, the movement of an individual tip cell was modelled by a stochastic differential equation (SDE), with a deterministic part modelling chemotaxis, and a stochastic part modelling random motion. Other examples of stochastic models of angiogenesis can be found in \cite{plank2003reinforced,plank2004lattice,capasso2009stochastic,das2010hybrid}. In \cite{anderson1998continuous}, both deterministic continuum and stochastic discrete models of angiogenesis were studied (see also \cite{chaplain2000mathematical}). The authors discretised a continuum model to obtain a stochastic model 
of cell 
movement on a lattice. 

Other approaches that have been used to model angiogenesis include cellular Potts models, where individual cells occupy subdomains on a regular lattice \cite{bauer2007cell}, phase-field models \cite{travasso2011tumor}, models with mechanical interactions \cite{holmes2000mathematical,tosin2006mechanics} and models including further biochemical effects \cite{sleeman2001partial,levine2001mathematicalBMB,levine2001mathematicalJMB,mac2004model,mac2006computational,qutub2009multiscale,jackson2010cell}. Some models of angiogenesis couple angiogenesis with tumour growth 
\cite{breward2003multiphase,orme1996mathematical,chaplain1996avascular,hahnfeldt1999tumor,alarcon2005multiple,owen2009angiogenesis,macklin2009multiscale,frieboes2010three,perfahl2011multiscale}. In such models, the tumour can grow and release AFs, which in turn induce angiogenesis. Models which couple tumour growth with angiogenesis can then be used to investigate treatment strategies \cite{sachs2001simple,mcdougall2002mathematical,d2004tumour,mcdougall2006mathematical,billy2009pharmacologically,jackson2000mathematical}.

With such a variety of approaches available to model angiogenesis, it may be difficult to know which one to choose. While a multiscale model may account for many biological effects, it may also contain many unknown parameters, and be computationally expensive. In practice, in a stochastic model, individual vessels might grow differently when an experiment or simulation is repeated. Even so, the resulting vasculatures might yield similar distributions of nutrients. The question is then whether one needs to know when and where a new blood vessel forms. A simple continuum model, which predicts how much vasculature forms after angiogenesis is induced, might yield sufficient information to determine the resulting oxygen distribution. One problem is that it is not easy to relate 
continuum 
models and 
complex multiscale models. One can simulate both models and compare the resulting vasculatures. However, for a detailed mathematical analysis, it is necessary to know how to relate the parameters that appear in each model.

In this paper, we develop a stochastic, lattice-based model of angiogenesis. As in \cite{anderson1998continuous}, we can relate our model to a continuum one and in this way compare simulations of the continuum and stochastic models, using directly related parameters. Hence, we can study situations in which the continuum model is a good approximation of the stochastic model, and when stochastic or discrete effects become important. In contrast to \cite{anderson1998continuous} we state explicit expressions for terms modelling anastomosis, sprouting and vessel regression, both in the stochastic and the continuum model. To include these effects in a stochastic model, we chose a mesoscopic compartment approach for our model, similar to compartment models used to model diffusing and reacting chemicals (see the review \cite{erban2007practical}). Events such as the emergence of a new tip cell, or the migration of a tip cell, are based on individual cells. However, within each box of our lattice, we count only the number of 
cells of a certain type. This makes it easier to derive explicitly the deterministic continuum limit of the stochastic model.
Furthermore, by focusing on tip migration alone, we demonstrate that different stochastic lattice models can generate the same continuum equations. However, the behaviour of the associated vessel densities may be dependent on the underlying stochastic model. Finally, we note that when using lattice-based models, the choice of the lattice spacing appears as a model parameter, but is not related to any naturally measurable parameter. Hence, it is important to understand how the model scales with the lattice constant. Furthermore, understanding this scaling behaviour is also crucial in taking a consistent continuum limit. For these reasons, we study the scaling of the model with the lattice constant in detail.

We begin the paper by presenting a mesoscopic, stochastic lattice model in section \ref{sec:stochasticModel}. In the associated subsections, we consider the different effects modelled, such as sprouting of a new vessel, tip cell migration, anastomosis or vessel regression. When considering cell migration, we discuss several possible modelling choices. In section \ref{sec:MFequations} we derive the mean field equations of the stochastic model of section \ref{sec:stochasticModel}, and study the scaling behaviour and the continuum limit of the model. In particular, we show that existing PDE models of angiogenesis can only be recovered from the stochastic model when a 
novel approach to random cell movement, as discussed in section \ref{sec:stochasticModelMovement}, is employed. In section \ref{sec:simulationsMovement}, we compare 
different approaches to modelling tip cell movement and the production of new vessels. We also compare these stochastic models to their continuum counterparts, and investigate the dependence of the models on the lattice constant. Then, in section \ref{sec:simulationFullModel}, the full stochastic model is compared to existing PDE models as derived in \cite{balding1985mathematical,byrne1995mathematical}.

\section{Mesoscopic stochastic model}\label{sec:stochasticModel}

In this section, we develop a stochastic, lattice-based model of angiogenesis, formulated on the mesoscale. We distinguish between two inter-related
types of cells, static vessel cells and motile tip cells. When tip cells move, new static vessel cells form behind them. New tips can sprout from existing vessels, and when tip cells encounter another tip or a vessel, anastomosis can occur and the associated tips disappear. Finally, static vessel cells can regress and are removed from the system.
These effects are incorporated into the model via different transition rates $\mathcal{T}$, which influence the master equation that defines the time evolution of the probability density function $P$ for the distribution of tip and vessel cells. The transition rates can depend on the concentration of an angiogenic factor, whose evolution is determined by a reaction-diffusion equation.

For simplicity, we formulate the model using a one-dimensional, Cartesian geometry, decomposing the domain $x\in [0,L]$ into $k_{max}$ equally sized boxes
of length $h$ so that $k_{max}h=L$.\footnote{We choose the coordinates such that $x=kh$ corresponds to the midpoint of box $k$. The precise location of the box midpoint is not essential for the stochastic model, but subtleties arise when combining the stochastic model with PDEs, see section \ref{sec:stochasticChemotaxis}, and when choosing boundary conditions, see appendix \ref{sec:app:BCs}.} We denote by $N_k$ and $R_k$ the number of tip and vessel cells, respectively, in box $k$. We assume that $h$ is considerably larger than a
typical cell size, so that each box typically contains several cells. Furthermore, within each box, we only count the number of cells of each type, and do not track the movement of individual cells within a box.
We remark that the generalisation of our model to higher spatial dimensions is straightforward, see appendix \ref{sec:app:higherDimensions}.

The cell numbers $N_k(t)$ and $R_k(t)$ are stochastic processes whose evolution is determined by the joint probability density function $P=P(\{\{N_{j}\},\{R_{j}\}\},t)$.
Here, $\{N_{j}\} = \{N_1,\dots, N_{k_{max}}\}$ denotes the set of tip cell numbers in all boxes, and likewise 
$\{R_{j}\} = \{R_1,\dots, R_{k_{max}}\}$ for the vessel cells. We usually drop the $t$ dependence in $N_k(t)$ and $R_k(t)$, so that $\{\{N_{j}\},\{R_{j}\}\}$ specifies a state in our stochastic 
model, and $P(\{\{N_{j}\},\{R_{j}\}\},t)$ is the probability that at time $t$ the system is in this state. If we assume that the system is Markovian, then the time evolution of the joint probability density function is described by a master equation (see \cite{van1992stochastic}), which can be written in general form as
\begin{align}
\label{eq:generalMasterEquation}
 \frac{d}{dt}P(\{\{N_{j}\},\{R_{j}\}\},t) = \sum_{\{N_{k}\},\{R_{k}\}}&\left(\mathcal{T}_{\{N_{j}\},\{R_{j}\}|\{N_{k}\},\{R_{k}\}}P(\{\{N_{k}\},\{R_{k}\}\},t) \right. \nonumber\\
 -&\left.\mathcal{T}_{\{N_{k}\},\{R_{k}\}|\{N_{j}\},\{R_{j}\}}P(\{\{N_{j}\},\{R_{j}\}\},t) \right).
\end{align}
In \eqref{eq:generalMasterEquation} we sum over the entire state space, and $\mathcal{T}_{\{N_{k}\},\{R_{k}\}|\{N_{j}\},\{R_{j}\}}$
denotes the transition rate from state $\{\{N_{j}\},\{R_{j}\}\}$ to a state $\{\{N_{k}\},\{R_{k}\}\}$. Intuitively, this means that  
if the system is in state $\{\{N_{j}\},\{R_{j}\}\}$ at a given time $t$,
then $\mathcal{T}_{\{N_{k}\},\{R_{k}\}|\{N_{j}\},\{R_{j}\}}dt$ is the probability that the system switches to state $\{\{N_{k}\},\{R_{k}\}\}$ at time $t+dt$, where $dt$ is an infinitesimal time interval. \\
Equation \eqref{eq:generalMasterEquation} is the general form for a master equation with two spatially varying species $N$ and $R$. Our model of angiogenesis is formulated by 
specifying the non-zero transition rates in Table \ref{tab:differentTransitionRates}, each of which represents a different biological process. 
 \begin{table}[h!]
\caption{Transition rates associated with angiogenesis model.}
\centering
    \begin{tabular}{c|c}
   	 \emph{Process}    & \emph{Transition Rate}   \\ \hline
		\text{tip migration} & $\mathcal{T}_{N_k-1,N_l+1,R_k+\delta_R,R_l|N_k,N_l,R_k,R_l}$\\
 		\text{and vessel production} & $l=k\pm 1$ \\ \hline
   		\text{tip birth due to sprouting} & $\mathcal{T}_{N_{k}+1|N_{k}}$  \\ \hline
    	\text{tip-vessel anastomosis} & $\mathcal{T}_{N_{k}-1|N_{k}}$ \\ \hline
    	\text{tip-tip anastomosis} & $\mathcal{T}_{N_{k}-2|N_{k}}$ \\ \hline
     	\text{vessel regression} & $\mathcal{T}_{R_{k}-1|R_{k}}$
    \end{tabular}
    \label{tab:differentTransitionRates}
\end{table}
The precise functional form of these transition rates will be discussed below.
For concision, when specifying the transition rates, we state only those variables which are involved in the transition. For example, a transition rate
$\mathcal{T}_{N_{k}+1|N_{k}}$, which describes the birth of a tip in box $k$, is an abbreviation for 
\[
 \mathcal{T}_{N_{k}+1|N_{k}} = \mathcal{T}_{\{N_1,\dots,N_{k}+1,\dots,N_{k_{max}}\},\{R_1,\dots,R_{k_{max}}\}|\{N_1,\dots,N_{k},\dots,N_{k_{max}}\},\{R_1,\dots,R_{k_{max}}\}}.
\]
The number of vessel cells $R_l$ is unchanged by this process, as is the number of tip cells in boxes $k=1,\dots,k-1,k+1,\dots,k_{max}$. Hence, these indices are omitted from the subscript of the transition rate.
 
Let $f(N_1,R_1\dots,N_k,R_k,\dots,N_{k_{max}},R_{k_{max}})$ be an arbitrary function of the state variable $R_k,N_k$. Then, we define the shift operators $E_{N_k}^\alpha, E_{R_k}^\alpha$ as follows:
\begin{align}
&E_{N_k}^\alpha f(N_1,R_1,\dots,N_k,R_k,\dots,N_{k_{max}},R_{k_{max}})\nonumber\\
&:=f(N_1,R_1\dots,N_k+\alpha,R_k,\dots,N_{k_{max}},R_{k_{max}}),\nonumber\\
&E_{R_k}^\alpha f(N_1,R_1\dots,N_k,R_k,\dots,N_{k_{max}},R_{k_{max}})\nonumber\\
&:=f(N_1,R_1\dots,N_k,R_k+\alpha,\dots,N_{k_{max}},R_{k_{max}}).
\end{align}
Hence, $E_{N_k}^\alpha$ shifts $N_k$ by $\alpha$. For positive integers, we have $E_{N_k}^\alpha= \left(E_{N_k}^{+1}\right)^\alpha$, and $E_{N_k}^{-1} = \left(E_{N_k}^{+1}\right)^{-1}$ is the inverse operator\footnote{Note that there are subtleties involved with the space on which the operator act. In this paper, we are interested in functions depending on the number of cells in the boxes, which are clearly non-negative, so care has to be taken when acting with $E_{N_k}^{-1}$ on functions with $N_k=0$.}.
Substituting the transition rates specified in Table \ref{tab:differentTransitionRates} into the general form of the master equation \eqref{eq:generalMasterEquation}, we arrive at the 
following master equation for the angiogenesis model:
\begin{align}\label{eq:generalMasterEquationAngiogenesis}
 &\frac{dP(\{\{N_{j}\},\{R_{j}\}\},t)}{dt} = \nonumber\\
&\sum_{k,l\in \nene{k}}(E_{N_k}^{+1}E_{N_l}^{-1}E_{R_k}^{-\delta_R}-1)\mathcal{T}_{N_{k}-1,N_{l}+1,R_{k}+\delta_R,R_{l}|N_{k},N_{l},R_{k},R_{l}}P\nonumber\\
&\text{(tip cell movement)}\nonumber\\
&+\sum_{k}(E_{N_k}^{-1}-1)\mathcal{T}_{N_{k}+1|N_{k}}P\quad\text{(sprouting)}\nonumber\\
&+\sum_{k}(E_{N_k}^{+1}-1)\mathcal{T}_{N_{k}-1|N_{k}}P\quad\text{(tip-vessel anastomosis)}\nonumber\\
&+\sum_{k}(E_{N_k}^{+2}-1)\mathcal{T}_{N_{k}-2|N_{k}}P\quad\text{(tip-tip anastomosis)}\nonumber\\
&+\sum_{k}(E_{R_k}^{+1}-1)\mathcal{T}_{N_{k}+1|N_{k}}P\quad\text{(vessel regression)},
\end{align}
where we drop the arguments of $P$ for concision. Expanding, for example, the last term, we have 
\begin{align}
&\sum_{k}(E_{R_k}^{+1}-1)\mathcal{T}_{N_{k}+1|N_{k}}P(N_1,\dots,N_k,\dots,N_{k_{max}},R_1,\dots,R_{k_{max}})\nonumber\\
&=\sum_{k}\left(\mathcal{T}_{N_{k}|N_{k}-1}P(N_1,\dots,N_k-1,\dots,N_{k_{max}},R_1,\dots,R_{k_{max}})\right.\nonumber\\
&-\left.\mathcal{T}_{N_{k}+1|N_{k}}P(N_1,\dots,N_k,\dots,N_{k_{max}},R_1,\dots,R_{k_{max}})\right),
\end{align}
which is clearly of the form of the terms in \eqref{eq:generalMasterEquation}.

In the transition rate which describes tip movement and the associated production of new vessel cells, $\delta_R=\delta_R(h)$ denotes the number of vessel cells produced behind a moving tip. This depends on the choice of the precise definition of the transition rate and on the lattice constant $h$ in a way discussed in section \ref{sec:MFequations}. We remark that tip birth, anastomosis and vessel pruning are assumed to be local terms, increasing or decreasing the cell number in a particular box. This requires the boxes to be reasonably large, so boundary effects such as anastomosis between vessels in neighbouring boxes are negligible. At the same time, $h$ should not be too large. Otherwise the cells would not be well mixed within a box, and we would need to account for the movement of a tip within a single box. While this would not change the tip cell distribution, it 
could lead to vessel production in a particular box. 

The transition rates for sprouting and tip movement depend on a third model variable, generic angiogenic factor,
whose concentration is denoted by $c$. 
%
so perhaps better to suppress dependence on $x$ and $t$ for now
A large number of angiogenic factors (AFs), including VEGF, have been identified \cite{risau1997mechanisms}, but, for simplicity, here we focus on a single, generic chemical, $c$. 
As we typically have far more molecules of the AF in our system
than numbers of tip or vessel cells, the noise due to fluctuations in the chemical concentration is expected to be
significantly smaller than that associated with the movement, proliferation and death of tip and vessel cells. Hence,
we will view the AF as deterministic and continuous, $c=c(x,t)$, where $x$ is related to the box index $k$ via $x=kh$, so $x\in[0,L]$. 

In principle, the time dependence of $c$ and, hence, the transition rates imply the appearance of additional terms in the master equation. However, these effects are neglected since the timescale of changes in the AF are much shorter than the timescales associated with changes in the blood vessels\footnote{More precisely, in the model of a corneal assay discussed in section \ref{sec:simulationFullModel} and modelled via PDEs in \cite{byrne1995mathematical}, the AF remains close to its steady state configuration and is only slightly influenced by the growing vasculature. Thus, we can assume that the transition rates affecting the endothelial cells remain constant between events associated with the stochastic model. 
This assumption would cease to hold if the source of the AF were itself dynamic and would change on a similar or faster timescale than the timescale of events effecting tip and vessel cells. For instance, we could include tumour cells as a new species in our model, which would act as a source of AF. The tumour, and hence the resulting distribution of AF, could change faster than the blood vessels. Hence, we cannot assume the AF is in a steady state between events affecting tip and vessel cells.}. 
Hence, the AF will be in a quasi-steady state between any two events that occur in the stochastic model for the vessel cells.
In appendix \ref{sec:app:stochasticChemical}, we discuss an alternative approach in which $c$ is treated as a stochastic variable.

In the subsections that follow we introduce functional forms for the transition rates in Table \ref{tab:differentTransitionRates}. We start in subsection \ref{sec:stochasticModelMovement} by discussing different approaches that can be used to model the movement of the tip cells. Then, in subsections \ref{sec:stochasticSprouting}, \ref{sec:stochasticAnastomosis} and \ref{sec:stochasticRegression}, we introduce transition rates that describe sprouting, anastomosis and regression. Throughout this section we keep the lattice size $h$ fixed. In section \ref{sec:MFequations} we derive scaling conditions for the parameters in terms of the lattice size $h$, which allow us to take a continuum limit.

\subsection{Mechanisms of tip cell movement and sprout production}\label{sec:stochasticModelMovement}
We now consider how to model tip cell movement. A key assumption is that new vessel cells are produced behind tip cells along the path that each tip cell describes. A similar assumption was made in earlier continuum models of angiogenesis \cite{balding1985mathematical,byrne1995mathematical}, where it was termed the "snail-trail" approach. We will discuss the relationship between our stochastic model and such continuum models in section \ref{sec:MFequations}. For simplicity, in what follows we ignore vessel maturation and blood flow.

Following \cite{stokes1991analysis,anderson1998continuous,plank2003reinforced}, we distinguish two principal mechanisms of tip cell movement: undirected, random movement and directed movement associated with chemotaxis. 
We will compare two approaches to modelling undirected random movement, both of which yield the same behaviour for the tip cell density in the deterministic limit, but yield different behaviour for the vessel density. The first approach is based on a conventional random walk, similar to that used to model chemical diffusion (see \cite{erban2007practical}), whereas the second one can be interpreted as a random walk with crowding effects.

Chemotaxis is modelled as a biased random walk: the probability of tip cell movement depends on the concentration of the chemoattractant in the boxes involved in the movement, leading to a preferred direction for migration.

\begin{figure}[h!]
\subfloat[before jump]
{
\includegraphics[width=0.48 \linewidth]{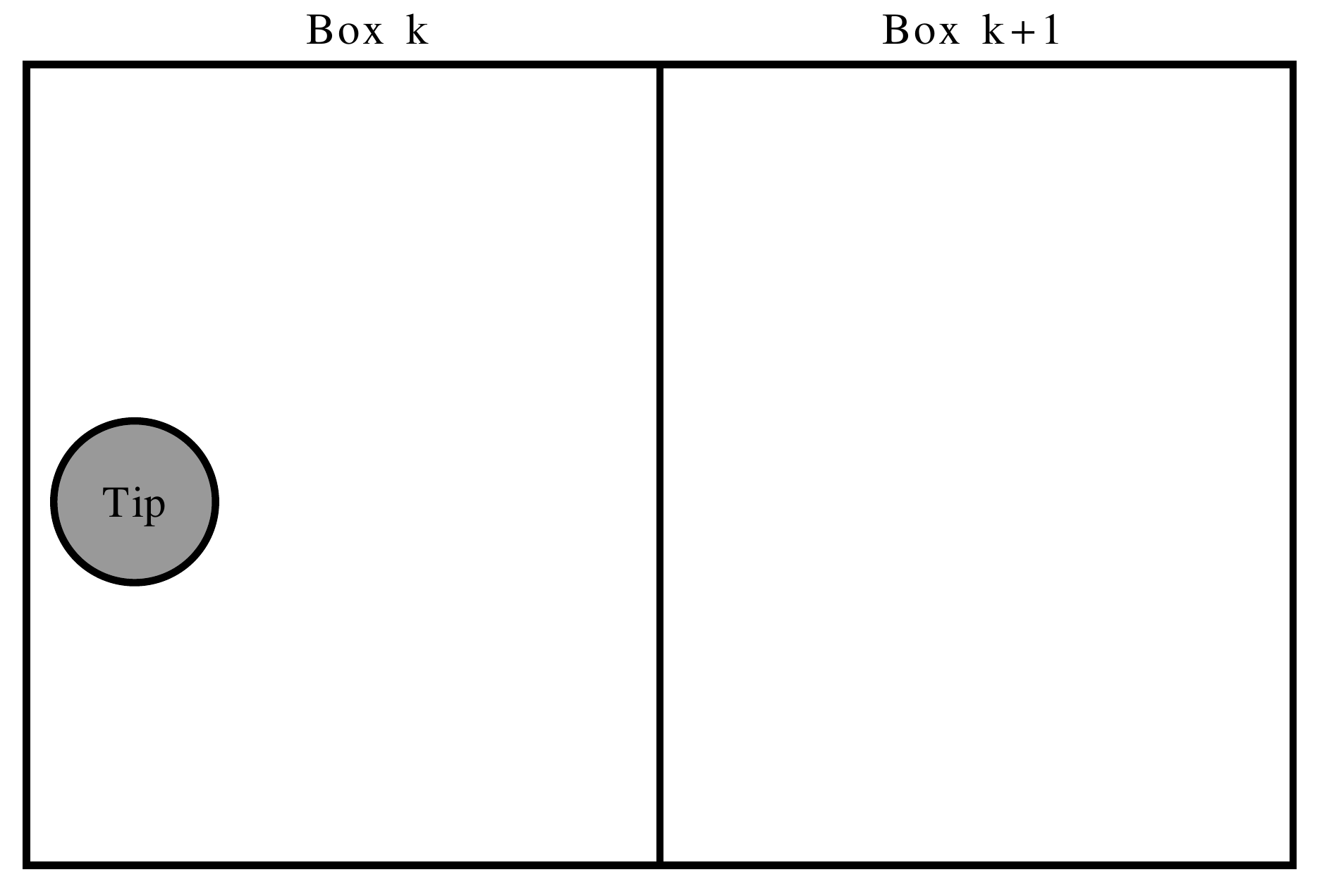}
 }
\subfloat[after jump]
{
\includegraphics[width=0.48 \linewidth]{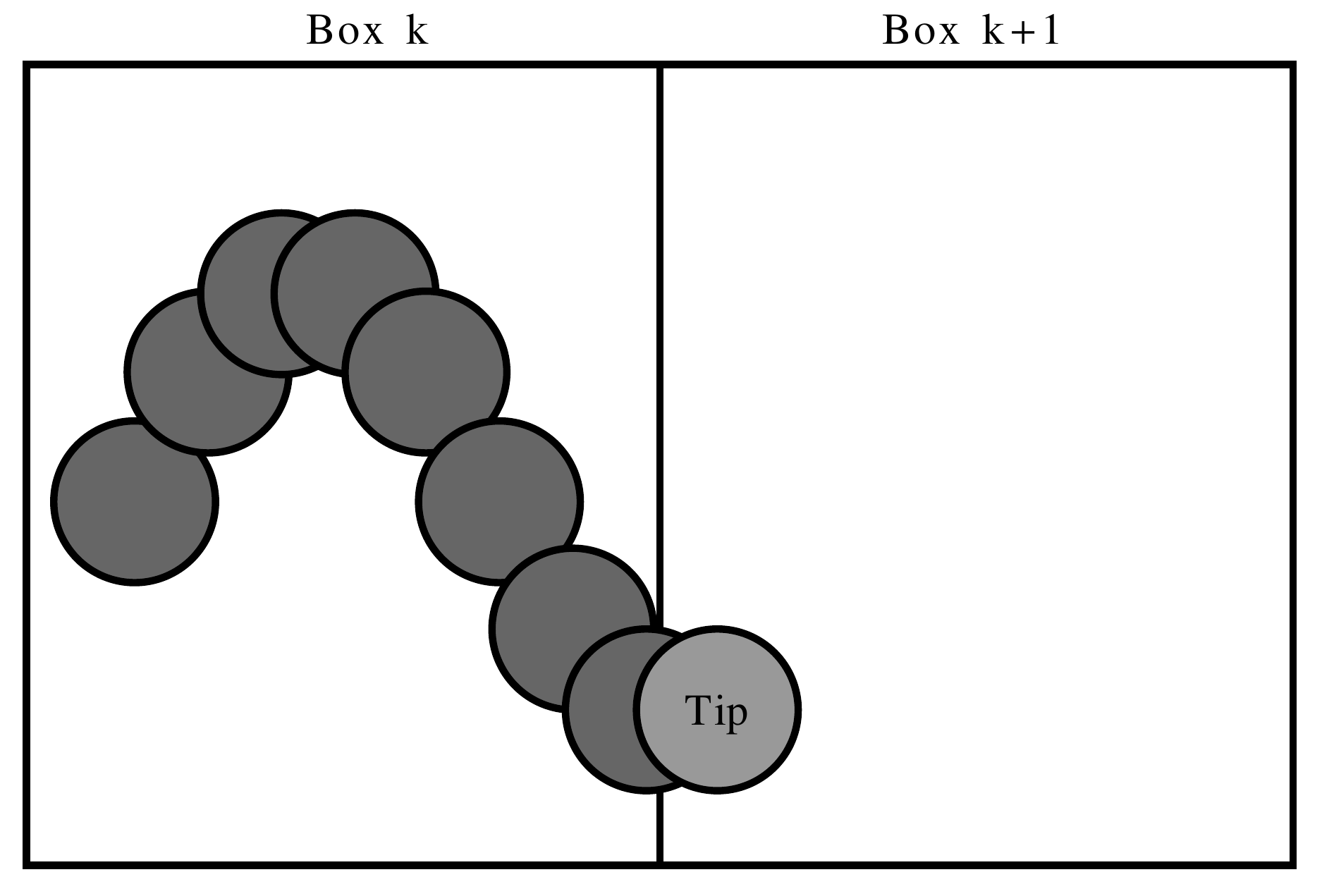}
}\\
\caption{\label{fig:jumpingTip} A tip cell migrates from box $k$ to box $k+1$ leaving a trail of vessel cells in box $k$. }
\end{figure}

For each type of tip cell movement under consideration, the transition rate can be written as
\beq
 \mathcal{T}_{N_k-1,N_l+1,R_k+\delta_R,R_l|N_k,N_l,R_k,R_l},\nonumber
\eeq
with $l=k\pm1$, and the processes may be depicted as in Figure \ref{fig:jumpingTip}.
When a tip cell moves from box $k$ to box $l=k\pm 1$, we assume that $\delta_R$ vessel cells are produced in the outgoing box $k$. We remark that we could have equally assumed that the new vessel cells were produced in the incoming box. For the types of movement under discussion, it will not matter whether the vessels are located in the incoming, outgoing or both boxes. 

\subsubsection{Random movement}\label{sec:randomMovement}
We will now consider two approaches for modelling undirected movement, both of which will lead to the diffusion equation in the macroscopic limit, as will be shown in section \ref{sec:MFrandom}.
\paragraph{Case 1:}
The simplest approach to modelling random movement of tip cells is to assume that the transition rates for the migration of a tip cell
out of box $k$ into any of its nearest neighbour boxes is proportional to the number of tip cells in box $k$. This means that the tip cells move independently of each other. Our principal modelling assumption 
for vessel production is that vessel cells will be produced behind a moving tip. This is analogous to the snail-trail approach used to develop continuum models of angiogenesis \cite{balding1985mathematical,byrne1995mathematical}. Let box $l$ be a nearest neighbour of box $k$, so $l=k\pm 1$. Then the transition rate governing the migration of a single tip cell from box $k$ to box $l$, which also leads to the production of a
number $\delta_R$ of vessel cells in box $k$, is given by
\beq\label{eq:transitionRateDiffusionStandard}
 \mathcal{T}^D_{N_k-1,N_l+1,R_k+\delta_R,R_l|N_k,N_l,R_k,R_l} = \stochDiff N_k.
\eeq
The transition rate is proportional to $N_k$, the number of tip cells in box $k$, the box from which the cell migrates. $\stochDiff$ is the coefficient of proportionality and has units $[\stochDiff]=\frac{1}{\text{time}}$, so the transition rate also has units $[\mathcal{T}]=\frac{1}{\text{time}}$. $\stochDiff$ is assumed to depend on the lattice constant $h$ alone, and is otherwise assumed to be constant. In particular, it does not depend on the cells, which means there is no interaction between the tip cells with other cells. This assumption can easily be relaxed.

\paragraph{Case 2:}
Here we account for crowding effects, assuming that a tip cell in box $k$ will only move to a neighbouring box if the number of tip cells is higher in box $k$ than in the neighbouring box. One might argue that tip cells themselves are not the main obstacle to movement but rather other parts of the tissue, such as normal tissue cells, cancerous cells or extracellular matrix (ECM). The approach chosen here has the advantage that the transition rate depends only on one species, namely the tip cells, and that the ansatz for the transition rate we make below will lead to the same macroscopic evolution equation for the tip cell densities as the ansatz in Case 1. As before, the number of vessel cells produced during tip migration is denoted $\delta_R$. 
However, the scaling dependence on $h$ is not assumed to be the same as for Case 1, and we shall comment on this later. The transition rate is given by
\beq\label{eq:transitionRateDiffusionDifference}
 \mathcal{T}^D_{N_k-1,N_l+1,R_k+\delta_R,R_l|N_k,N_l,R_k,R_l} = \stochDiff\pospart{N_k-N_l},
\eeq
where
\[
 \pospart{x}:=max(x,0).
\]
We could reasonably take other positive, increasing functions of $N_k-N_l$, or, more generally, positive functions of the two arguments $N_k$ and $N_l$ which increase with $N_k$ and decrease with $N_l$. Such monotonicity would imply that tip cells only compete for space and, as such, ignores potentially cooperative behaviour, such as attractive signalling between tip cells, in which case we would need also to allow for increasing functions of $N_l$. The advantage of the current choice for \eqref{eq:transitionRateDiffusionDifference} is that it is piecewise linear and leads to the same mean field equation and continuum limit as transition rate \eqref{eq:transitionRateDiffusionStandard}(see section \ref{sec:MFrandom}\footnote{We can interpret this model as a locally averaged version of Case 1: in Case 1, tip cells can also move into a neighbouring box with a higher number of tip cells, but on average, more tip cells would move from the box with the higher tip cell number to the one with lower number. In Case 2, only the net movement is taken into account.}).

\subsection{Chemotaxis}\label{sec:stochasticChemotaxis}
Recall that chemotaxis is the directed movement of a cell in response to spatial gradients of a diffusible chemical. In the case of tip cells, an important chemoattractant is VEGF \cite{carmeliet2011molecular}. The simplest models of chemotaxis assume a linear dependence of the transition rate on the spatial gradient of the chemoattractant. If the chemical is discretised on the same lattice as the cells, then by gradient we mean the difference in chemical concentrations between neighbouring lattice sites. (See \cite{stevens1997aggregation,hillen2009user} for reviews of chemotactic models). A simple transition rate for chemotaxis is given by
\beq\label{eq:transitionRateChemotaxis}
 \mathcal{T}^\chi_{N_k-1,N_l+1,R_k+\delta_R,R_l|N_k,N_l,R_k,R_l} = \stochChemo N_k \pospart{c_l-c_k},
\eeq
where $c_k$ is the AF concentration in box $k$. The chemotactic coefficient $\stochChemo$ depends on the lattice constant $h$, and is assumed constant and non-negative
(for a chemorepellent, we would have $\stochChemo<0$). We remark that it is straightforward to consider more general forms of $\stochChemo$, for example, $\stochChemo=\stochChemo(N,R,c)$. Further, \eqref{eq:transitionRateChemotaxis} is always non-negative, as required for a transition rate.

So far, we have not specified how $c$ itself evolves. It could be specified externally, or, as part of the model, undergo random movement. As the number of molecules of the AF is typically much larger than the number of tip cells, we expect stochastic fluctuations in $c$ will be much smaller than those associated with the tip cells. Hence, we model the AF as a continuous and deterministic variable, with $c = c(x,t)$ evolving according to a reaction-diffusion equation of the form:
\beq\label{eq:PDEAFgeneral}
\frac{\partial c}{\partial t} = D_c\nabla^2c + g(N,R,c),
\eeq
where $g(N,R,c)$ represents the net rate of production of $c$ ($g\equiv$ (sources) - (sinks)).
We shall discuss in appendix \ref{sec:app:stochasticChemical} how to treat $c$ as a stochastic variable.

There are several ways in which the concentration $c_k$ that appears in equation \eqref{eq:transitionRateChemotaxis} and varies discretely in space can be related to the continuous function $c(x,t)$ obtained by solving \eqref{eq:PDEAFgeneral}. If $x=kh$, then we can fix $c_k = c(kh,t)$. Alternatively, we could sample $c_k$ at the midpoints, rather than the endpoints, of the discrete boxes. We could also choose $c_k$ to be the average concentration in box $k$, $c_k = \frac{1}{h}\int_{box}c(x,t)dx$. In what follows, we assume $c$ is smooth and varies slowly on the lengthscale $h$. Hence, the different definitions of $c_k$ will be equivalent up to terms of higher order in $h$ and, without loss of generality, we define $c_k = c(kh,t)$, with $kh$ being the midpoint of box $k$. Likewise, when solving equation \eqref{eq:PDEAFgeneral}, tip and vessel cell numbers are interpreted locally so that if $x\in [k-\frac{h}{2},k+\frac{h}{2})$, then the source and sink terms contributing at position $x$ are given by $g(N_k,R_k,c(x,t))$. This interpretation of $c_k$ is consistent with centering the boxes associated with the stochastic model.

We have made explicit the dependence of $c$ on the tip and vessel cells, as only these cells are modelled explicitly in the current paper. In practice, the source could, for instance, be a tumour. If tumour cells were modelled explicitly, then $g$ would depend on the tumour cells. In the simulations in section \ref{sec:simulationFullModel} the source of AF is assumed to be localised to the boundary of the domain $x=0$ and the interior of the domain is source-free. The function $g$ will then only contain sink terms so that, for example, we may consider
\beq\label{eq:AFSourceTerm}
g(N,R,c)(x,t) = -\lambda c - a_1 H(c(x,t)-{\hat c})\frac{N_k(t)}{h}c(x,t),
\eeq
where, as before, $kh=x$ and $H(x)$ represents the Heaviside step function
($H(x)=1$ for $x\geq0$ and $H(x)=0$ for $x<0$). We assume that $c$ is degraded at a constant rate $\lambda$ and that it binds irreversibly to tip cells with constant of proportionality $a_1$ provided that $c > {\hat c}$. Further motivation for this functional form for $g$ is provided in section \ref{sec:stochasticSprouting}. We remark that other functional forms for $g$ could straight-forwardly be implemented.

\subsection{Combining random and directed movement}\label{sec:TransitionRateTotalMovement}
We will now consider a simple model of cell movement which combines undirected random movement with chemotaxis. We treat these effects separately so that the full model simply comprises two transition rates $T^D_{N_k-1,N_l+1,R_k+\delta_R^{\stochDiff},R_l|N_k,N_l,R_k,R_l}$ and $T^\chi_{N_k-1,N_l+1,R_k+\delta_R^{\stochChemo},R_l|N_k,N_l,R_k,R_l}$,
where $T^D$ is the transition rate for random motion which is given by \eqref{eq:transitionRateDiffusionStandard} for Case 1, and \eqref{eq:transitionRateDiffusionDifference} for Case 2, and $T^\chi$ denotes the transition rate for chemotaxis, \eqref{eq:transitionRateChemotaxis}. 
We remark that we distinguish $\delta_R^{\stochDiff}$ and $\delta_R^{\stochChemo}$, as they are, in general, different. However, if  $\delta_R^{\stochDiff} = \delta_R^{\stochChemo}$, then we could combine the two transition rates into one (we will show in section \ref{sec:MFequations} that this situation arises for Case 2).
The simplest approach is to say that the total transition rate for a jump is given by
\beq\label{eq:transitionDiffPlusChemo1}
 \mathcal{T}_{N_k-1,N_l+1,R_k+\delta_R,R_l|N_k,N_l,R_k,R_l} = \stochDiff\pospart{N_k-N_l} + \stochChemo N_k \pospart{c_l-c_k}.
\eeq
Here, we take the positive parts of the discrete gradients of $c$ and $N$ individually, so there is no cancellation when the two terms oppose each other. Hence, a tip cell can migrate down a strong gradient of $c$, as long as the discrete gradient of $N$ also points downwards.

Alternatively, we could combine the transition rates such that
\beq\label{eq:transitionDiffPlusChemo2}
 \mathcal{T}_{N_k-1,N_l+1,R_k+\delta_R,R_l|N_k,N_l,R_k,R_l} = \pospart{\stochDiff(N_k-N_l) + \stochChemo N_k (c_l-c_k)}.
\eeq
In this case a jump occurs only when the combined random and chemotactic term is postive. We shall see in section \ref{sec:MFcombinedMovement} that transition rates \eqref{eq:transitionDiffPlusChemo1} and \eqref{eq:transitionDiffPlusChemo2} lead to slightly different continuum models. This will clarify the sense in which chemotactic and random (diffusive) fluxes can reinforce or oppose each other.

\subsection{Sprouting}\label{sec:stochasticSprouting}

\begin{figure}[h!]
\subfloat[before sprouting]
{
\includegraphics[width=0.48\linewidth]{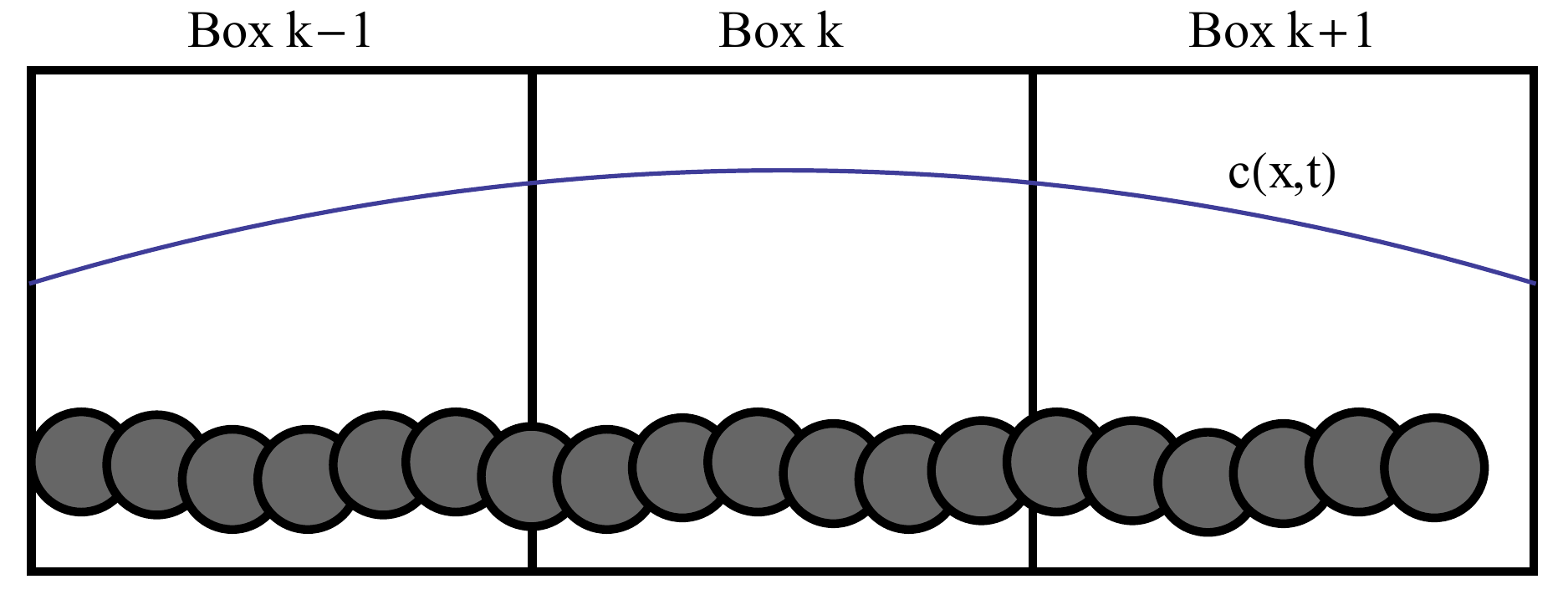}
}
\subfloat[after sprouting]
{
\includegraphics[width=0.48\linewidth]{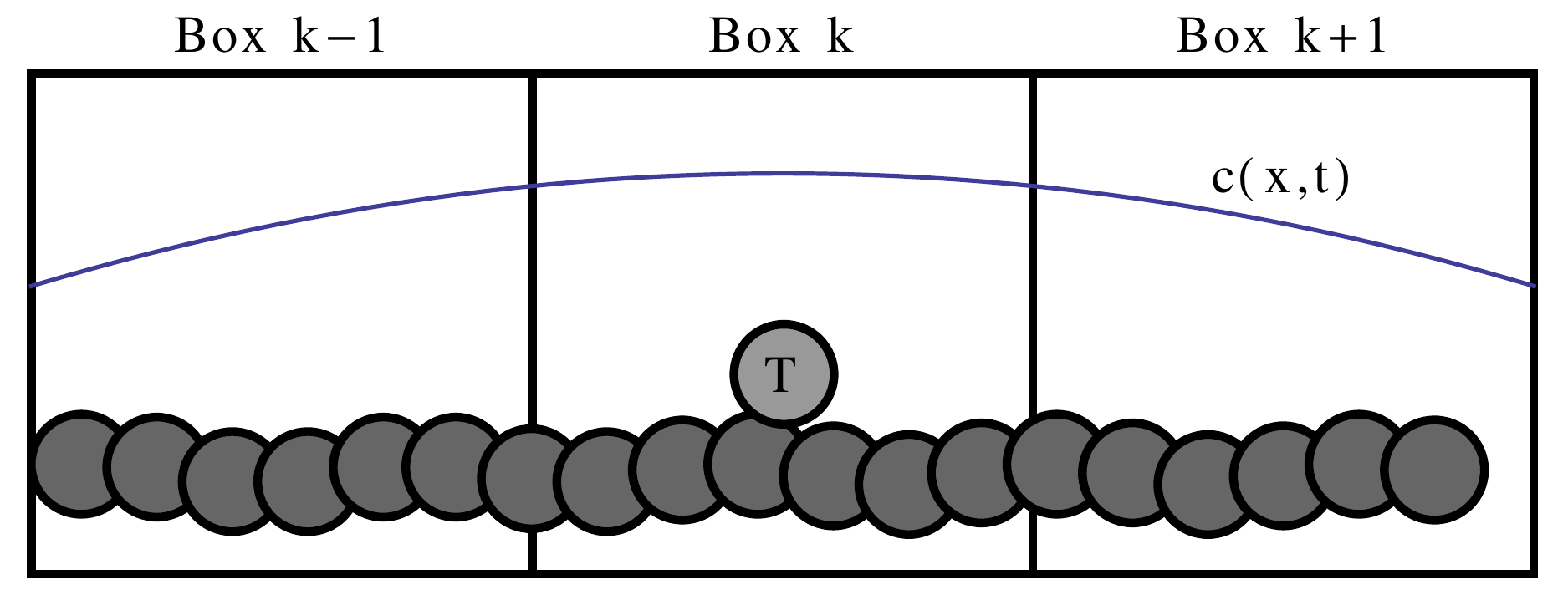}
}
\caption{\label{fig:sprouting} In response to the AF $c$, a new tip cell (T) emerges from an existing vessel.}
\end{figure}

The probability of sprouting is assumed to depend on the local concentration of $c_k$. Such behaviour has been shown for VEGF (see \cite{gerhardt2003vegf}). Sprouting from an existing vessel in our model is depicted in Figure \ref{fig:sprouting}.
We suppose that, as the box size of our model is larger than the size of a single cell, the newly formed tip cell is created in the same box as that occupied by the vessel from which it sprouts. This means that the transition rate has the structure $\mathcal{T}_{N_{k+1}|N_k}$. The simplest ansatz for this transition rate is that the rate is proportional to the number of vessel cells in box $k$ and to the concentration of the AF $c$. Alternatively, new tip cells may emerge from an existing tip cell. Assuming such events are mediated by the same AF $c$, we can incorporate this effect into transition rate $\mathcal{T}_{N_{k+1}|N_k}$ via a term proportional to the product $N_kc_k$. Following \cite{byrne1995mathematical}, we assume further that tip cell proliferation only happens when $c$ exceeds a threshold value, $\hat{c}$. Combining the above assumptions we obtain the following expression for the transition rate:
\beq\label{eq:transitionRateSprouting}
\mathcal{T}_{N_{k+1}|N_k} = {\tilde a_0} R_k c_k +{\tilde a_1} N_k c_k H(c_k-\hat{c}).
\eeq
While in general the coefficients ${\tilde a_0},{\tilde a_1}$ may depend on the lattice scaling, we will show in section \ref{sec:MFsproutingAnastomosisRegression} that ${\tilde a_0}$ and ${\tilde a_1}$ are independent of $h$.

\subsection{Anastomosis}\label{sec:stochasticAnastomosis}

\begin{figure}[h!]
\subfloat[before anastomosis]
{
\includegraphics[width=0.48\linewidth]{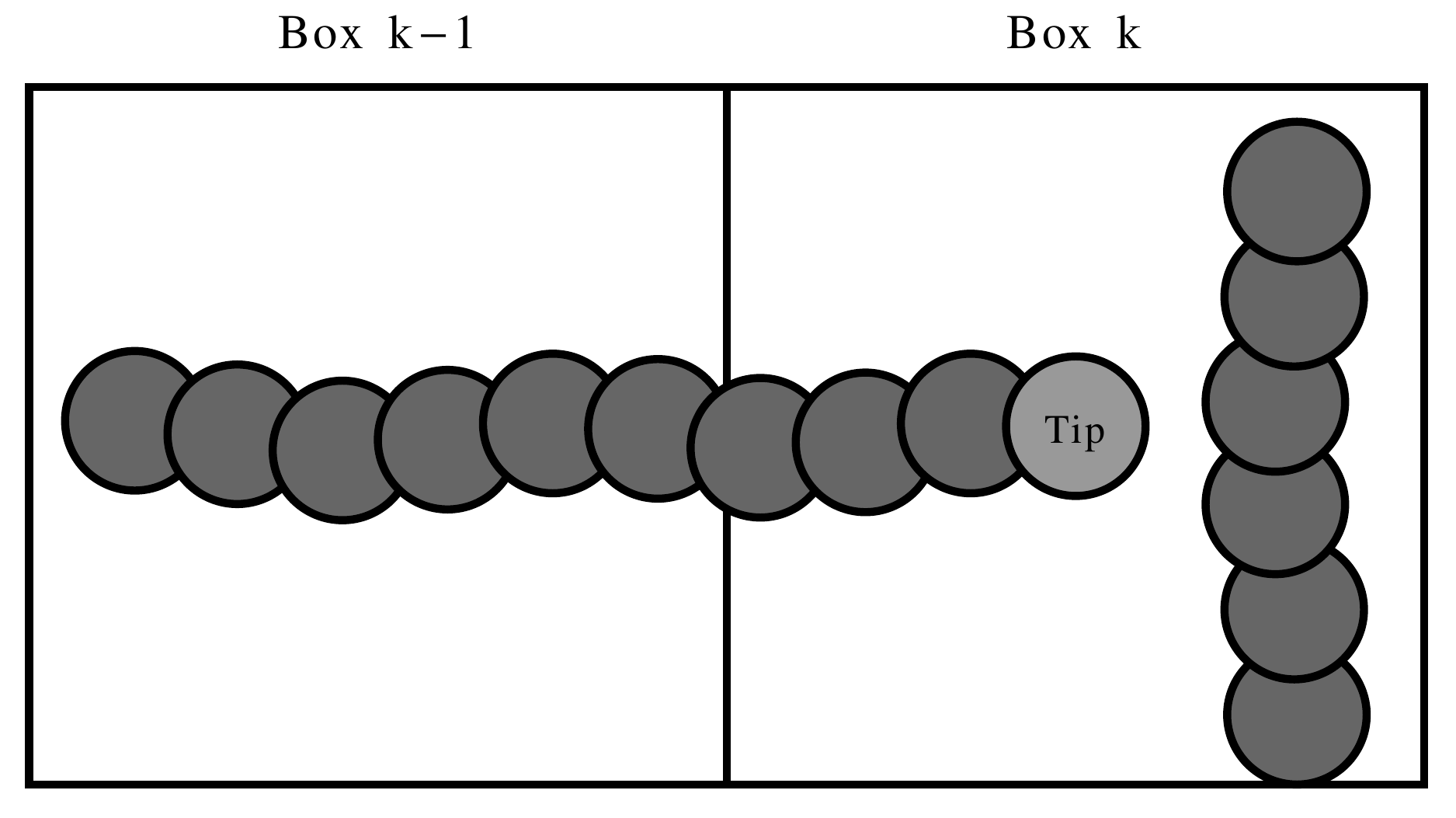}
}
\subfloat[after anastomosis]
{
\includegraphics[width=0.48\linewidth]{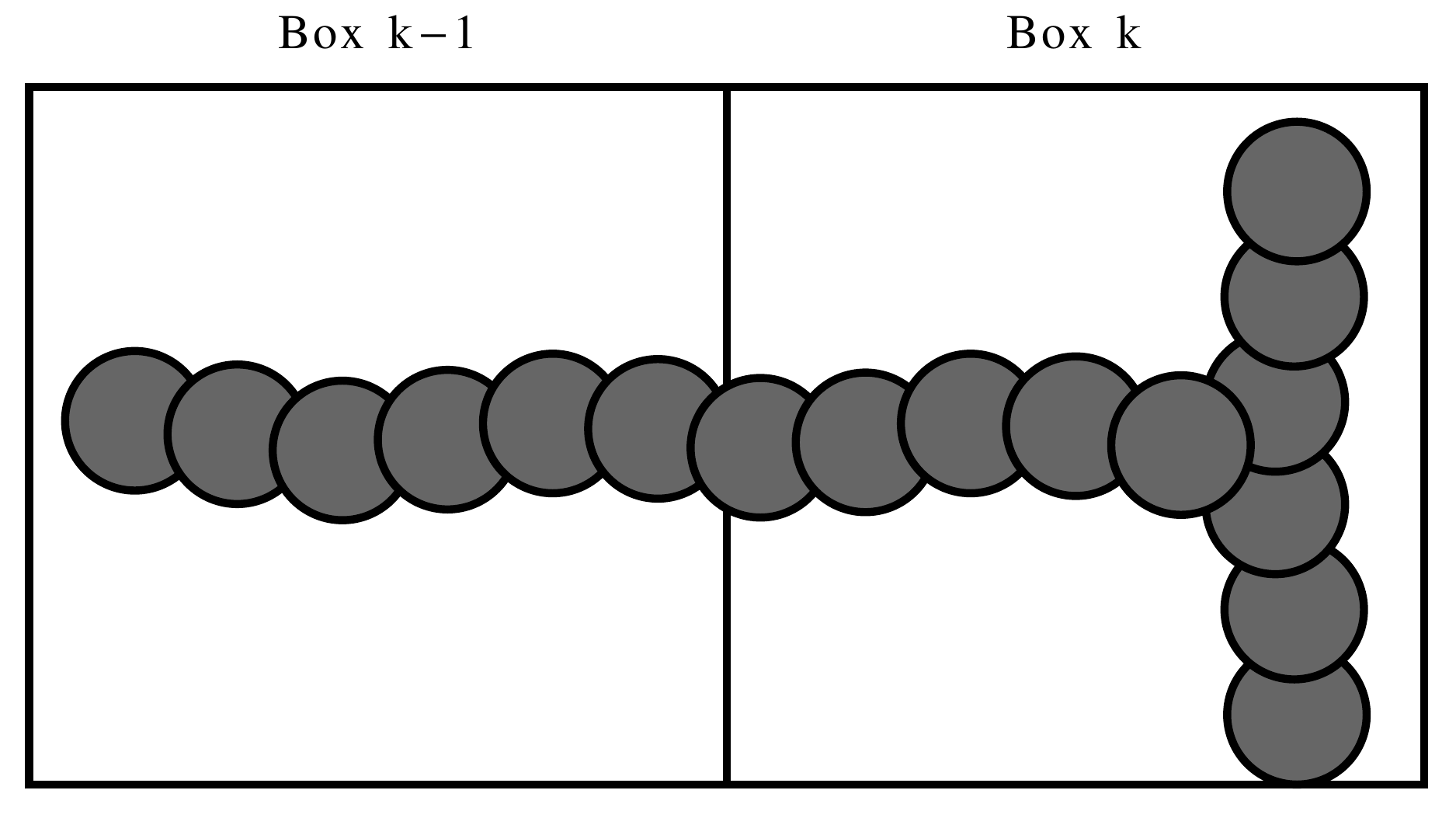}
}
\caption{\label{fig:anastomosis}A schematic diagram of anastomosis: when a tip cell fuses with a vessel, the tip cell disappears. }
\end{figure}

When a new vessel encounters another vessel, a new connected loop can form. This process is called anastomosis and is depicted in Figure \ref{fig:anastomosis}.
We view anastomosis as an event which is localised in a specific box and whose probability of occurrence is proportional to the number of tip cells and the number of vessel cells in that box. In what follows, it will be convenient to assume that when anastomosis takes place, the tip cell disappears from the system. The transition rate for anastomosis is then given by
\beq\label{eq:transitionRateAnastomosis}
\mathcal{T}_{N_{k-1}|N_k} = {\tilde \beta_1} N_k R_k,
\eeq
where ${\tilde \beta_1}$ depends on the lattice constant $h$, as will be discussed in section \ref{sec:MFsproutingAnastomosisRegression}. Anastomosis will also occur when two tip cells meet and the two sprouts behind them connect. Such events are modelled by transition rates of the form
\beq\label{eq:transitionRateAnastomosisTipTip}
\mathcal{T}_{N_{k-2}|N_k} = {\tilde \beta_2} N_k (N_k-1).
\eeq
Since there will typically be more vessel cells in a given box than tip cells, we assume that the probability of tip-tip anastomosis events is much smaller than tip-vessel events. A small value of ${\tilde \beta_2}$ would act as an additional sink term for tips and hence decrease the total number of tips. 
However, in the absence of suitable experimental data and given that tip-tip anastomosis seems typically less likely than tip-vessel anastomosis, we simplify our model by setting ${\tilde \beta_2}=0$ in what follows.

\subsection{Vessel regression}\label{sec:stochasticRegression}

There are two main reasons for vessel regression: Unperfused sprouts regress if they do not anastomose within a fixed time period after they form. Secondly, perfused vessels are pruned away if there is low shear-stress on their walls due to small blood flow \cite{resnick2003fluid}. In our model, we do not take blood flow into account, so we do not distinguish between perfused and unperfused vessels. Hence, both processes lead to a loss of vessel cells, $R_k$. In general, vessel regression is a complex process involving several AFs and interactions with other cell types such as pericytes and smooth muscle cells \cite{holash1999new}. For simplicity, we model vessel regression by the transition rate
\beq
\label{eq:transitionRateRegression1}
\mathcal{T}_{R_{k-1}|R_k} = {\tilde \gamma} R_k,
\eeq
where the positive constant ${\tilde \gamma}$ describes the rate of vessel regression. Since our model does not take into account vessel length or the age of a new vessel sprout, for simplicity we take ${\tilde \gamma}$ constant. Our modelling framework makes it easy to generalise the above transition rate by including more complex functional forms for the transition rate above.

%
\section{Mean field equations, scaling and continuum limit}\label{sec:MFequations}

In this section, we derive the mean field equations for the mesoscopic stochastic models developed in section \ref{sec:stochasticModel} and the scaling dependence of the model on the lattice constant $h$, and subsequently take the continuum limit. We compare the resulting equations to established continuum models of angiogenesis. The first step, finding the mean field equations, amounts to determining the equations governing the time evolution of the means $\mean{N_k}$ and $\mean{R_k}$ of our state variables $N_k$ and $R_k$ for $k=1,\dots,k_{max}$. The means are defined in the standard way as
\begin{align}
\mean{N_{k}} &= \sum_{\{N_{j}\},\{R_{j}\}}N_k P(\{\{N_{j}\},\{R_{j}\}\},t),\nonumber\\
\mean{R_{k}} &= \sum_{\{N_{j}\},\{R_{j}\}}R_k P(\{\{N_{j}\},\{R_{j}\}\},t).
\end{align}
Then, the time evolution of the mean values is given by
\begin{align}\label{eq:generalMFequation}
\frac{d \mean{N_{k}}}{d t} &= \sum_{\{N_{j}\},\{R_{j}\}}N_k \frac{dP(\{\{N_{j}\},\{R_{j}\}\},t)}{dt},\nonumber\\
\frac{d\mean{R_{k}}}{d t} &= \sum_{\{N_{j}\},\{R_{j}\}}R_k \frac{dP(\{\{N_{j}\},\{R_{j}\}\},t)}{dt}.
\end{align}
We then substitute in the expression for $\frac{dP(\{\{N_{j}\},\{R_{j}\}\},t)}{dt}$ from equation \eqref{eq:generalMasterEquation}. In general, the transition rates entering the master equation are nonlinear. To obtain a closed set of equations for the time evolution of the means, we 
also assume that the mean field approximation is valid. This approximation consists of substituting for moments arising in \eqref{eq:generalMFequation} the product of means. For example, $\mean{N_{k_1}^{n_1}N_{k_2}^{n_2}R_{k_3}^{n_3}R_{k_4}^{n_4}}=\mean{N_{k_1}}^{n_1}\mean{N_{k_2}}^{n_2}\mean{R_{k_3}}^{n_3}\mean{R_{k_4}}^{n_4}$ for positive integers $n_1,n_2,n_3$ and $n_4$. More rigorously, we perform a system size expansion of our model using the method of van Kampen \cite{van1992stochastic}, and obtain the mean field equations at the leading order of the expansion. Note also that the means $\mean{N_{k}}$ and $\mean{R_{k}}$ depend explicitly on time through the probability density function $P$. Hence, these equations will form $2 k_{max}$ coupled ordinary differential equations (ODEs) for the variables $\mean{N_1},\dots,\mean{N_{k_{max}}},\mean{R_1},\dots,\mean{R_{k_{max}}}$.

We now derive general expressions for the mean field equations and the continuum limit, exploiting the general structure of the transition rates which appeared in the stochastic model of angiogenesis in section \ref{sec:stochasticModel}, but which are not dependent on the specific functional form of the transition rates. We write the mean field equations as 
\begin{align}\label{eq:generalMFequationAllTerms}
\frac{d \mean{N_{k}}}{d t} &= \mathcal{E}^{N_k}_M+\mathcal{E}^{N_k}_S+\mathcal{E}^{N_k}_A+\mathcal{E}^{N_k}_R,\nonumber\\
\frac{d\mean{R_{k}}}{d t} &= \mathcal{E}^{R_k}_M+\mathcal{E}^{R_k}_S+\mathcal{E}^{R_k}_A+\mathcal{E}^{R_k}_R,
\end{align}
where the symbol $\mathcal{E}$ denotes a particular event, the superscript denotes the cell type and box to which the particular event contributes, and the subscripts $M,S,A$ and $R$ denote movement, sprouting, anastomosis and regression, respectively.
The terms in \eqref{eq:generalMFequationAllTerms} are of the form
\begin{align}\label{eq:meanFieldAll}
\mathcal{E}^{N_k}_M &= \sum_{\{N_{j}\},\{R_{j}\}}\sum_{l\in \nene{k}}(\mathcal{T}_{N_{l}-1,N_{k}+1,R_{l}+\delta_R,R_k |N_{l},N_{k},R_{l},R_{k}}\nonumber\\
 &-\mathcal{T}_{N_{k}-1,N_{l}+1,R_{k}+\delta_R,R_{l} |N_{k},N_{l},R_{k},R_{l}})P,\nonumber\\
 \mathcal{E}^{R_k}_M &= \sum_{\{N_{j}\},\{R_{j}\}}\sum_{l\in \nene{k}}\delta_R\mathcal{T}_{N_{k}-1,N_{l}+1,R_{k}+\delta_R,R_{l} |N_{k},N_{l},R_{k},R_{l}}P,\nonumber\\
 \mathcal{E}^{N_k}_S &= \sum_{\{N_{j}\},\{R_{j}\}} \mathcal{T}_{N_{k}+1|N_{k}}P,\qquad \mathcal{E}^{R_k}_S = 0, \nonumber\\
  \mathcal{E}^{N_k}_A &= -\sum_{\{N_{j}\},\{R_{j}\}} \mathcal{T}_{N_{k}-1|N_{k}}P,\qquad \mathcal{E}^{R_k}_A = 0,\nonumber\\
   \mathcal{E}^{N_k}_R &= 0,\qquad \mathcal{E}^{R_k}_R = -\sum_{\{N_{j}\},\{R_{j}\}} \mathcal{T}_{R_{k}-1|R_{k}}P.
\end{align}
The derivation of these terms can be found in appendix \ref{sec:app:derivationMeanField}. The structure does not depend on the precise functional form of the transition rates, but only on how they affect the state variables $N_k, R_k$. The precise functional dependence will be discussed below.

\subsection{Scaling and Continuum limit}

From the mean field equations for the discretised mean fields $\mean{N_k}$ and $\mean{R_k}$ for $k=1,\dots,k_{max}$, we will now take the continuum limit. As before, we perform all calculations in spatial dimension $d=1$, so we transform the discrete box index $k$ to a new variable $x\in [0,L]$, with $L=hk_{max}$. We can think of the continuum limit as taking the limits $h\to 0$, $k_{max}\to\infty$, with $L$ kept fixed. In order to do this, we need to determine how our model parameters scale with $h$.  

We now discuss and compare the continuum limit for the various parts of the model discussed in section \ref{sec:stochasticModel}. The continuum variables are the tip cell density $n(x,t)$, the vessel cell density $\rho(x,t)$ and the concentration of AF $c(x,t)$.
Note that the AF $c$ was already treated as continuous in section \ref{sec:stochasticModel}. Hence, in the continuum limit of the stochastic model, $c$, $n$ and $\rho$ are all deterministic, continuum variables whose evolution is governed by PDEs. To translate between tip and vessel cell densities in a continuum model, and cell numbers, we define
\begin{align}
 n(x,t)=\frac{\mean{N_k(t)}}{h}, \quad \rho(x,t)=\frac{\mean{R_k(t)}}{h},\nonumber
\end{align}
where $x=kh$ corresponds to the midpoint of box $k$. Note that in dimensions $d>1$, we should replace $\frac{1}{h}$ by the inverse of the volume of a $d$-dimensional hypercube, $\frac{1}{h^d}$. We write the continuum equations in the general form
\begin{align}\label{eq:PDEgeneralAllTerms}
 \frac{\partial n(x,t)}{\partial t} &= \epsilon^{n}_M+\epsilon^{n}_S+\epsilon^{n}_A+\epsilon^{n}_R,\nonumber\\
 \frac{\partial \rho(x,t)}{\partial t} &= \epsilon^{\rho}_M+\epsilon^{\rho}_S+\epsilon^{\rho}_A+\epsilon^{\rho}_R,
\end{align}
with the event terms $\epsilon$ corresponding to the event terms in the discrete mean field equations \eqref{eq:generalMFequationAllTerms}.
In the following subsection, we will go through the individual model components, as we did in section \ref{sec:stochasticModel}, and discuss for each case the mean field equations, the scaling behaviour and the continuum limit. 

\subsection{Random movement}\label{sec:MFrandom}

We contrast the mean field equations for the models based on the different transition rates describing the movement of tip cells and subsequent production of vessel cells. We begin by discussing the rates given by \eqref{eq:transitionRateDiffusionStandard} and \eqref{eq:transitionRateDiffusionDifference}, both of which describe random movement.

\paragraph{Case 1:}

To derive the contribution to the mean field equations from the model with transition rate \eqref{eq:transitionRateDiffusionStandard}, $\mathcal{T}_{N_k-1,N_l+1,R_k+\delta_R,R_l|N_k,N_l,R_k,R_l} = \stochDiff N_k$, we simply substitute this transition rate into \eqref{eq:meanFieldTipsGeneral} and \eqref{eq:meanFieldVesselsGeneral} and obtain:
\begin{align}\label{eq:meanFieldDiffusionStandard}
\mathcal{E}^{N_k}_M &= \stochDiff\left(\mean{N_{k+1}} + \mean{N_{k-1}} - 2 \mean{N_k}\right),&\quad \mathcal{E}^{R_k}_M &= b^R_1\mean{N_k}.
\end{align}
These equations are valid for $k=2,\dots,k_{max}-1$, and we introduced $b^R_1 = 2\delta_R \stochDiff$. To perform the continuum limit $h \to 0$, we need to determine how the parameters $\stochDiff$, $\delta_R$ and $b^R_1$ scale with $h$. We start by investigating the number of vessel cells $\delta_R$ produced behind a moving tip. 
If tip cells move linearly, then a jump would represent a path of average length of the lattice spacing $h$. Hence, if $\mu$ denotes the typical diameter of a vessel cell, during a jump, $\delta_R = \frac{h}{\mu}$ cells would be produced. However, our mathematical model defined by
 \eqref{eq:transitionRateDiffusionStandard} is a mesoscopic model of tip cells following Brownian motion. It is a well known fact that the fractal dimension of the path of Brownian motion in dimensions $d=2$ or higher is $2$ (see \cite{falconer2007fractal}). As the vessels are produced behind the moving tip, the vessel length and hence the number of vessel cells produced should have the same scaling behaviour as the tip path length. This implies we have $\delta_R \propto h^{2}$. The constant of proportionality is the typical vessel cell size $\mu$, so  
\beq
\delta_R = \frac{h^2}{\mu^{2}}.\nonumber
\eeq
Similarly, Brownian motion implies a scaling of $\stochDiff\propto\frac{1}{h^2}$, so we set
 \beq
 \stochDiff=\frac{D}{h^2},
 \eeq
where $D$ is the macroscopic diffusion coefficient. This implies that the constant $b^R_1 = 2 \stochDiff \delta_R = \frac{2D}{\mu^2} $ does not depend on $h$.

The continuum limit of \eqref{eq:meanFieldDiffusionStandard} gives contributions to \eqref{eq:PDEgeneralAllTerms} as follows:
\begin{align}\label{eq:PDEDiffusionStandard}
 \epsilon^n_M &= D\Delta n,&  \epsilon^\rho_M &= b^R_1 n.
\end{align}
Here, we have written the second spatial derivative as the one-dimensional Laplace operator, $\Delta = \nabla^2 =\frac{\partial^2}{\partial x^2}$. Written this way, \eqref{eq:PDEDiffusionStandard} will remain valid in higher dimensions. Only the 
constant $b^R_1$ will require a slight modification: in $d$ dimensions, it will be given by $b^R_1 =  \frac{2dD}{\mu^2}$.
We see that our scaling arguments lead to well defined PDEs in the continuum limit. The corresponding boundary conditions will be discussed, together with the boundary conditions of the stochastic model, in appendix \ref{sec:app:BCs}. The tip cell densities in \eqref{eq:PDEDiffusionStandard} follow the diffusion equation, and the vessel density at any point in space changes with time at a rate proportional to the local density of tip cells. Again, we emphasise that it is not clear {\it a priori} that the continuum limit \eqref{eq:PDEDiffusionStandard} yields biologically relevant results, as in the original stochastic model, we should not have a lattice constant $h$ which is smaller than the typical cell size. We will compare simulations of the continuum and the stochastic equations in section \ref{sec:simulationsMovement} to confirm when the continuum limit is a reasonable approximation of our system.

\paragraph{Case 2:}
Similarly to Case 1, transition rate \eqref{eq:transitionRateDiffusionDifference}, $\mathcal{T}_{N_k-1,N_l+1,R_k+\delta_R,R_l|N_k,N_l,R_k,R_l} = \stochDiff \pospart{(N_k-N_l)}$, leads to contributions to the mean field equations \eqref{eq:generalMFequationAllTerms}
\begin{align}\label{eq:meanFieldDiffusionDifference}
\mathcal{E}^{N_k}_M  &= \stochDiff\left(\mean{N_{k+1}} + \mean{N_{k-1}} - 2 \mean{N_k}\right),\nonumber\\
\mathcal{E}^{R_k}_M &=  \delta_R \stochDiff \left(\pospart{\mean{N_k}-\mean{N_{k+1}}} + \pospart{\mean{N_k}-\mean{N_{k-1}}}\right).
\end{align}
As for Case 1, these equations are valid in the bulk of the domain, $k=2,\dots, k_{max}-1$. Note that the mean field equation for the tips is the same as in \eqref{eq:meanFieldDiffusionStandard}, whereas the equation for the mean number of vessels in box $k$ now depends on the difference in the concentration of tips in neighbouring boxes, in contrast to \eqref{eq:meanFieldDiffusionStandard}, where the dependence was only on the number of tips in the same box, $\mean{N_{k}}$.

To understand the scaling of the number of vessels $\delta_R$ produced by a jumping tip cell, imagine one has a locally constant discrete gradient of tip cells. Transition rate \eqref{eq:transitionRateDiffusionDifference} implies that tip cells will only be migrating in one direction, namely, up the gradient. This means that the tip path, and thus also the length of the growing vessel, will be locally linear and scale linearly with $h$. Hence, no fractal path as in Case 1 will form. The number of vessel cells produced is then
\beq
\delta_R =\frac{h}{\mu}.
\eeq
The coefficient in front of the transition rate, $\stochDiff$, scales in the same way as in Case 1, 
\beq
\stochDiff = \frac{D}{h^2}.
\eeq
This is to be expected as the mean field equations for the tip cells \eqref{eq:meanFieldDiffusionDifference} are identical to the mean field equations \eqref{eq:meanFieldDiffusionStandard} in Case 1. 

The continuum limit of \eqref{eq:meanFieldDiffusionDifference} leads to a contribution to \eqref{eq:PDEgeneralAllTerms} of the form
\begin{align}\label{eq:PDEDiffusionDifference}
  \epsilon^n_M &= D\Delta n,\nonumber\\
  \epsilon^\rho_M &=  b^R_2\left|\nabla n\right|_1.
\end{align}
Here, we have used $\left|\nabla n\right|_1 = \left|\frac{\partial n}{\partial x}\right|$ in one spatial dimension. Again, the result written in terms of $\nabla$ remains valid in higher dimensions, as will be shown in appendix \ref{sec:app:higherDimensions}. We introduced a new constant $b^R_2 = \stochDiff \delta_R h = \frac{D}{\mu}$, which is independent of $h$. In the derivation, we have used $\pospart{f}-\pospart{-f}=|f|$, and the subscript $|f|_1$ denotes the $L_1$ norm, again indicating the correct generalisation to higher spatial dimensions.

As for Case 1, the continuum limit for the evolution of the mean of the densities of tips gives the diffusion equation, whereas the production of vessels depends on the norm of the gradient of tip density. Hence, on the macroscale when the deterministic continuum limit is valid, tips will behave in exactly the same way for this model as in Case 1. However, the way vessels are produced is quite different. Imagine the situation that the tip density is constant, but non-zero. In Case 1, there will still be vessel production, as tips will still move. Only on average one has as many left-moving as right-moving tips, so there is no change in macroscopic tip density. For Case 2 however, there is no tip movement at all, and hence no vessel production. We will investigate the differences between the two cases, as well as the difference between the stochastic models and their continuum limits, in more detail in section \ref{sec:simulationsMovement}.

\subsection{Chemotaxis}\label{sec:MFchemotaxis}

The derivation of the contribution to the mean field equations from chemotaxis, based on transition rate \eqref{eq:transitionRateChemotaxis}, $\mathcal{T}_{N_k-1,N_l+1,R_k+\delta_R,R_l|N_k,N_l,R_k,R_l} = N_k \pospart{\stochChemo(c_l-c_k)}$, follows in the same fashion from \eqref{eq:meanFieldTipsGeneral} and \eqref{eq:meanFieldVesselsGeneral} as the mean field equations for random movement for Cases 1 and 2. We obtain in \eqref{eq:generalMFequationAllTerms}
\begin{align}\label{eq:meanFieldChemotaxis}
\mathcal{E}^{N_k}_M  &= -\stochChemo\mean{N_{k+1}}(c_{k+1} + c_{k-1} - 2 c_k)\nonumber\\
&+\stochChemo \left(\mean{N_{k+1}}-\mean{N_{k}}\right)\pospart{c_k-c_{k+1}}+\stochChemo\left(\mean{N_{k-1}}- \mean{N_{k}}\right)\pospart{c_{k} - c_{k-1}},\nonumber\\
\mathcal{E}^{R_k}_M &=  \delta_R \stochChemo \mean{N_k}\left(\pospart{c_{k-1} - c_k}+\pospart{c_{k+1} - c_k}\right).
\end{align}
The scaling arguments for the number of vessel cells produced, $\delta_R$, are similar to those for Case 2 of the random movement transition rate: a locally constant concentration gradient of the AF $c$ implies linear movement of the tip cells up the gradient, so the tip cell path length will scale linearly with $h$. Hence, the number of vessel cells produced will be
\beq
\delta_R =\frac{h}{\mu}.
\eeq
Likewise, locally linear movement of tip cells implies that $\stochChemo = \frac{\chi}{h^2}$, where $\chi$ does not depend on $h$. This way, the continuum limit of \eqref{eq:meanFieldChemotaxis} is well defined, and we obtain a contribution to \eqref{eq:PDEgeneralAllTerms} of the form
\begin{align}\label{eq:PDEchemotaxis}
\epsilon^n_M&= -\nabla(\chi n\nabla c),\nonumber\\
\epsilon^\rho_M &= b^R_3\left| n\nabla c\right|_1.
\end{align}
The evolution of $c$ is determined via equation \eqref{eq:PDEAFgeneral}.
In \eqref{eq:PDEchemotaxis}, we have introduced $b^R_3=\delta_R \stochChemo h =\frac{\chi}{\mu}$. We remark that \eqref{eq:PDEchemotaxis} looks structurally similar to the mean field equations for Case 2 of the random movement model, \eqref{eq:PDEDiffusionDifference}: The time evolution of the tip density equals the negative divergence of a flux, and the time evolution of the vessel densities is proportional to the norm of the gradient of the tip flux. In equation \eqref{eq:PDEDiffusionDifference}, this flux is the diffusive flux
\beq\label{eq:diffusionFlux}
J^D = -\nabla (D n),
\eeq
whereas in \eqref{eq:PDEchemotaxis} it is the chemotactic flux
\beq\label{eq:chemotaxisFlux}
J^\chi = \chi n\nabla  c.
\eeq
Therefore, both \eqref{eq:PDEDiffusionDifference} and \eqref{eq:PDEchemotaxis} have the structure
\begin{align}
\epsilon^n_M &= -\nabla J,\nonumber\\
\epsilon^\rho_M &= b |J|,\nonumber
\end{align}
with $b$ and $J$ chosen accordingly. 

\subsection{Combined Diffusion and Chemotaxis}\label{sec:MFcombinedMovement}
\paragraph{Case 1:}
As we argued in section \ref{sec:TransitionRateTotalMovement}, if we take both diffusion Case 1 and chemotaxis into account as modes of movement, due to different scaling behaviour we should consider chemotaxis and undirected randomness as two separate physical effects represented in the model by two different transition rates, \eqref{eq:transitionRateDiffusionStandard} and \eqref{eq:transitionRateChemotaxis}. The mean field equations, as well as the continuum equations, will then simply be the sum of the mean field and continuum equations corresponding to \eqref{eq:transitionRateDiffusionStandard} and \eqref{eq:transitionRateChemotaxis}, i.e. equations \eqref{eq:meanFieldDiffusionStandard}, \eqref{eq:PDEDiffusionDifference} and \eqref{eq:meanFieldChemotaxis}, \eqref{eq:PDEchemotaxis}, respectively. 
We thus obtain the total contribution to the mean field equations from movement,
\begin{align}\label{eq:meanFieldTotalMovementStandard}
\mathcal{E}^{N_k}_M  &= \stochDiff(\mean{N_{k+1}} + \mean{N_{k-1}} - 2 \mean{N_k})-\stochChemo(\mean{N_{k}}(c_{k+1} + c_{k-1} - 2 c_k)\nonumber\\
&+\stochChemo (\mean{N_{k+1}}-\mean{N_{k}})\pospart{c_k-c_{k+1}}+\stochChemo(\mean{N_{k-1}} - \mean{N_{k}})\pospart{c_{k} - c_{k-1}},\nonumber\\
\mathcal{E}^{R_k}_M &=  b^R_1\mean{N_k} + \frac{b^R_3}{h} \mean{N_k}\left(\pospart{c_{k-1} - c_k}+\pospart{c_{k+1} - c_k}\right),
\end{align}
and the contribution to the continuum equations \eqref{eq:PDEgeneralAllTerms}
\begin{align}\label{eq:PDEtotalMovementStandard}
\epsilon^n_M  &=  D\Delta n-\nabla(\chi n\nabla c),\nonumber\\
\epsilon^\rho_M  &=b^R_1 n+ b^R_3| n\nabla c|_1.
\end{align}
\paragraph{Case 2:}
If we would like to combine chemotaxis with diffusion Case 2, we have several choices. Simply considering the two transition rates \eqref{eq:transitionRateDiffusionDifference} and \eqref{eq:transitionRateChemotaxis} independently, or combining them into one transition rate in a linear way, as done in equation \eqref{eq:transitionDiffPlusChemo1}, will lead to contributions to the mean field equations 
\begin{align}\label{eq:meanFieldTotalMovementDifference1}
\mathcal{E}^{N_k}_M  &= \stochDiff(\mean{N_{k+1}} + \mean{N_{k-1}} - 2 \mean{N_k})-\stochChemo(\mean{N_{k}}(c_{k+1} + c_{k-1} - 2 c_k)\nonumber\\
&+\stochChemo (\mean{N_{k+1}}-\mean{N_{k}})\pospart{c_k-c_{k+1}}+\stochChemo(\mean{N_{k-1}} - \mean{N_{k}})\pospart{c_{k} - c_{k-1}},\nonumber\\
\mathcal{E}^{R_k}_M &= \frac{b^R_2}{h} (\pospart{\mean{N_k}-\mean{N_{k+1}}} +\pospart{\mean{N_k}-\mean{N_{k-1}})}  \nonumber\\
&+ \frac{b^R_3}{h} \mean{N_k}\left(\pospart{c_{k-1} - c_k}+\pospart{c_{k+1} - c_k}\right),
\end{align}
and to the corresponding contributions to the continuum equations 
\begin{align}\label{eq:PDETotalMovementDifference1}
\epsilon^n_M  &=  D\Delta n-\nabla(\chi n\nabla c),\nonumber\\
\epsilon^\rho_M  &=b^R_2|\nabla n|_1 + b^R_3| n\nabla c|_1.
\end{align}
On the other hand, starting with transition rate \eqref{eq:transitionDiffPlusChemo2} we obtain contributions to the mean field equations 
\begin{align}\label{eq:meanFieldTotalMovementDifference2}
\mathcal{E}^{N_k}_M  &= \pospart{\stochDiff(N_{k+1}-N_k) + \stochChemo N_{k+1} (c_{k}-c_{k+1})}\nonumber\\
&+\pospart{\stochDiff(N_{k-1}-N_{k}) + \stochChemo N_{k-1} (c_k-c_{k-1})}\nonumber\\
&-\pospart{\stochDiff(N_k-N_{k+1}) + \stochChemo N_k (c_{k+1}-c_k)}\nonumber\\
&-\pospart{\stochDiff(N_k-N_{k-1}) + \stochChemo N_k (c_{k-1}-c_k)}\nonumber\\
\mathcal{E}^{R_k}_M &= \delta_R \left(\pospart{\stochDiff(N_k-N_{k+1}) + \stochChemo N_k (c_{k+1}-c_k)}\right.\nonumber\\
&\left. +\pospart{\stochDiff(N_k-N_{k-1}) + \stochChemo N_k (c_{k-1}-c_k)}\right),
\end{align}
and corresponding contributions to the continuum equations 
\begin{align}\label{eq:PDETotalMovementDifference2}
\epsilon^n_M  &=  D\Delta n - \nabla(\chi n\nabla c),\nonumber\\
\epsilon^\rho_M  &=|b^R_2\nabla n - b^R_3 n\nabla c|_1.
\end{align}
For each combinations of the random and chemotactic transition rates \eqref{eq:PDEtotalMovementStandard},\eqref{eq:PDETotalMovementDifference1} and \eqref{eq:PDETotalMovementDifference2}, we obtain the same equation for the evolution of the tip densities. This equation, which simply has the interpretation that the change of tip density in time is the negative of the divergence of the total flux of tip cells, i.e. the combined diffusive and chemotactic flux, has been assumed in many continuum models of angiogenesis \cite{balding1985mathematical,chaplain1993model,byrne1995mathematical,anderson1998continuous}.
Note that, as before, both equations \eqref{eq:PDETotalMovementDifference1} and \eqref{eq:PDETotalMovementDifference2} should be supplemented by the reaction-diffusion PDE for the angiogenic factor, equation \eqref{eq:PDEAFgeneral}.
The equation describing the time evolution of the vessel densities is different for Cases 1 and 2, as the tip density in Case 2 enters via its gradient, whereas in Case 1 it enters as a simple linear factor. Case 2 leads to equations known as the snail-trail model \cite{balding1985mathematical,byrne1995mathematical}. Note the subtle difference between the two possible combinations of chemotaxis and diffusion in Case 2:
In equation \eqref{eq:PDETotalMovementDifference1}, the rate of vessel production involves the sum of the norm of the diffusive flux and the norm of the chemotaxis flux, whereas in \eqref{eq:PDETotalMovementDifference2} we obtain the norm of the sum of the fluxes. This means that if the diffusive and chemotactic flux oppose each other, equation \eqref{eq:PDETotalMovementDifference2} will lead to a lower rate of vessel cell production than equation \eqref{eq:PDETotalMovementDifference1}.

Note that early continuum models of tip migration and vessel formation \cite{balding1985mathematical,byrne1995mathematical} ignore these subtleties of the norm. Indeed, in those papers, we find that the vessel densities evolve structurally according to
\beq\label{eq:PDEvesselsOldPapers}
\frac{\partial \rho}{\partial t} = b^R_2 \frac{\partial n}{\partial x} - b^R_3 n\frac{\partial c}{\partial x}
\eeq
In the simulations performed in \cite{balding1985mathematical,byrne1995mathematical} initial conditions are chosen such that the fluxes initially point in the same direction, so that all terms are initially positive. Hence the appearance of the norm in equation \eqref{eq:PDETotalMovementDifference1} or equation \eqref{eq:PDETotalMovementDifference2} makes initially no difference. However, different boundary and initial conditions can lead to disagreement between different evolution equations for the vessel densities, as we will confirm in section \ref{sec:SimDiffusionAndChemotaxis}. Furthermore, in \eqref{eq:PDEvesselsOldPapers}, the vessel density can potentially become negative.

\subsection{Sprouting, anastomosis and regression}\label{sec:MFsproutingAnastomosisRegression}

Sprouting, anastomosis and regression in the stochastic model (sections \ref{sec:stochasticSprouting}, \ref{sec:stochasticAnastomosis}, \ref{sec:stochasticRegression}) were all built upon local transition rates, which only change the cell content of a single box without dependence on the contents of other boxes. So in each case, the transition rate is of the form
\beq
\mathcal{T}_{N_k\pm \delta_N,R_k\pm \delta_R|N_k,R_k} = f(N_k,R_k),\nonumber
\eeq
where $f$ is a generic function dependent only on the local state variables $N_k$ and $R_k$.\footnote{We do not make explicit the dependence on the deterministic AF $c$, as, similar to the tip cell movement terms, all stochastic events discussed here will occur on a much slower timescale than the timescale governing $c$.}
Imagine that we double the lattice spacing $h\to 2h$, and in each new box of size $2h$ there are twice as many cells of each type as before, so the cell densities remain constant. Hence, twice as many stochastic events will take place in a given time period, so the transition rate for the model with lattice size $2h$ will be twice as large as the transition rate for the model with lattice size $h$. Here, we have made a crucial modelling assumption for our mesoscopic compartment model, which is that the cells in each box are well mixed and that boundary effects such as anastomosis events between two cells in neighbouring lattice sites are negligible.

The above argument implies that all the local transition rates for sprouting, anastomosis and regression should scale like
\beq
\mathcal{T}_{N_k\pm1,R_k\pm1|N_k,R_k} = h \tilde{f}\left(\frac{N_k}{h},\frac{R_k}{h}\right).\nonumber
\eeq
Here, $\tilde{f}$ is a function which depends only on the intensive variables $\frac{N_k}{h},\frac{R_k}{h}$, i.e. variables which do not change with system size. These are the discrete tip and vessel cell densities, rather than the tip or vessel cell numbers. The factor of $h$ in front of the transition rate ensures that the total transition rate is extensive, which means it grows linearly with system size. 

Equipped with this insight, we can now deduce the scaling of the parameters involved in the transition rates in sections \ref{sec:stochasticSprouting}, \ref{sec:stochasticAnastomosis}, \ref{sec:stochasticRegression}. For sprouting, we had $\mathcal{T}_{N_{k}+1|N_k} = {\tilde a_0} R_k c_k + {\tilde a_1}N_k c_k H(c_k-\hat{c})$. Recall that the AF $c$ is measured in terms of concentration, so it is already an intensive variable. Hence, ${\tilde a_0}$ and ${\tilde a_1}$ do not depend on $h$. We then obtain the contribution to the mean field equation,
\begin{align}
 \mathcal{E}^{N_k}_S &= {\tilde a_0}\mean{R_k} c_k + {\tilde a_1}\mean{N_k} c_kH(c_k-\hat{c}),\quad \mathcal{E}^{R_k}_S=0 \nonumber
\end{align}
and the contribution to the PDE is
\begin{align}
 \epsilon^n_A &= a_0\rho c_k + a_1n c_kH(c_k-\hat{c}),\quad \epsilon^\rho_A = 0. \nonumber
\end{align}
Here, we simply put $a_0={\tilde a_0}$ and $a_1={\tilde a_1}$. Similarly, we find the scaling relations ${\tilde \beta_1}=\frac{\beta}{h}$, ${\tilde \gamma}=\gamma$, where $a_0,a_1,\beta$ and $\gamma$ are $h$-independent, so we obtain contributions to the mean field equation from anastomosis and regression
\begin{align}
 \mathcal{E}^{N_k}_A &= {\tilde \beta_1}\mean{N_k} \mean{R_k},&\quad \mathcal{E}^{R_k}_A &= 0, \nonumber\\
 \mathcal{E}^{N_k}_R &= 0,&\quad \mathcal{E}^{R_k}_R &= -{\tilde \gamma}\mean{R_k}, \nonumber
\end{align}
and similarly contributions to the PDE
\begin{align}
 \epsilon^{n}_A &= \beta n \rho,&\quad \epsilon^\rho_A&= 0, \nonumber\\
 \epsilon^{n}_R &= 0,&\quad \epsilon^{\rho}_R &= -\gamma\rho. \nonumber
\end{align}
%


\section{Comparison of different mechanisms of movement}\label{sec:simulationsMovement}
This section will focus exclusively on the study of different approaches to modelling tip movement and the subsequent production of new vessel cells. We will ignore all other aspects of the full angiogenesis model, namely sprouting, anastomosis and vessel regression. Our purpose is to understand the implications of choosing the different transition rates describing the movement of tip cells and subsequent vessel production, which were outlined in section \ref{sec:stochasticModelMovement}, and had the common structure $\mathcal{T}_{N_k-1,N_l+1,R_k+\delta_R,R_l|N_k,N_l,R_k,R_l}$.
A second purpose of this section is to compare the behaviour of the stochastic models with their corresponding deterministic continuum limits. In general, we would assume that a continuum model only makes sense when we have a large number of cells, such that collectively on a larger scale the distribution of cells resembles a continuum. Moreover, one might question whether a typical biological setup of angiogenesis is such that we have sufficiently many tip cells to justify modelling them as a continuum. However, we will show that in some cases, even when the stochastic model for tip movement leads to noisier behaviour in terms of cell migration than that predicted by the corresponding continuum model, the resulting production of vessel cells will be similar for both models. Furthermore, we find situations where the results of the continuum model are similar to the average of many runs of the stochastic model. On the other hand, we will also encounter situations where the continuum and stochastic models completely disagree, even when 
averaging over many simulations in the stochastic model, and both models produce markedly different amounts 
of blood vessels. These latter situations signal the breakdown in the validity of the deterministic continuum approximation, and require the use of the stochastic model.

Throughout this section, we perform simulations with no initial static vessel cells. This might be questioned biologically, as when we start with tip cells, we should also have accompanying vessel cells. The reason for assuming that initially there are no vessel cells is simply that, as long as we ignore anastomosis, regression and sprouting, the vessel cells actually decouple from the dynamics of the model. Hence, any choice of the initial vessel cells would not change the behaviour of the system, and it is easier to see how the model behaves when there are no vessel cells initially. Biologically, we can imagine that such a situation corresponds to seeding individual tip cells in an {\it in-vitro} environment, without parent vessels. 
We will first focus on pure diffusion, that is, we assume we have a homogeneous distribution of the angiogenic factor, so chemotaxis does not contribute to cell movement.
Then we focus on pure chemotaxis, which means we have a strong chemotactic gradient such that diffusion is negligible. However, we emphasise that the choices of initial and boundary conditions, as well as model parameters, in this section are guided by the need to understand the dynamics of tip movement and vessel production in the model, rather than accurately to represent the biology. Indeed, all the parameters appearing in this section can be removed by non-dimensionalisation. Hence we do not attempt to choose realistic values. Only in section \ref{sec:simulationFullModel} will we study the dynamics of the full model and, with that, aim to capture the essential qualitative and quantitative features of 
angiogenesis.
Details on the techniques used for solving the stochastic model and the governing PDEs can be found in appendix \ref{sec:app:Numerics}.

\subsection{Random movement}\label{sec:SimRandom}

We now compare simulations of the stochastic model based on the two simple transition rates \eqref{eq:transitionRateDiffusionStandard} and \eqref{eq:transitionRateDiffusionDifference} describing undirected random movement, together with their respective deterministic continuum limits \eqref{eq:PDEDiffusionStandard} and \eqref{eq:PDEDiffusionDifference}. We consider several sets of initial and boundary conditions to test our model.

\subsubsection{Vascular invasion with Dirichlet boundary conditions}

We choose a domain of length $1$, so $x \in [0,1]$.  The scenario in this section is that we start with some number, $N^0$, of tips in the leftmost box. We choose Dirichlet boundary conditions, which means in the stochastic model that we keep the number of cells in the left and right boundary boxes fixed. The biological interpretation of this condition is that on the left boundary we have a parent vessel from which new tips emerge, and to the right we have a tumour into which we assume the tips disappear. One might argue that sprouting should stop some time after the vessels have formed. However, here we are not considering a detailed model of angiogenesis, so for now we focus on the mathematical behaviour of the movement part of the model, which is a vital component of the full, more biologically realistic model. Alternatively, this boundary condition can be interpreted as introducing new tip cells at the boundary of a petri dish at a constant rate.

The PDEs we are solving are equations \eqref{eq:PDEDiffusionStandard} and \eqref{eq:PDEDiffusionDifference} for random movement for Cases 1 and 2, respectively. The setup described above translates into the following initial and boundary conditions:
\begin{align}\label{eq:ICBCPDEInvasion}
n(0,x)&=\frac{N^0}{h}H(h-x),& \rho(0,x)&=0, , \nonumber\\
n(t,0)&=\frac{N^0}{h}, & n(t,1)&=0, \quad t\geq 0, \nonumber\\
\end{align}
where $0<h<1$ is the lattice constant of the stochastic model, and $H$ is the Heaviside function, which we smooth in the PDE by setting $H(x)=\frac{1+tanh(\omega x)}{2}$, with $\omega>>1$ chosen such that we have a close approximation to the Heaviside function while still obtaining stable numerical results. The initial and boundary conditions in the stochastic model are chosen to be compatible with the initial and boundary conditions for the PDE, and take the form
\begin{align}\label{eq:ICBCstochasticInvasion}
N_{1}(t=0) &= N^0, \quad N_{k}(t=0) = 0, \quad k=2,\dots,k_{max}, \nonumber\\
R_k(t=0) &= 0, \quad k=1,\dots,k_{max}, \nonumber\\
N_1(t),& N_{k_{max}}(t), R_1(t),R_{k_{max}}(t) \text{ fixed in time}.
\end{align}
The smoothing of the Heaviside function can lead to an insignificant rounding error in the initial condition of $N_2(t=0)$. The following parameters\footnote{Fixing $\delta_R$ means we use different cell scales $\mu$ when comparing simulations based on the different cases of the transition rate. Again, we emphasise that, as the vessel cells decouple from the dynamics in our test case ignoring sprouting, anastomosis and regression, fixing either $\mu$ or $\delta_R$ only leads to a rescaling in the total amount of vessel production. $\delta_R$ should be an integer, so it is easier to enforce this in the simulation by simply fixing it, rather than calculating it from the lattice constant $h$ and the cell size $\mu$.} were chosen:
\begin{align}
 D &=1, & \delta_R &= 8,& h&=0.05,& k_{max} &= 21.\nonumber
\end{align}
Recall that as the vessel cells decouple from the dynamics in our investigations ignoring sprouting, anastomosis and regression, the choice of $\delta_R$ does not really matter. Likewise, $D$ can be absorbed by rescaling the spatial or temporal coordinate. We note that the PDE for $n$ has an analytic solution, which can be found by Fourier expansion. 
\begin{figure}[h!]
\subfloat[Tips $n(t,x)$ for Case 1]
{
\includegraphics[width=0.48 \textwidth]{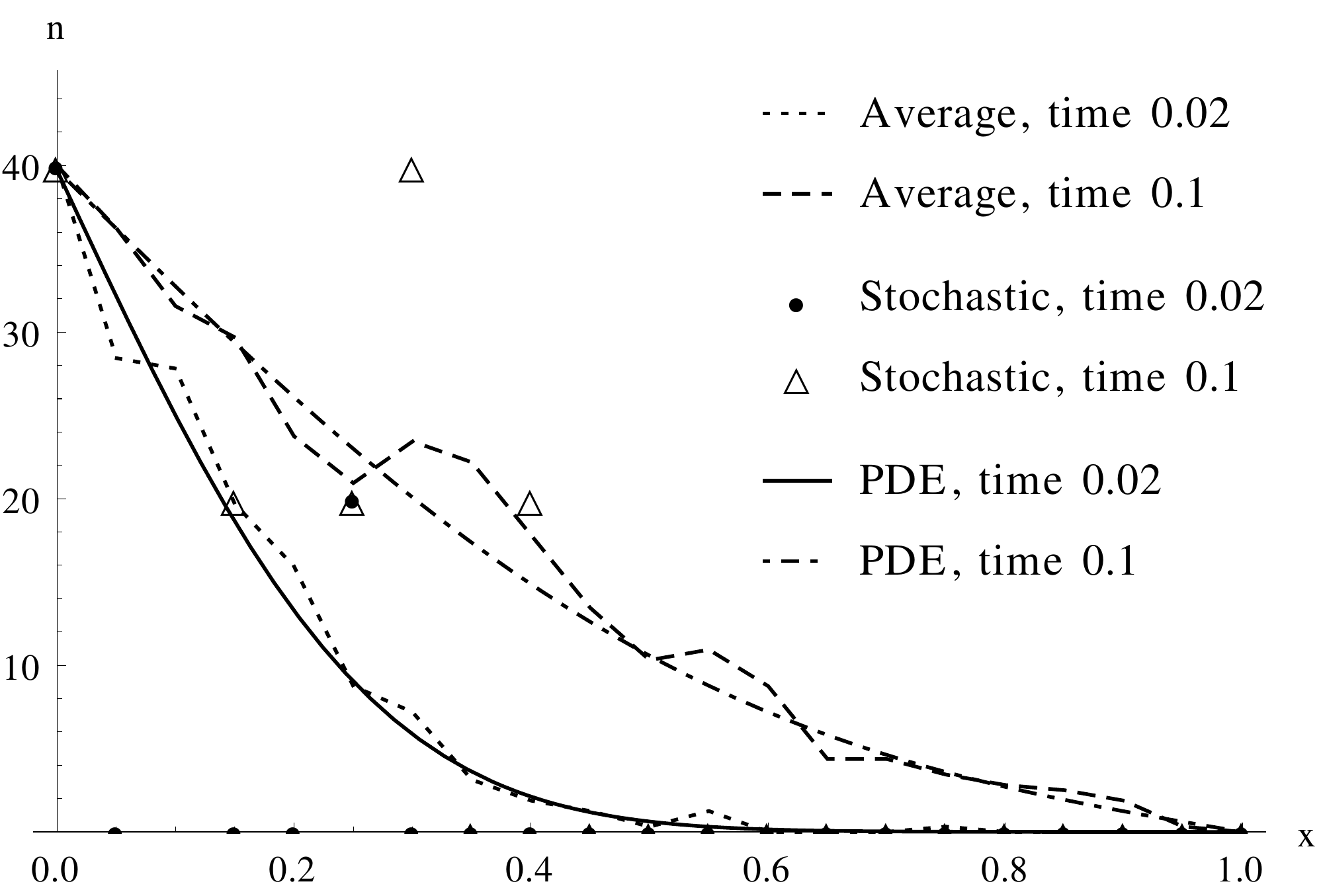}
\label{fig:vascularInvasionCase1TipsLow}
}
\subfloat[Vessels $\rho(t,x)$ for Case 1]
{
\includegraphics[width=0.48 \textwidth]{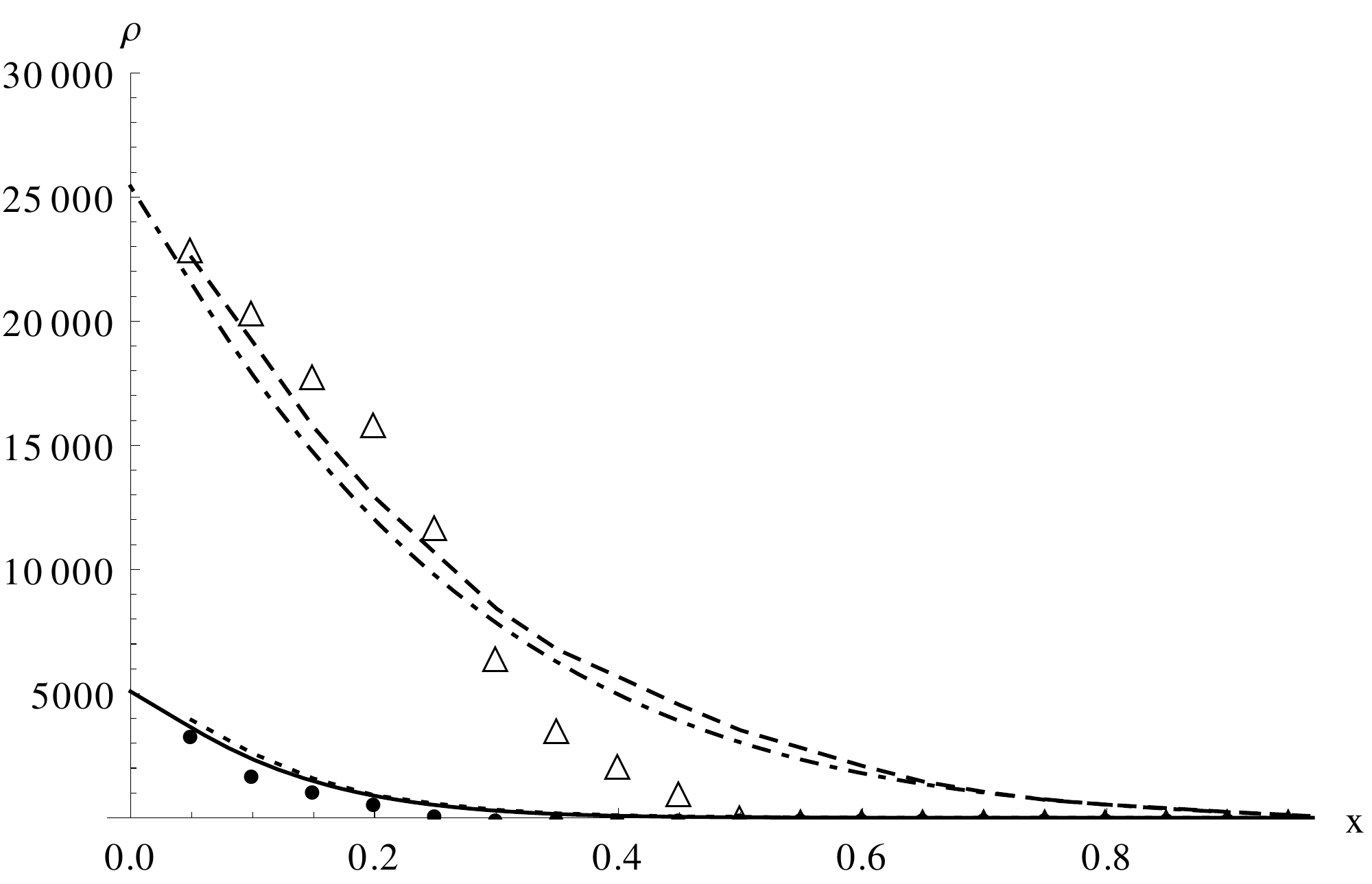}
\label{fig:vascularInvasionCase1VesselsLow}
}\\
\subfloat[Tips $n(t,x)$ for Case 2]
{
\includegraphics[width=0.48 \textwidth]{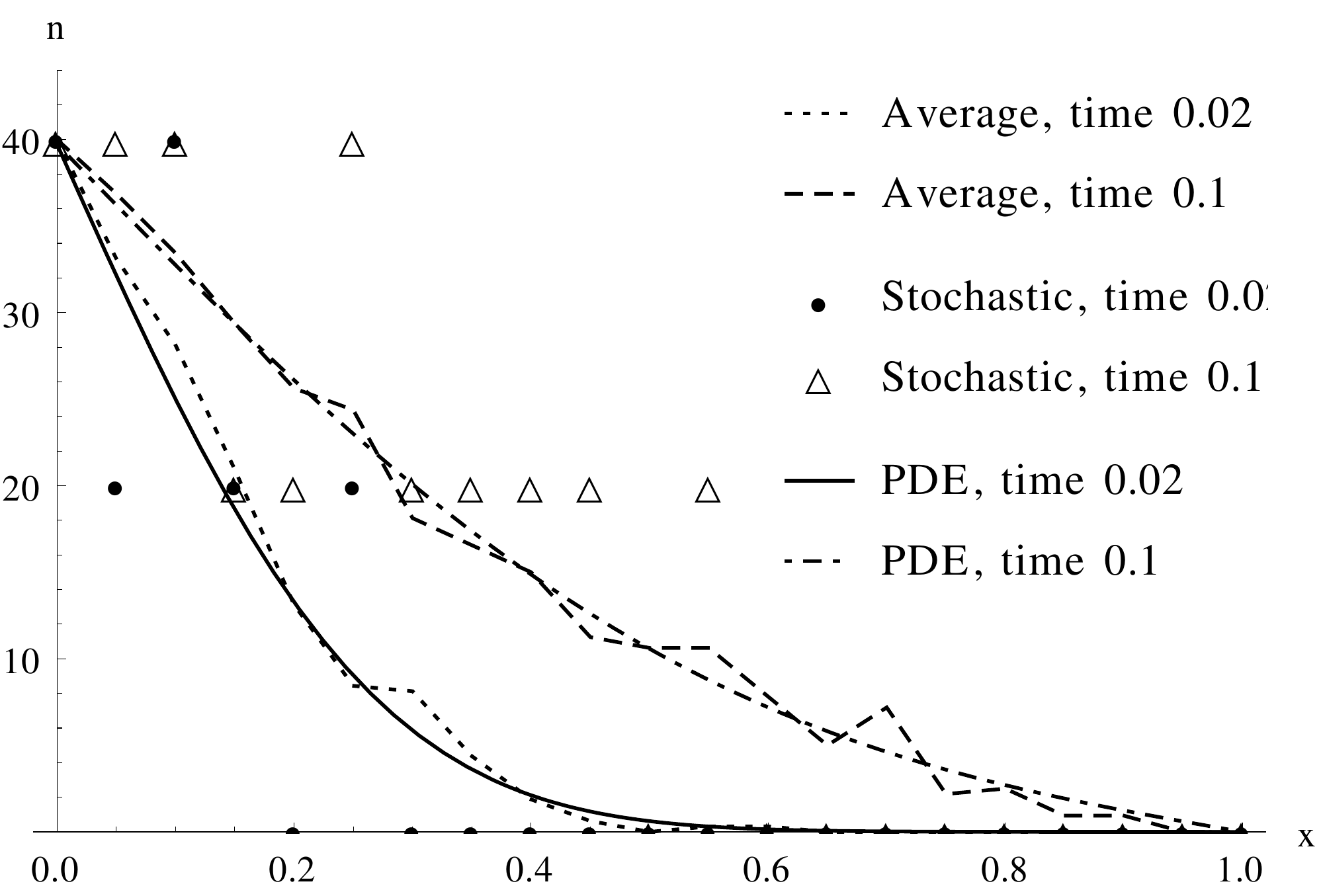}
\label{fig:vascularInvasionCase2TipsLow}
}
\subfloat[Vessels $\rho(t,x)$ for Case 2]
{
\includegraphics[width=0.48 \textwidth]{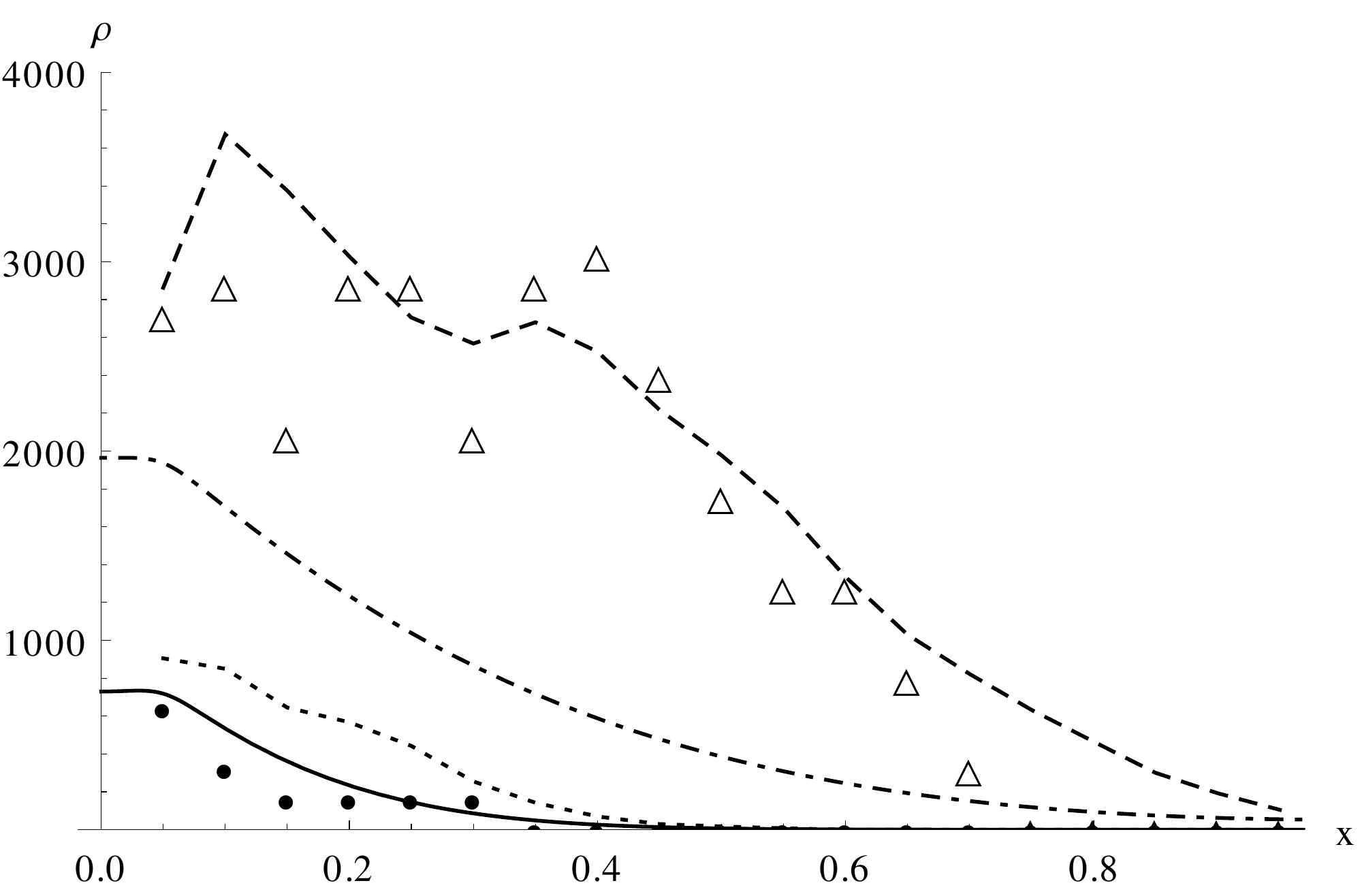}
\label{fig:vascularInvasionCase2VesselsLow}
}
\caption{
\label{fig:vascularInvasionLow}
Comparison of tip migration and subsequent vessel production of a single realisation of the stochastic model \eqref{eq:transitionRateDiffusionStandard} or \eqref{eq:transitionRateDiffusionDifference} in Case 1 and 2, respectively (filled dots and triangles), the average of 64 realisations (dashed and dotted lines), and the PDE \eqref{eq:PDEDiffusionStandard} or \eqref{eq:PDEDiffusionDifference} (Solid and dashed-dotted line). Initial and boundary conditions as in \eqref{eq:ICBCstochasticInvasion}, \eqref{eq:ICBCPDEInvasion} for initially $N^0=2$ tip cells in the leftmost box. In each case the spatial distribution of tips or vessels, respectively, is shown at two times.
} 
\end{figure}
\begin{figure}[h!]
\subfloat[Tips $n(t,x)$ for Case 1]
{
\includegraphics[width=0.48 \textwidth]{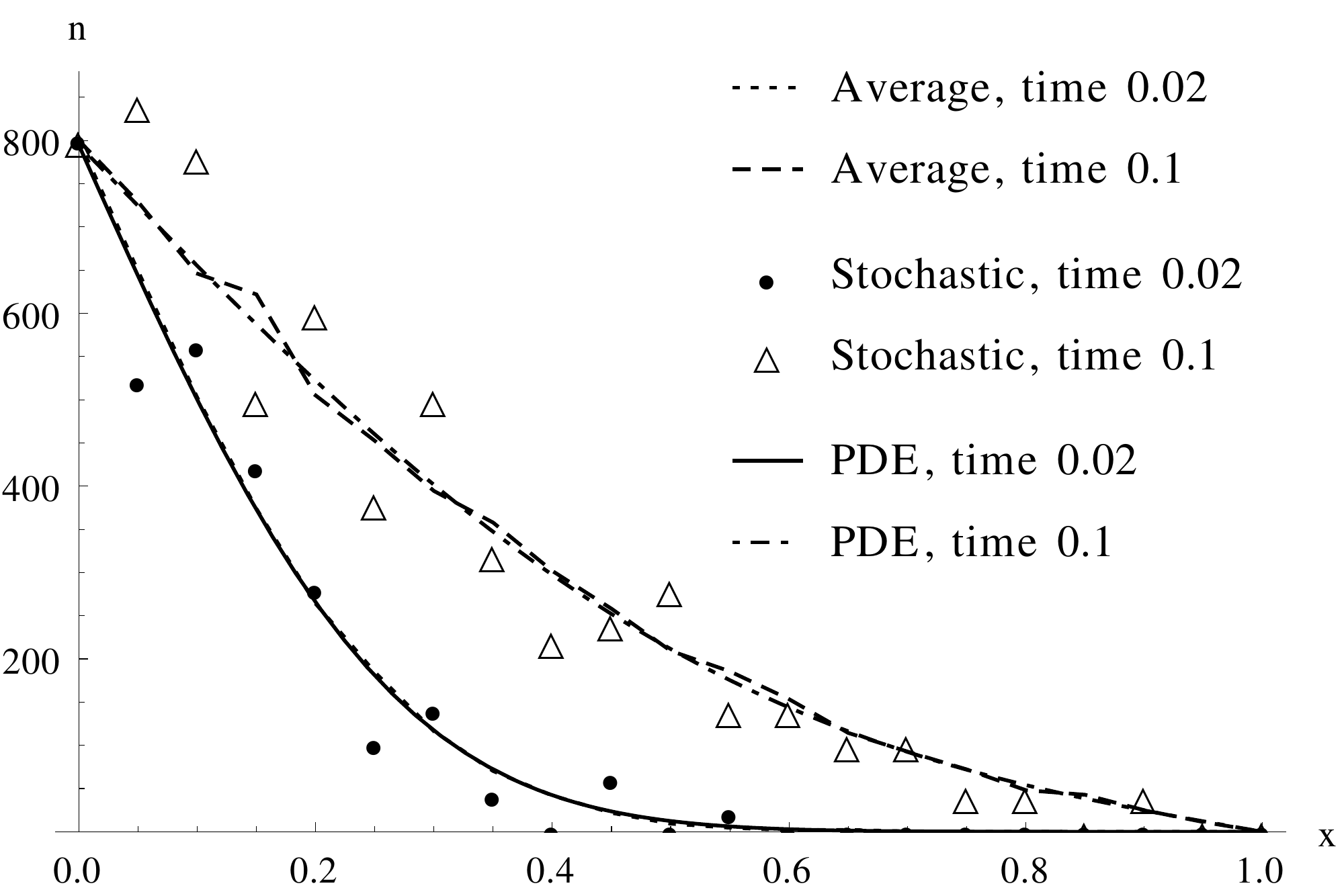}
\label{fig:vascularInvasionCase1TipsHigh}
}
\subfloat[Vessels $\rho(t,x)$ for Case 1]
{
\includegraphics[width=0.48 \textwidth]{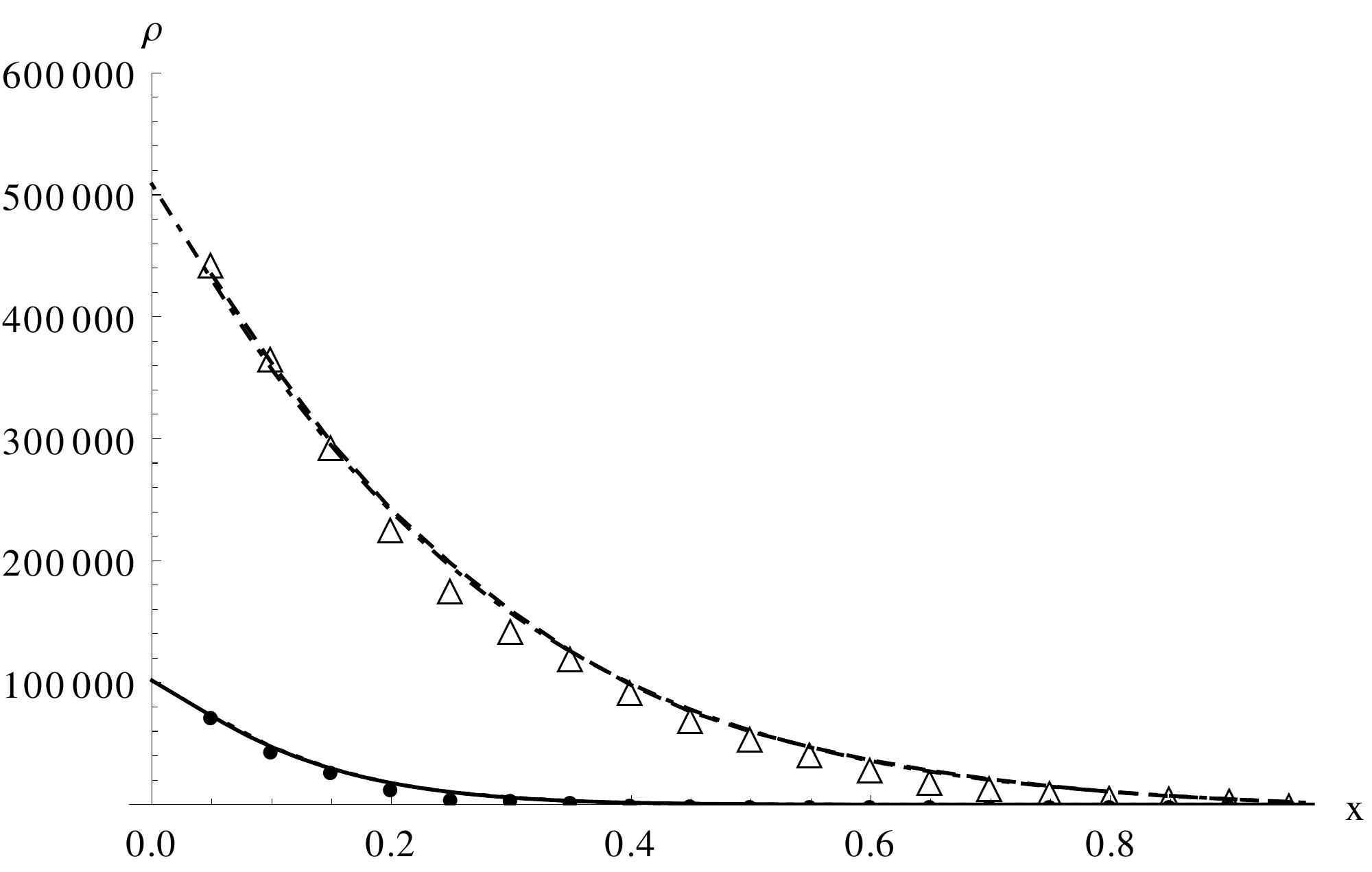}
\label{fig:vascularInvasionCase1VesselsHigh}
}
\\
\subfloat[Tips $n(t,x)$ for Case 2]
{
\includegraphics[width=0.48 \textwidth]{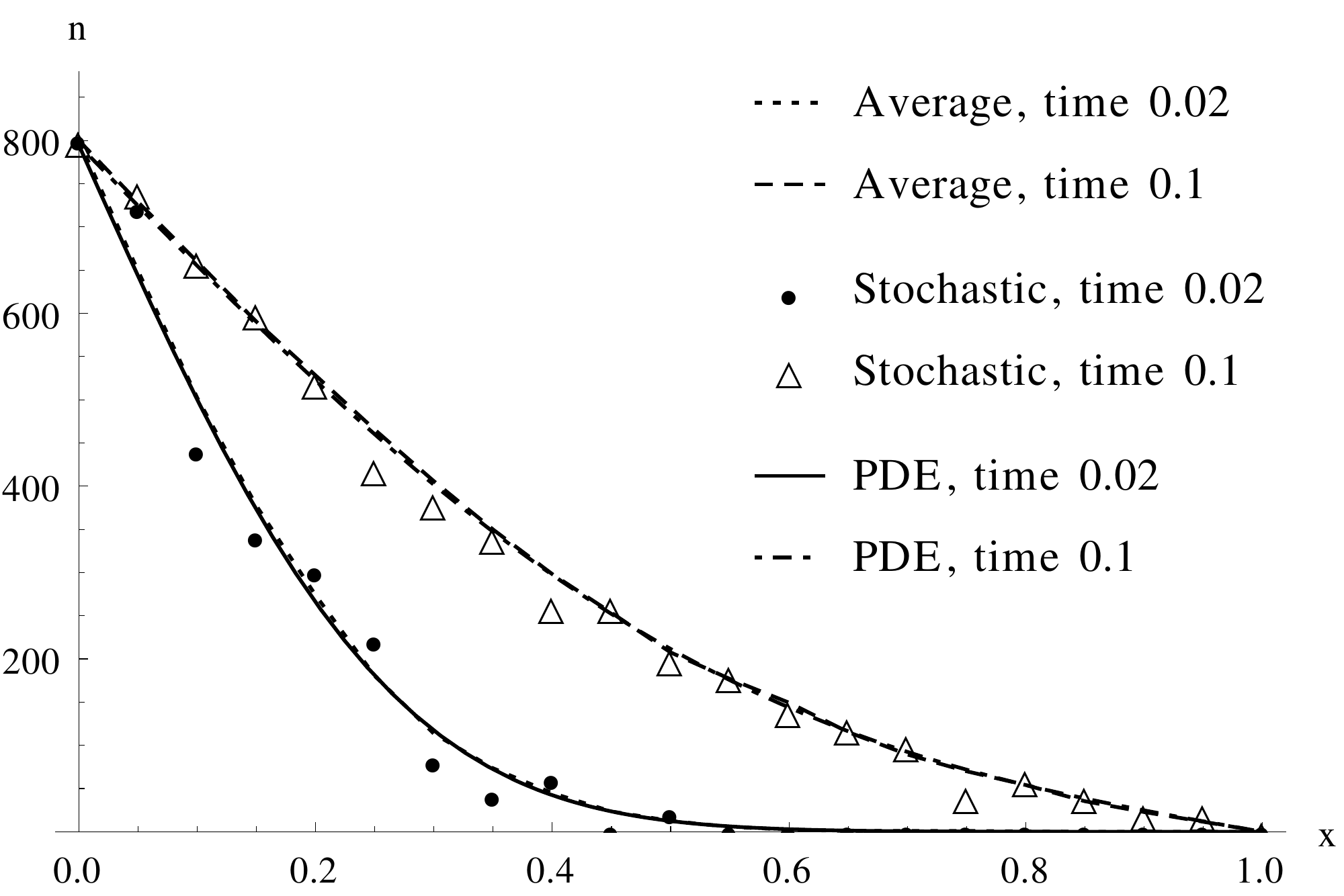}
\label{fig:vascularInvasionCase2TipsHigh}
}
\subfloat[Vessels $\rho(t,x)$ for Case 2]
{
\includegraphics[width=0.48 \textwidth]{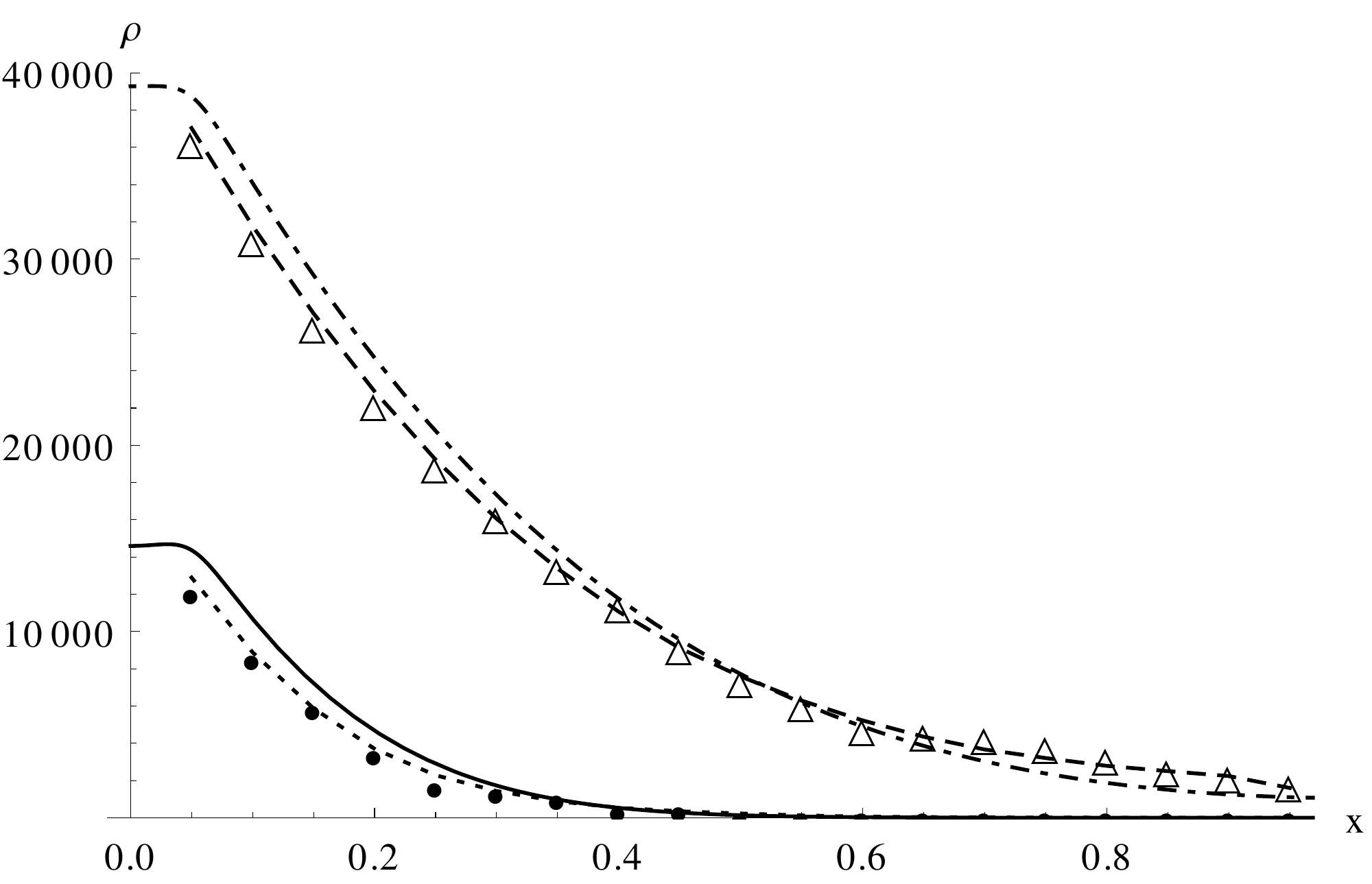}
\label{fig:vascularInvasionCase2VesselsHigh}
}
\caption{
\label{fig:vascularInvasionHigh}
Comparison of tip migration and subsequent vessel production of a single realisation of the stochastic model \eqref{eq:transitionRateDiffusionStandard} or \eqref{eq:transitionRateDiffusionDifference} in Case 1 and 2, respectively, the average of 64 realisations, and the PDE \eqref{eq:PDEDiffusionStandard} or \eqref{eq:PDEDiffusionDifference}. Initial and boundary conditions as in \eqref{eq:ICBCstochasticInvasion} and \eqref{eq:ICBCPDEInvasion} for initially $N^0=40$ tip cells in the leftmost box. In each case the spatial distribution of tips or vessels respectively is shown at two times.
} 
\end{figure}
Figures \ref{fig:vascularInvasionLow} and \ref{fig:vascularInvasionHigh} contrast the time evolution of the spatial tip and vessel cells distribution for random movement Case 1, i.e. the stochastic model defined by $\mathcal{T}_{N_k-1,N_l+1,R_k+\delta_R,R_l|N_k,N_l,R_k,R_l}=\frac{D}{h^2}N_k$ and the corresponding PDE \eqref{eq:PDEDiffusionStandard}, to random movement Case 2, i.e. 
$\mathcal{T}_{N_k-1,N_l+1,R_k+\delta_R,R_l|N_k,N_l,R_k,R_l}=\frac{D}{h^2}\pospart{N_k-N_l}$ and the corresponding PDE
\eqref{eq:PDEDiffusionDifference}. In Figure \ref{fig:vascularInvasionLow} we performed the simulations with the initial number of tip cells $N^0=2$ in the initial conditions \eqref{eq:ICBCstochasticInvasion}, whereas in Figure \ref{fig:vascularInvasionHigh} we have $N^0=40$. This corresponds to $n(0,0)=\frac{N^0}{h} = 40$ and $n(0,0)=800$, respectively. Note that in all figures, the results from the stochastic simulations have been rescaled by $h$ to compare with the solutions of the PDE, which are given in terms of cell densities.

Focusing on the profile of the tips, we notice that for $N^0=2$, both in Case 1, Figure \subref*{fig:vascularInvasionCase1TipsLow} and Case 2, Figure \subref*{fig:vascularInvasionCase2TipsLow}, the results from individual stochastic simulations look quite different from the solution of the PDE, but when we average over 64 realisations of the stochastic model, we obtain good agreement with the PDE. However, the resulting vessel profiles look quite different for Cases 1 and 2. Contrasting Figures \subref*{fig:vascularInvasionCase1VesselsLow} and \subref*{fig:vascularInvasionCase2VesselsLow}, we see in Case 1 far more vessel cells are produced. This is not surprising, as the transition rate in Case 1 is always greater than, or equal to, the one for Case 2, implying more tip cell movement and hence a higher rate of vessel production. Furthermore, we see that in Case 1 the average of 64 stochastic realisations agrees well with the result of the PDE, and even a single realisation of the stochastic model is in much closer agreement with the PDE than the corresponding result for tips, as shown in Figure \subref*{fig:vascularInvasionCase1TipsLow}. This result can be explained as follows: the movement of tip cells is quite noisy, but this noise in Case 1 averages out. The total number of vessel cells is found by summing over all tip migration events, where the noise cancels. The biological interpretation is that even when we do not know exactly when and where a single tip cell migrates, the resulting blood vessels form in a robust way with little noise. The situation is quite different in Case 2, Figure \subref*{fig:vascularInvasionCase2VesselsLow}. Both a single realisation as well as the average of 64 realisations of the stochastic model disagree markedly from the result of the PDE. We confirmed this disagreement does not disappear when averaging over significantly more realisations. The reason is as follows: When deriving the PDE from the mean field equations, \eqref{eq:meanFieldDiffusionDifference}, we assume that we can approximate the finite differences between means of neighbouring sites by derivatives. The solution of the PDE is then always monotonically decreasing in $x$ for the chosen boundary conditions. However, in the stochastic model with the chosen initial and boundary conditions, $N_k$ can only take three values, $N_k=0,1,2$, and the solution is, in general, not monotonically decreasing but fluctuating. As the probability for tip cell movement, and hence vessel production, depends on the difference of tip cells in neighbouring boxes, more spatial fluctuations of tip cells imply a higher chance of movement. Hence, the stochastic model in Case 2 will typically produce more vessel cells than the corresponding PDE. We also note that the state where $N_k=N_l$ $\forall k,l$, which is homogeneous in the distribution of tip cells, is an absorbing state of the stochastic model defined by transition rate \eqref{eq:transitionRateDiffusionDifference}. Hence, at low cell numbers we are always close to such a state, and do not necessarily expect that the mean field approximation is valid. This is in contrast to Case 1 defined by transition rate \eqref{eq:transitionRateDiffusionStandard}, which is linear, so the system is self-averaging. This means the average over an ensemble of $N$ cells is the same as the average over $N$ random pathes of a single cell.

This difference between the PDE and stochastic models disappears when we consider a larger initial number of tip cells, $N^0=40$, as shown in Figure \ref{fig:vascularInvasionHigh}. Here, also in Case 2, Figure \subref*{fig:vascularInvasionCase2VesselsHigh}, there is good agreement between the PDE and the stochastic models (both for a single realisation and the average of 64 realisations). This is because the tip profile, as shown in Figure \subref*{fig:vascularInvasionCase2TipsHigh}, is now roughly monotonous in the stochastic model, and hence the approximation by a differentiable function obtained from a PDE is justified. The vessel profile for Case 1, Figure \subref*{fig:vascularInvasionCase1VesselsHigh} shows even better agreement between PDE and stochastic simulation. The corresponding tip profile obtained from the stochastic model, Figure \subref*{fig:vascularInvasionCase1TipsHigh}, shows more noise than for Case 2, Figure \subref*{fig:vascularInvasionCase2TipsHigh}. This is again explained by the fact that the transition rate for Case 2 is always smaller than, or equal to, that for Case 1 and hence produces less total movement of tip cells, but the same net directed movement.

\subsubsection{Lattice constant dependence}
\begin{figure}[h!]
\subfloat[Tips $n(t,x)$ for h=0.1]
{
\includegraphics[width=0.48\linewidth]{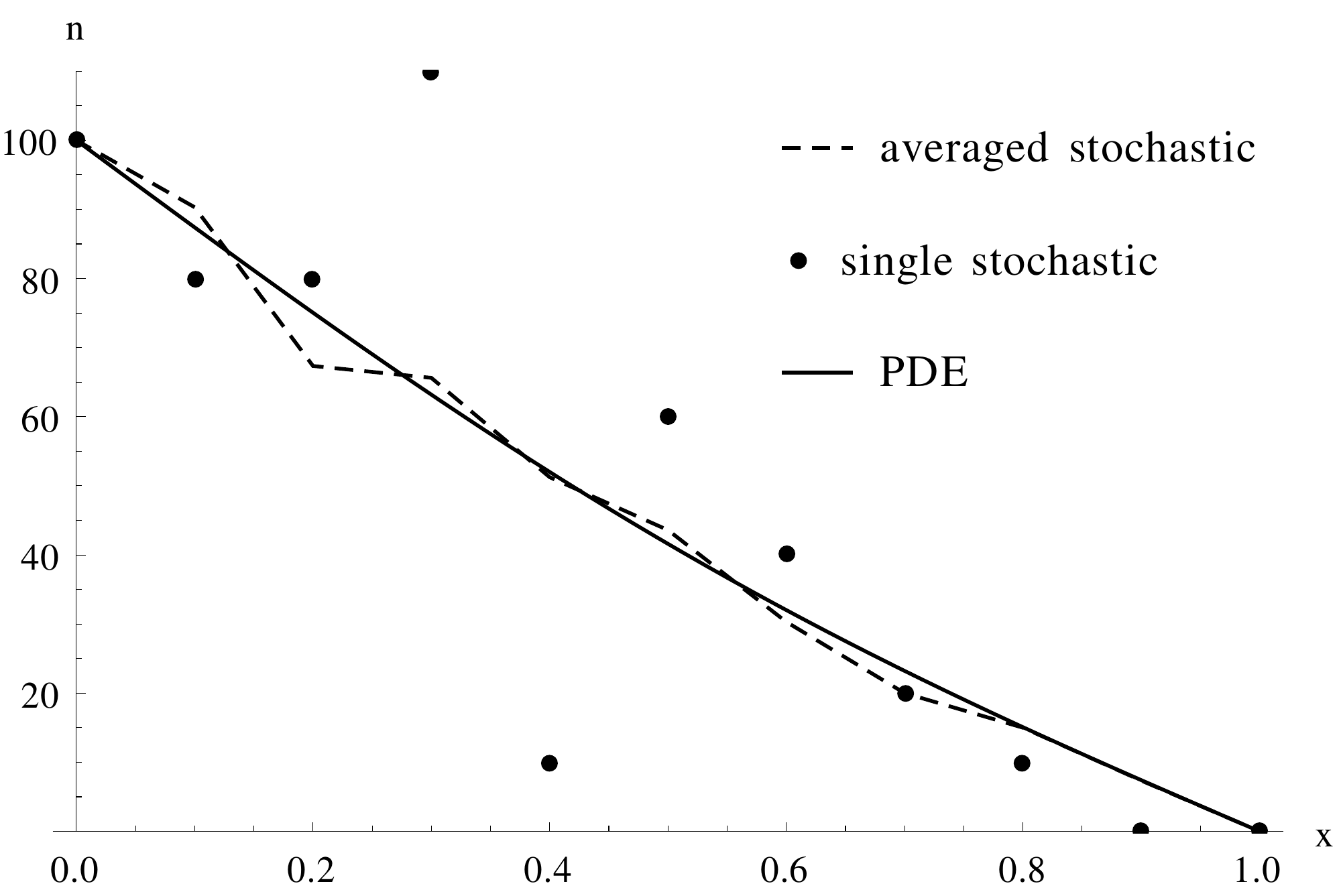}
}
\subfloat[Vessels $\rho(t,x)$ for h=0.1]
{
\includegraphics[width=0.48\linewidth]{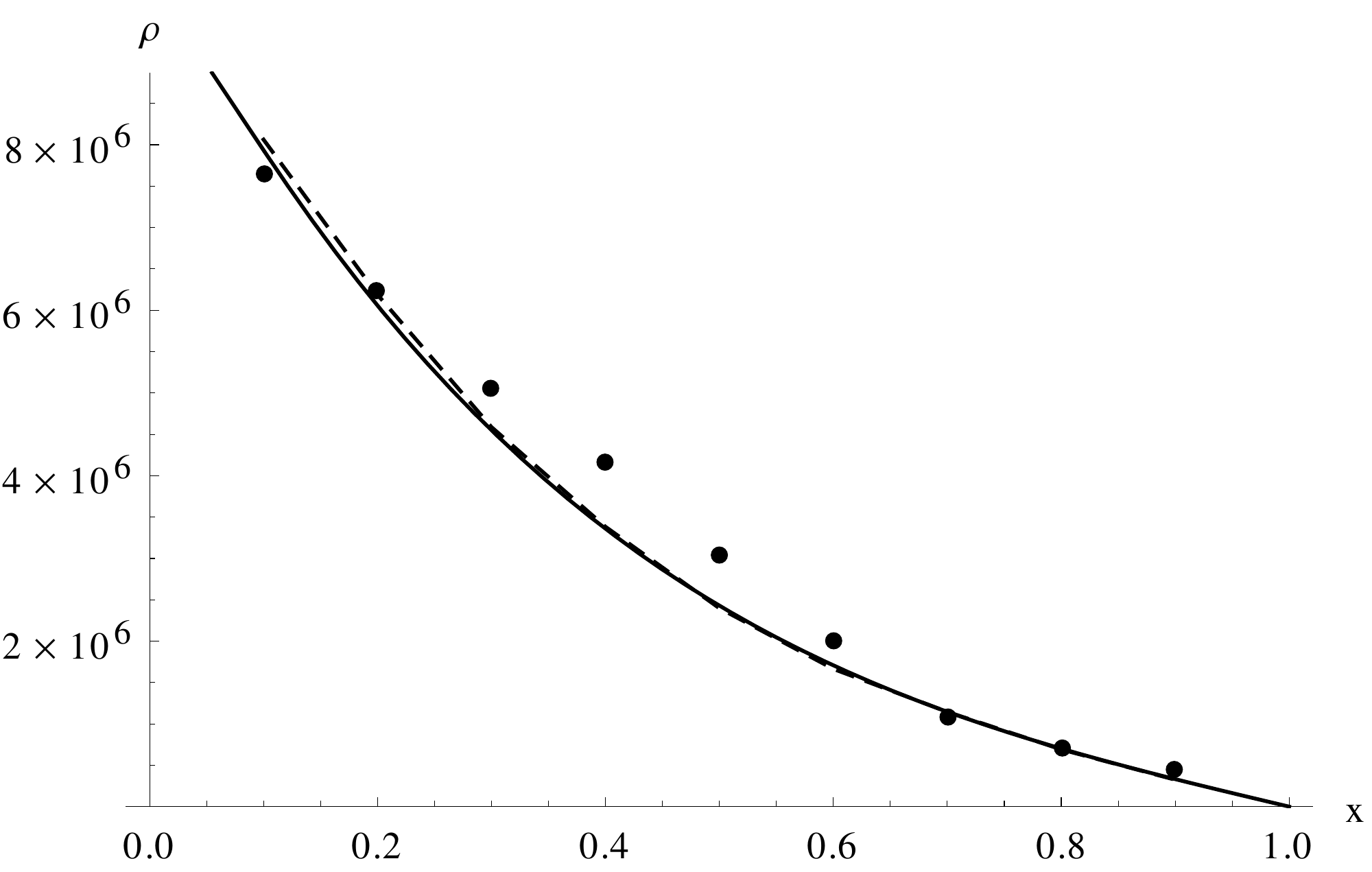}
}
\\
\subfloat[Tips $n(t,x)$ for h=0.05]
{
\includegraphics[width=0.48\linewidth]{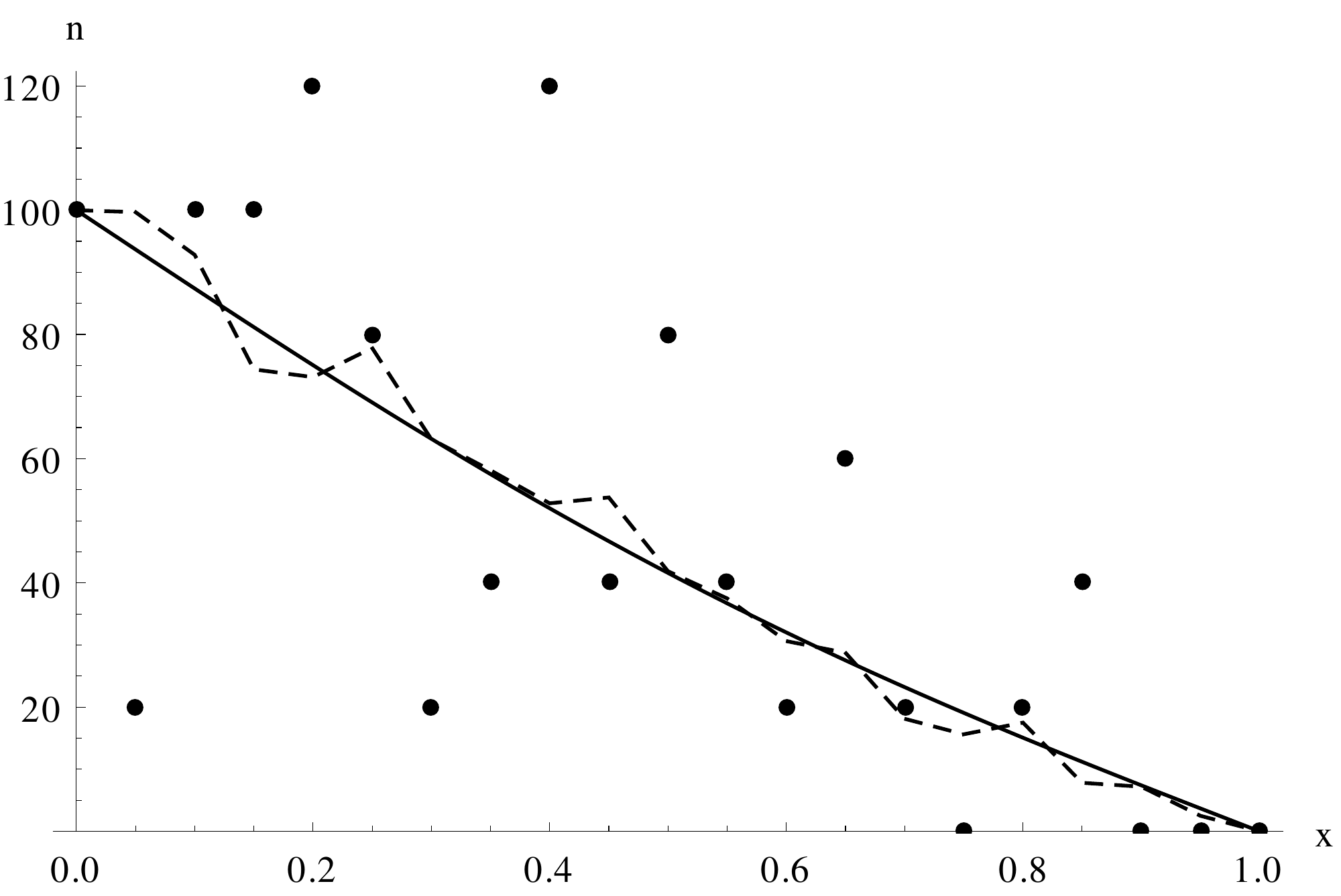}
}
\subfloat[Vessels $\rho(t,x)$ for h=0.05]
{
\includegraphics[width=0.48\linewidth]{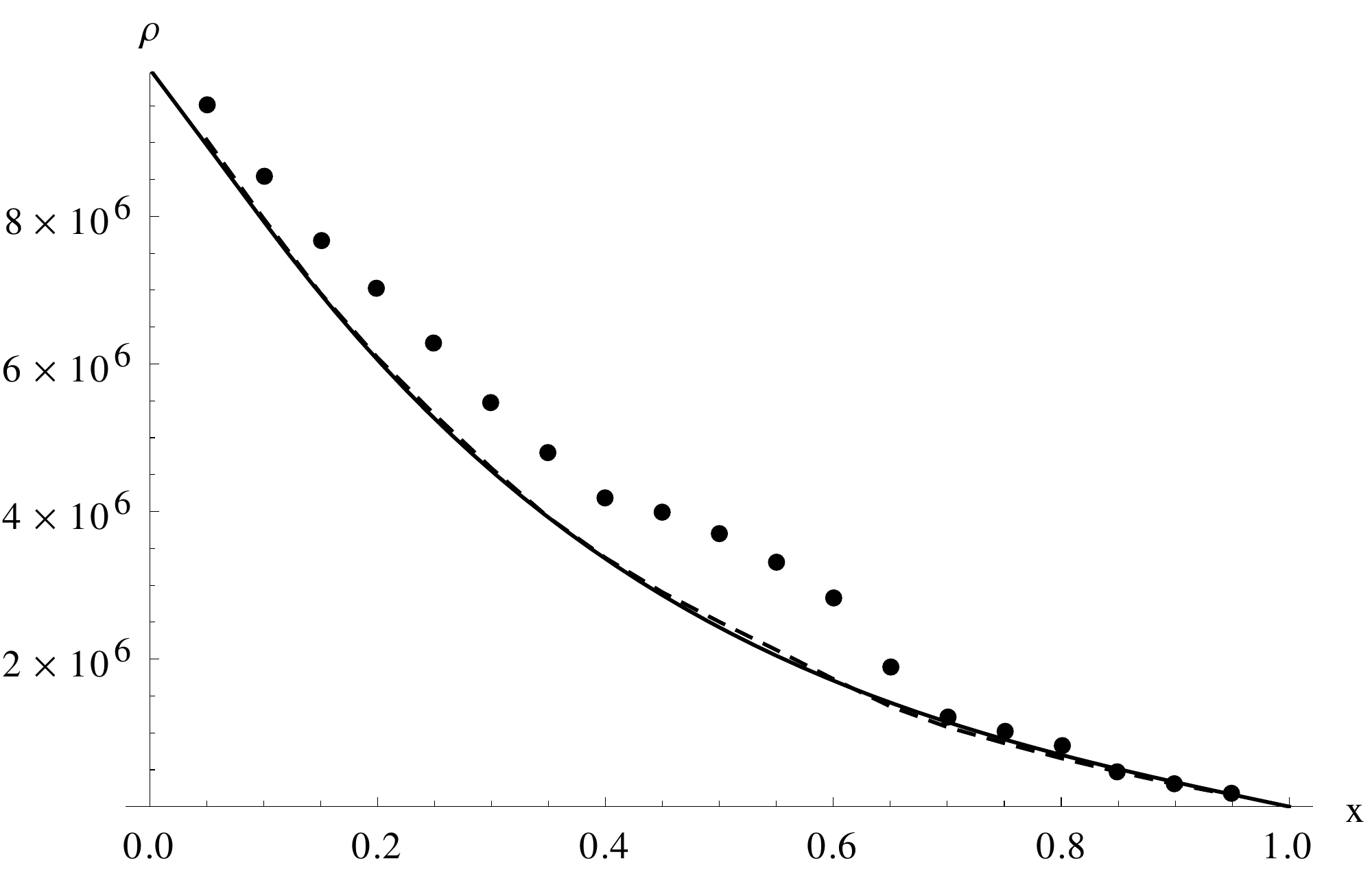}
}
\\
\subfloat[Tips $n(t,x)$ for h=0.02]
{
\includegraphics[width=0.48\linewidth]{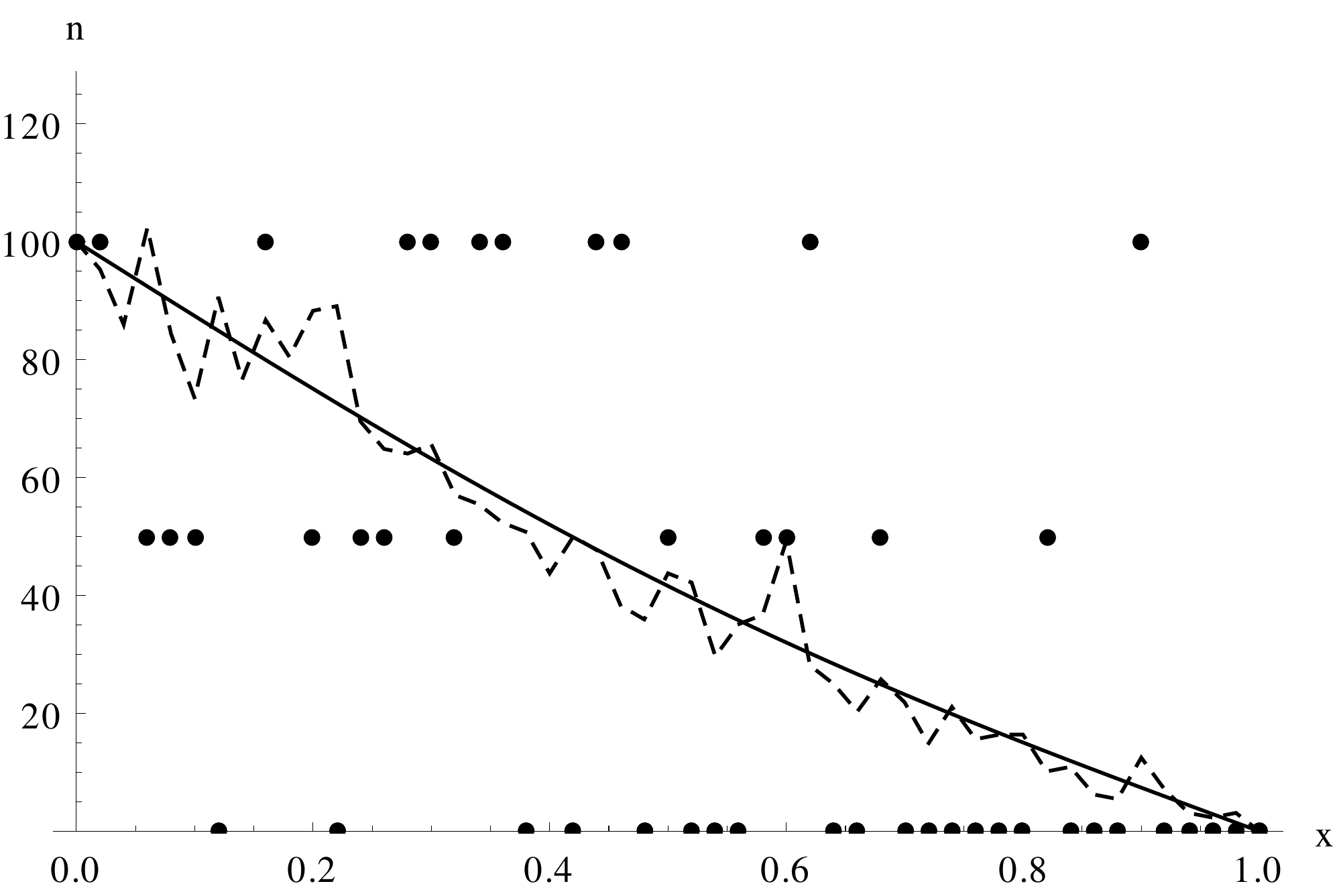}
}
\subfloat[Vessels $\rho(t,x)$ for h=0.02]
{
\includegraphics[width=0.48\linewidth]{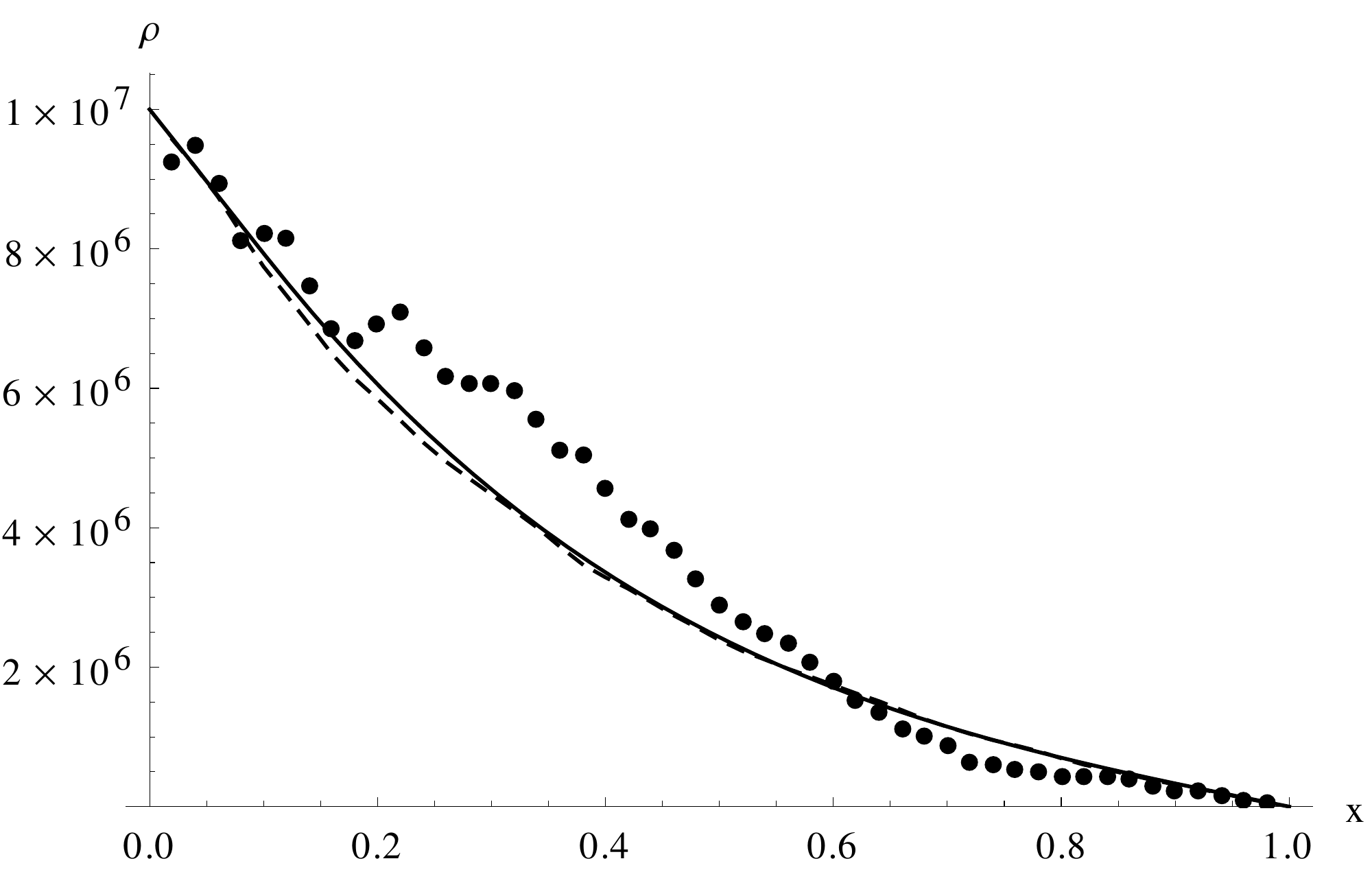}
}
\caption{
\label{fig:vascularInvasionCase1LatticeDependence}
Tip cells migrating randomly according to Case 1, transition rate \eqref{eq:transitionRateDiffusionStandard} and PDE \eqref{eq:PDEDiffusionStandard}, and corresponding vessel production, with initial and boundary conditions given in \eqref{eq:ICBCPDELatticeDependence} and \eqref{eq:ICBCstochasticLatticeDependence}. All snapshots taken at time $t=0.2$.
} 
\end{figure}

We now investigate how the behaviour of the models depends on the lattice constant $h$. We fix the typical vessel size to $\mu=0.002$, vary $h=0.1,0.05,0.02$ and adjust $\delta_R$ accordingly. We also keep the tip cell density constant, and correspondingly adjust the number of cells per box, according to the size of the box. We choose initial and boundary conditions for the PDE to be
\begin{align}\label{eq:ICBCPDELatticeDependence}
n(0,x)&= 100 H(0.1-x) , &\rho(0,x)&=0, \nonumber\\
n(t,0)&=100, & n(t,1)&=0,
\end{align}
so the left half of the domain is filled initially. For the stochastic model we have
\begin{align}\label{eq:ICBCstochasticLatticeDependence}
N_k&= 100 h,\quad k=1,\dots,\frac{k_{max}}{10}, \nonumber\\
N_k&= 0 ,\quad k=\frac{k_{max}}{10}+1,\dots,k_{max}, \nonumber\\
R_k&=0 ,\quad k=1,\dots,k_{max}. 
\end{align}
As discussed in the previous section, there can be an insignificant rounding error due to the smoothing of the Heaviside function. Figure \ref{fig:vascularInvasionCase1LatticeDependence} shows the results for random movement Case 1.
Looking at the profile of the tip cell densities, we see that the deviation of the stochastic simulation from the result of the continuum equation is larger for smaller values of the lattice constant. This can be understood as we fix the tip cell density, so for smaller lattice constants, there are fewer cells per box. However, this higher noise for smaller lattice constants in the tip cell distribution does not translate into higher noise for the vessel densities.

\begin{figure}[h!]
\subfloat[Tips $n(t,x)$ for h=0.1]
{
\includegraphics[width=0.48\linewidth]{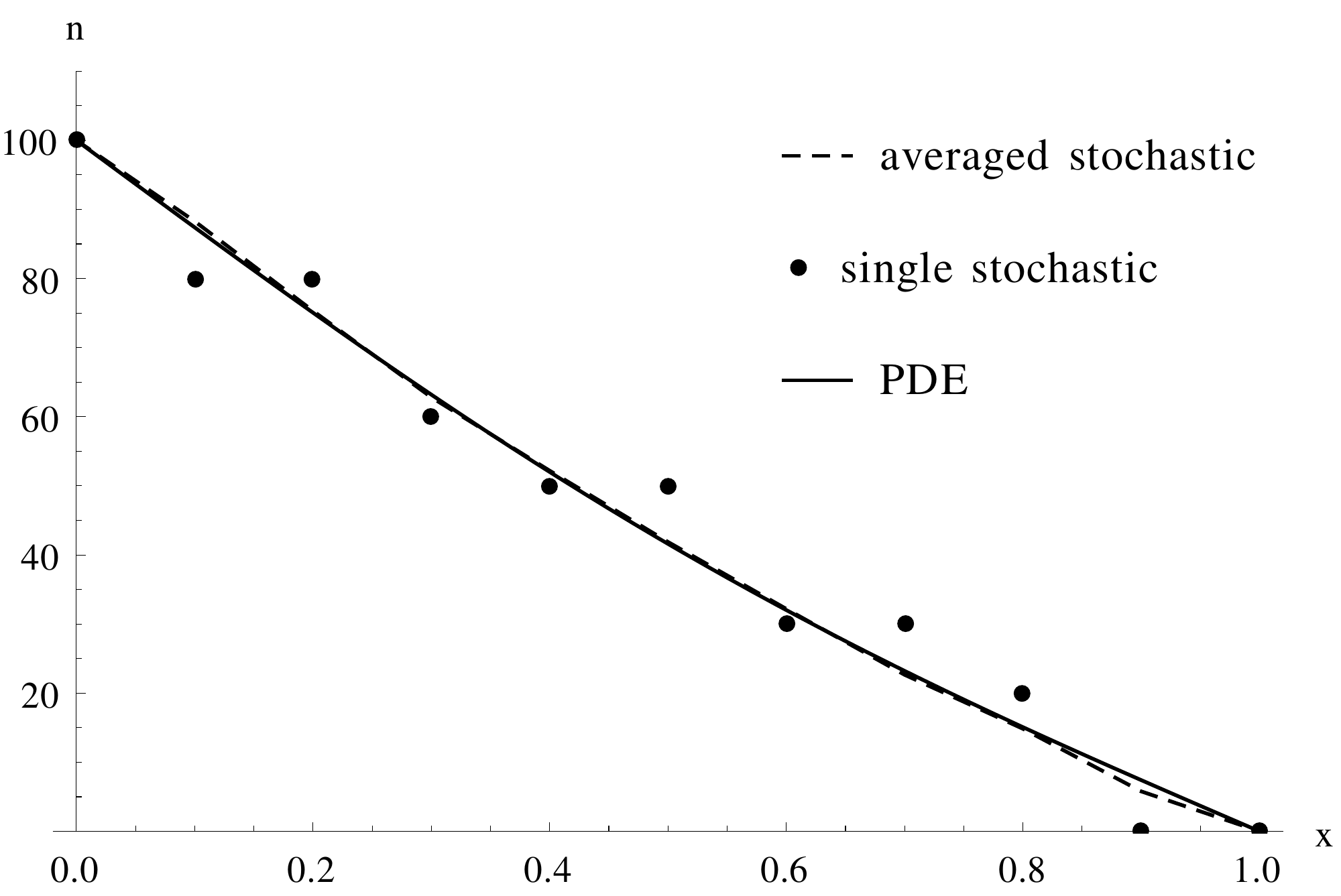}
}
\subfloat[Vessels $\rho(t,x)$ for h=0.1]
{
\includegraphics[width=0.48\linewidth]{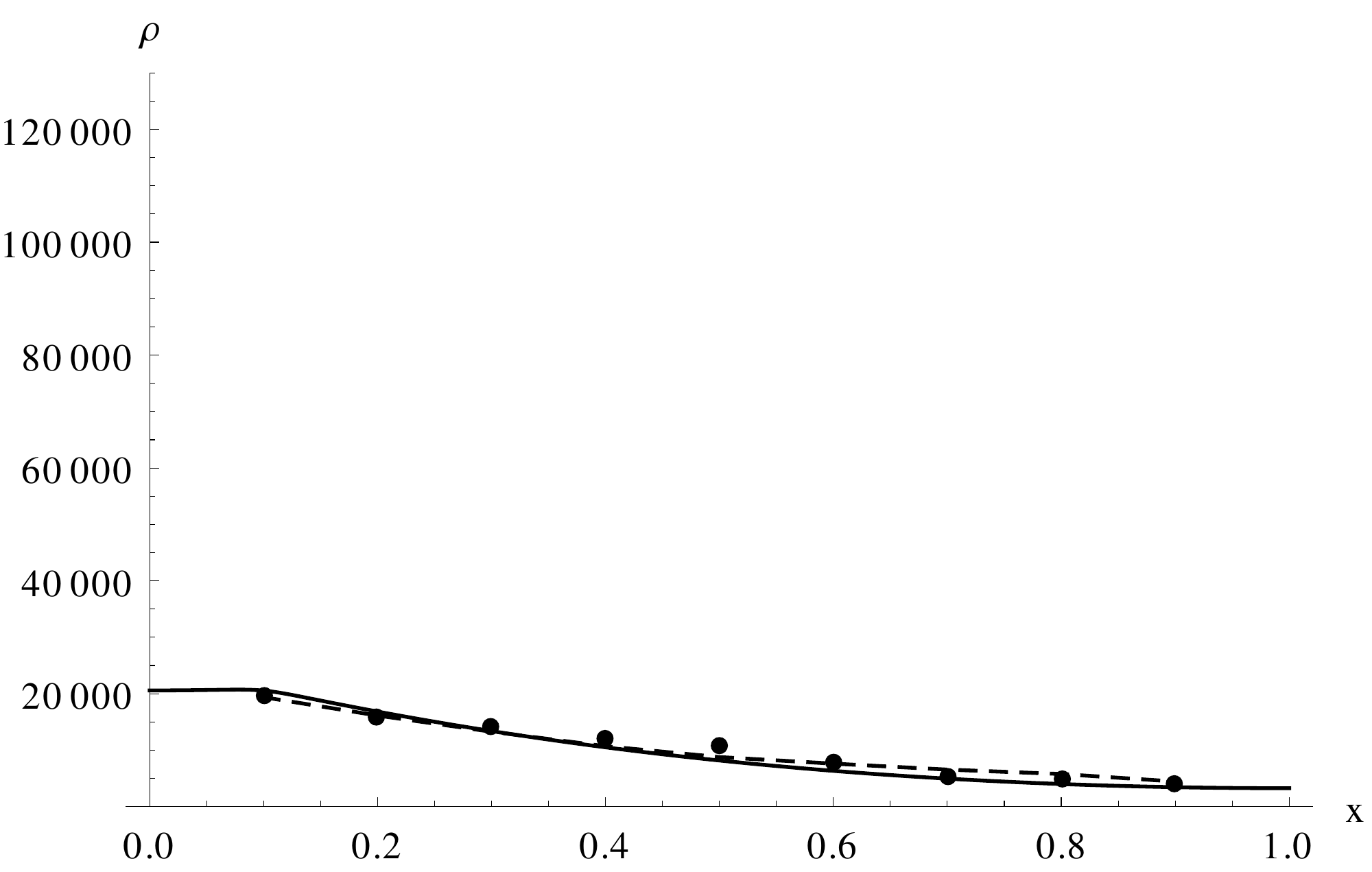}
\label{fig:vascularInvasionCase2LatticeDependenceVessel0p1}
}
\\
\subfloat[Tips $n(t,x)$ for h=0.05]
{
\includegraphics[width=0.48\linewidth]{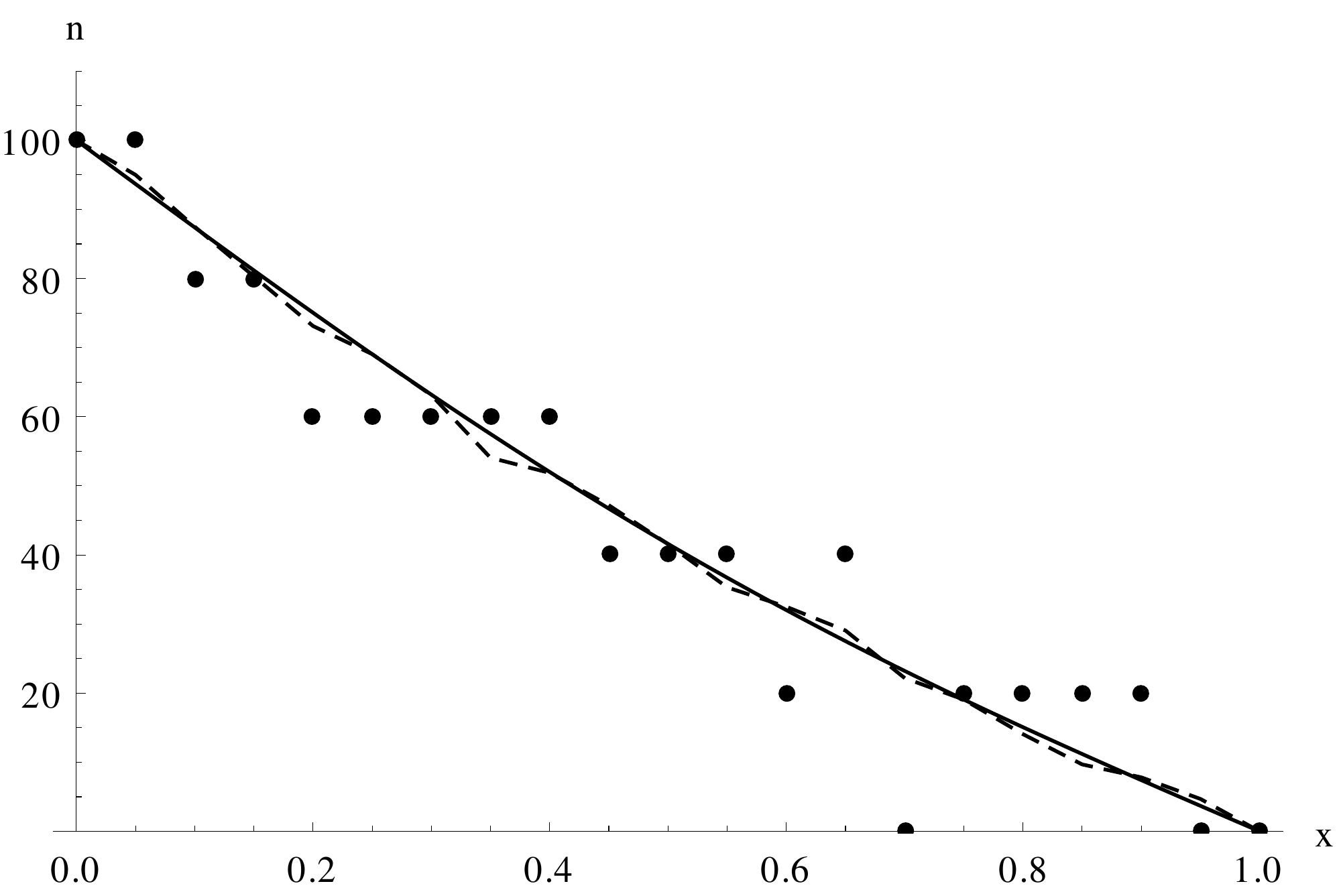}
}
\subfloat[Vessels $\rho(t,x)$ for h=0.05]
{\label{fig:vascularInvasionCase2LatticeDependenceVessel0p05}
\includegraphics[width=0.48\linewidth]{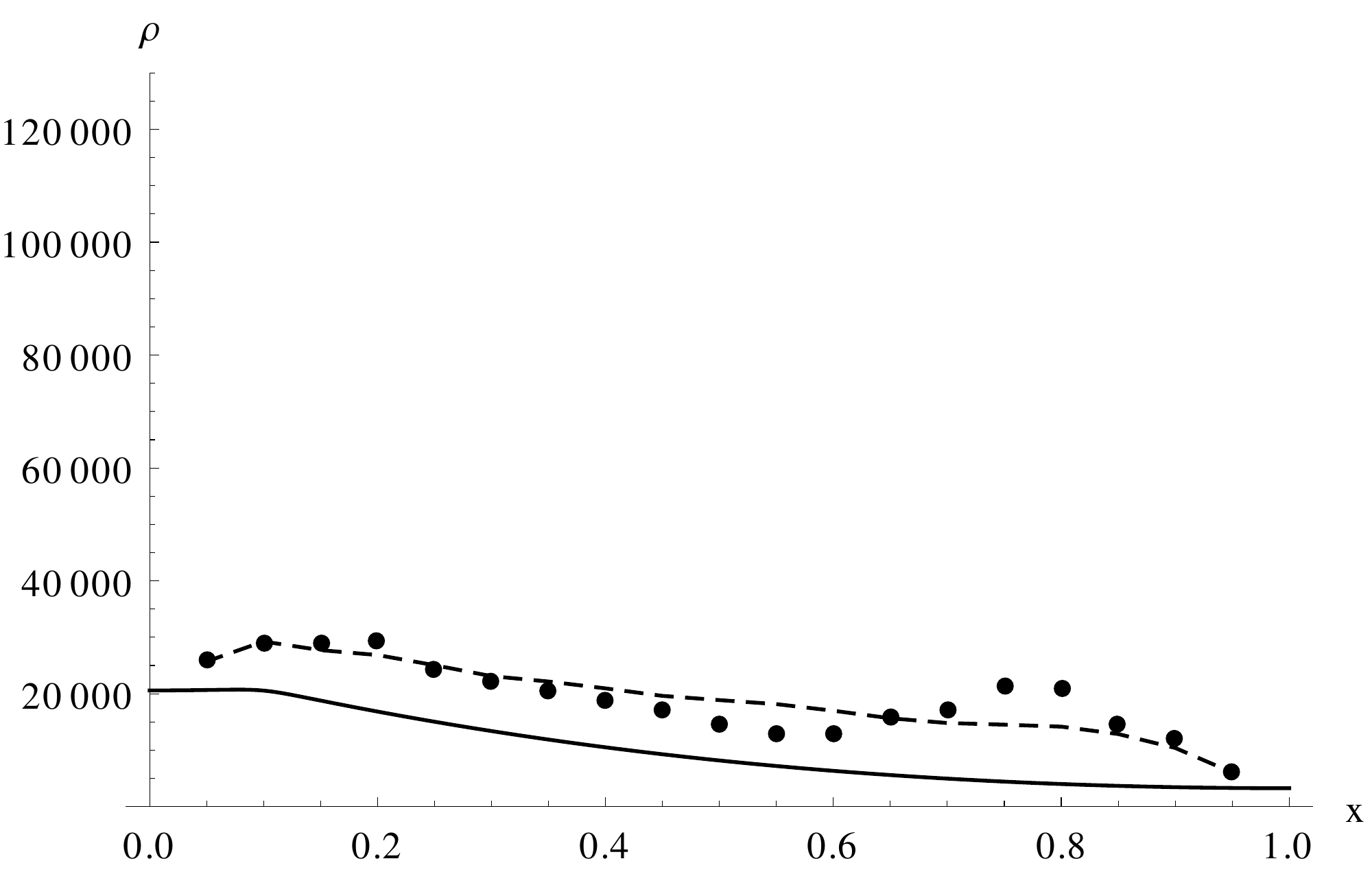}
}
\\
\subfloat[Tips $n(t,x)$ for h=0.02]
{\label{fig:vascularInvasionCase2LatticeDependenceTips0p02}
\includegraphics[width=0.48\linewidth]{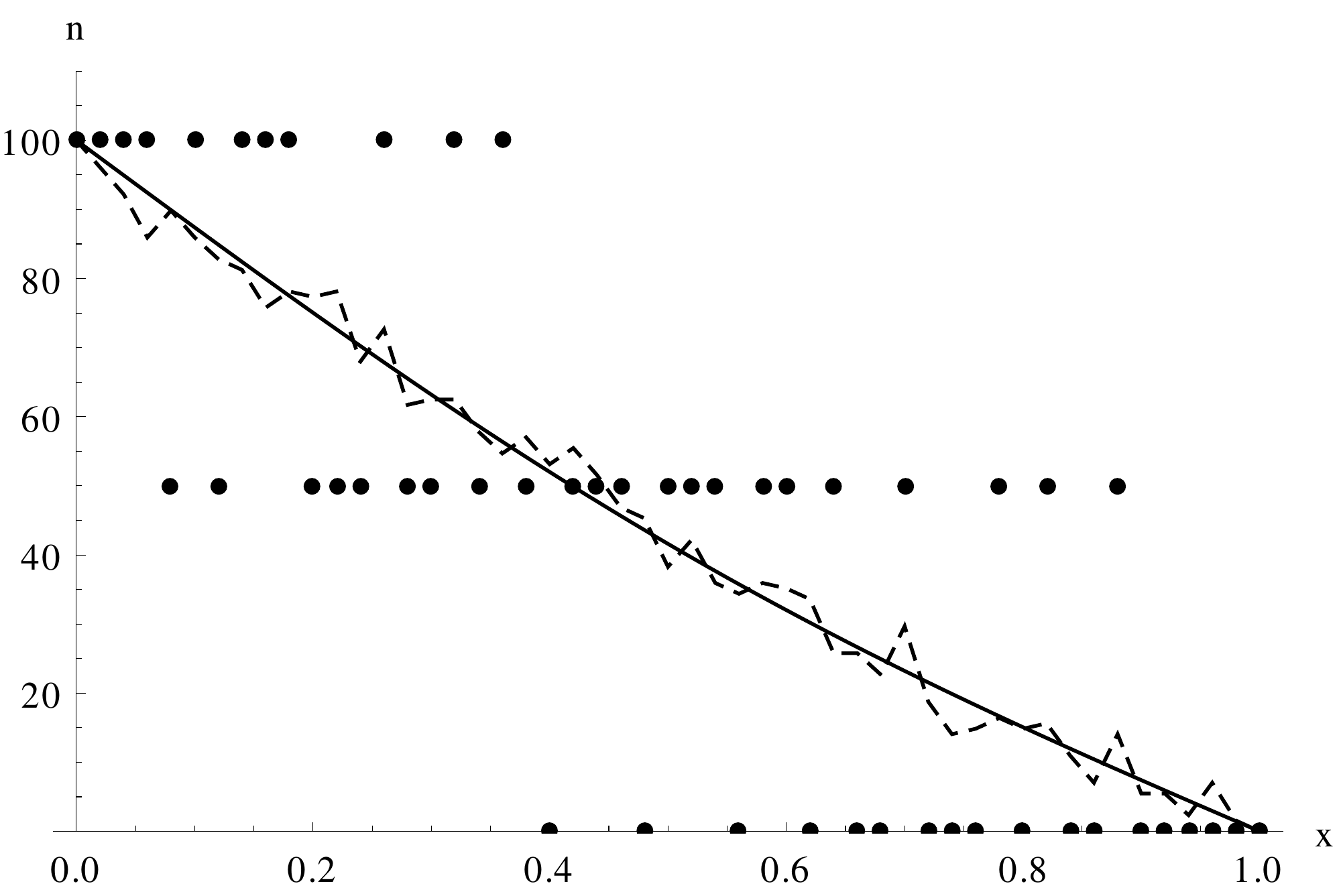}
}
\subfloat[Vessels $\rho(t,x)$ for h=0.02]
{\label{fig:vascularInvasionCase2LatticeDependenceVessel0p02}
\includegraphics[width=0.48\linewidth]{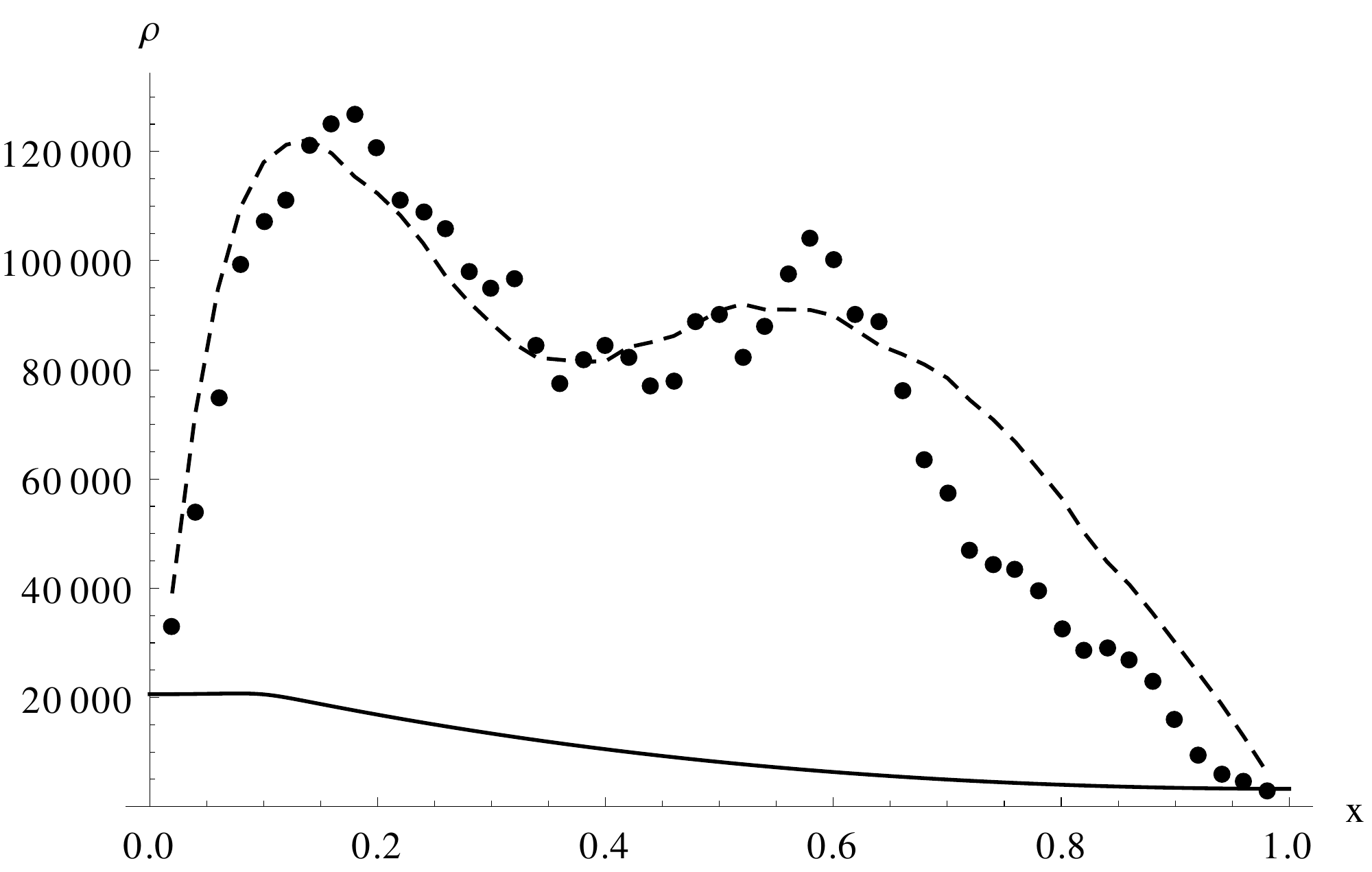}
}
\caption{\label{fig:vascularInvasionCase2LatticeDependence}
Tip cells migrating randomly according to Case 2, transition rate \eqref{eq:transitionRateDiffusionDifference} and PDE \eqref{eq:PDEDiffusionDifference}, and corresponding vessel production, with initial and boundary conditions given in \eqref{eq:ICBCPDELatticeDependence} and \eqref{eq:ICBCstochasticLatticeDependence}.
} 
\end{figure}

The corresponding results based on random movement for Case 2 are shown in Figure \ref{fig:vascularInvasionCase2LatticeDependence}.
As for Case 1, a smaller lattice constant leads to more noise, because there are fewer cells per box. However, in contrast to Case 1, here the higher noise translates into a higher rate of vessel production relative to the continuum model. Indeed, for a lattice constant $h=0.1$, Figure~\subref*{fig:vascularInvasionCase2LatticeDependenceVessel0p1} shows very good agreement between the continuum and stochastic models, whereas for a lattice constant of $h=0.02$ the stochastic model produces considerably more vessel cells than the continuum model, as can be seen in Figure~\subref*{fig:vascularInvasionCase2LatticeDependenceVessel0p02}. This behaviour is in contrast to the usual relation between a continuum model and its finite difference approximation: they typically agree when the lattice constant is small. As explained in the previous subsection, the difference form of the transition rate \eqref{eq:transitionRateDiffusionDifference} implies more movement of tip cells and vessel production when the tip profile is noisy compared to the profile obtained from the PDE. The noise is reduced the more cells one has per box. For a fixed number of total cells, a higher number of cells per box is obtained when choosing a larger lattice constant.

Note also the oscillatory pattern in the vessel profile in Figure \subref*{fig:vascularInvasionCase2LatticeDependenceVessel0p02}. This is an artefact associated with the difference form of the transition rate \eqref{eq:transitionRateDiffusionDifference} that is apparent at low tip cell numbers. As we can see from the tip profile in Figure \subref*{fig:vascularInvasionCase2LatticeDependenceTips0p02}, the spatial domain splits into three subdomains: in the left third, there are typically $2$ tips per box; in the middle third, there is typically only $1$ tip per box; and, in the right third of the domain, there are typically no tips per box. Due to the difference form of the transition rate, tips cannot migrate between boxes with the same number of tips in them. Mathematically, each subdomain is in, or close to, an absorbing state of the stochastic model. Hence, movement of tips is concentrated at the interfaces between the subdomains, and we obtain two peaks in Figure \subref*{fig:vascularInvasionCase2LatticeDependenceVessel0p02}. In general, we have checked that we obtain $N^0$ peaks if we have $N^0$ tip cells at the left boundary, and $0$ at the right, as then $N^0+1$ subdomains form in the tip profile. However, the larger $N^0$, the more the subdomains are blurred by stochastic fluctuations, so the $N^0$ peaks in the vessel profile will be less pronounced the larger $N^0$ is chosen.

\subsection{Chemotaxis}\label{sec:SimChemotaxis}

\begin{figure}[h!]
\subfloat[Tips $n(t,x)$]
{
\includegraphics[width=0.48 \linewidth]{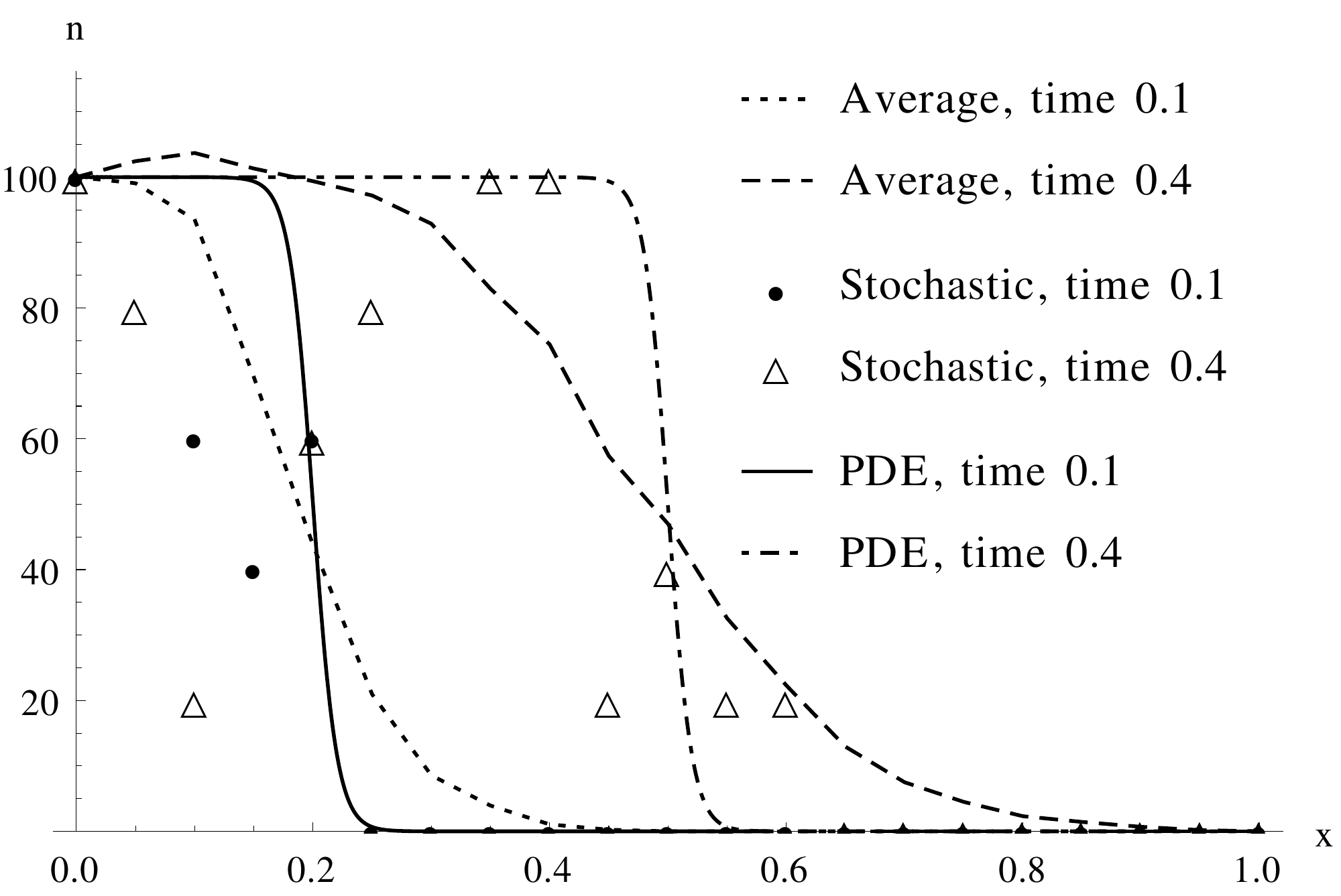}
}
\subfloat[Vessels $\rho(t,x)$]
{
\includegraphics[width=0.48 \linewidth]{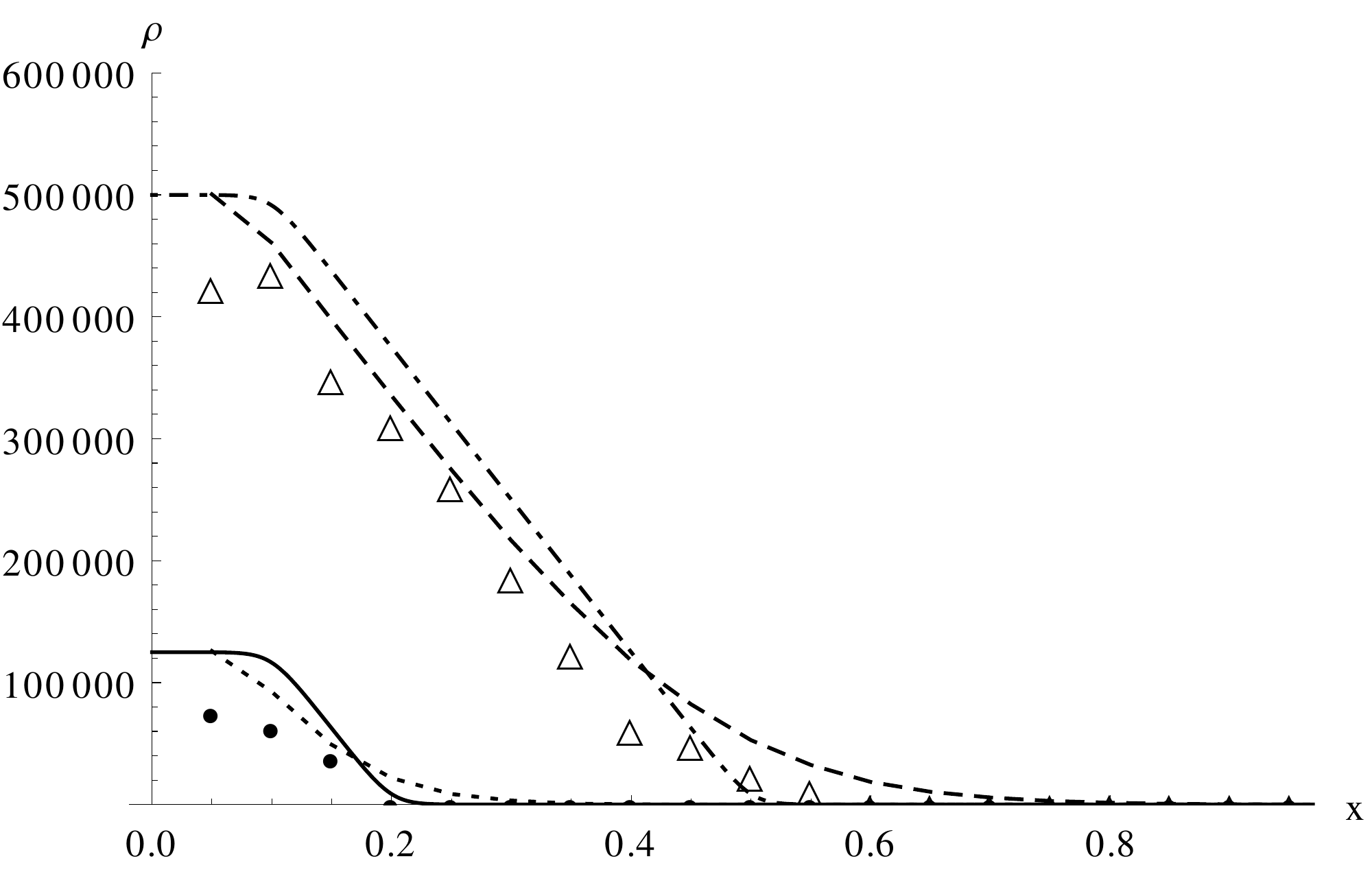}
}
\caption{
\label{fig:vascularInvasionChemotaxis}
Tip cells migrating by chemotaxis, according to transition rate \eqref{eq:transitionRateChemotaxis} with initial conditions as in \eqref{eq:ICBCstochasticLatticeDependence} and \eqref{eq:ICBCPDELatticeDependence} with initially $N^0=5$ in the first $2$ boxes. The lattice constant is $h=0.05$. The averaging of the stochastic model is over 64 realisations. In each case the spatial distribution of tips or vessels respectively is shown at two times.} 
\end{figure}

We repeat the analysis from the previous subsection in the case of chemotaxis, where the stochastic model is defined by transition rate \eqref{eq:transitionRateChemotaxis}, and the corresponding continuum model is given by \eqref{eq:PDEchemotaxis}. On a lattice with $21$ sites and spacing $h=0.05$, we use the initial and boundary conditions as in \eqref{eq:ICBCstochasticLatticeDependence} and \eqref{eq:ICBCPDELatticeDependence} with $N^0=5$ tip cells. As the PDE involves only first order spatial derivatives, we only fix the left boundary.
As well as sprouting, anastomosis and vessel regression, we also neglect the dynamics of the angiogenic factor. We assume that the AF is supplied from a source on the right-hand side of the domain, that it is removed via a sink on the left-hand boundary, and that the AF is already distributed in a steady state. Hence, without loss of generality, we choose 
$\chi c(t,x) = x$ at all times (the value of $\chi$ is arbitrary for the pure chemotaxis model and can be absorbed by suitable non-dimensionalisation). The results of the simulations for chemotaxis are shown in Figure \ref{fig:vascularInvasionChemotaxis}.
As for the random movement Case 1, we observe a marked difference in the tip profile between a single stochastic realisation and the PDEs, but relatively good agreement for the total number of vessels produced. Note the difference in qualitative shape of the vessel profile compared to both cases of random movement, Figures \ref{fig:vascularInvasionLow} and \ref{fig:vascularInvasionHigh}: the spatial profile is piecewise linear and grows at a constant rate in regions where we have a constant level of tip cells (the left part of the domain). This is due to the choice of a constant gradient of the angiogenic factor.

\begin{figure}[h!]
\subfloat[Tips $n(t,x)$ for h=0.1]
{
\includegraphics[width=0.48 \linewidth]{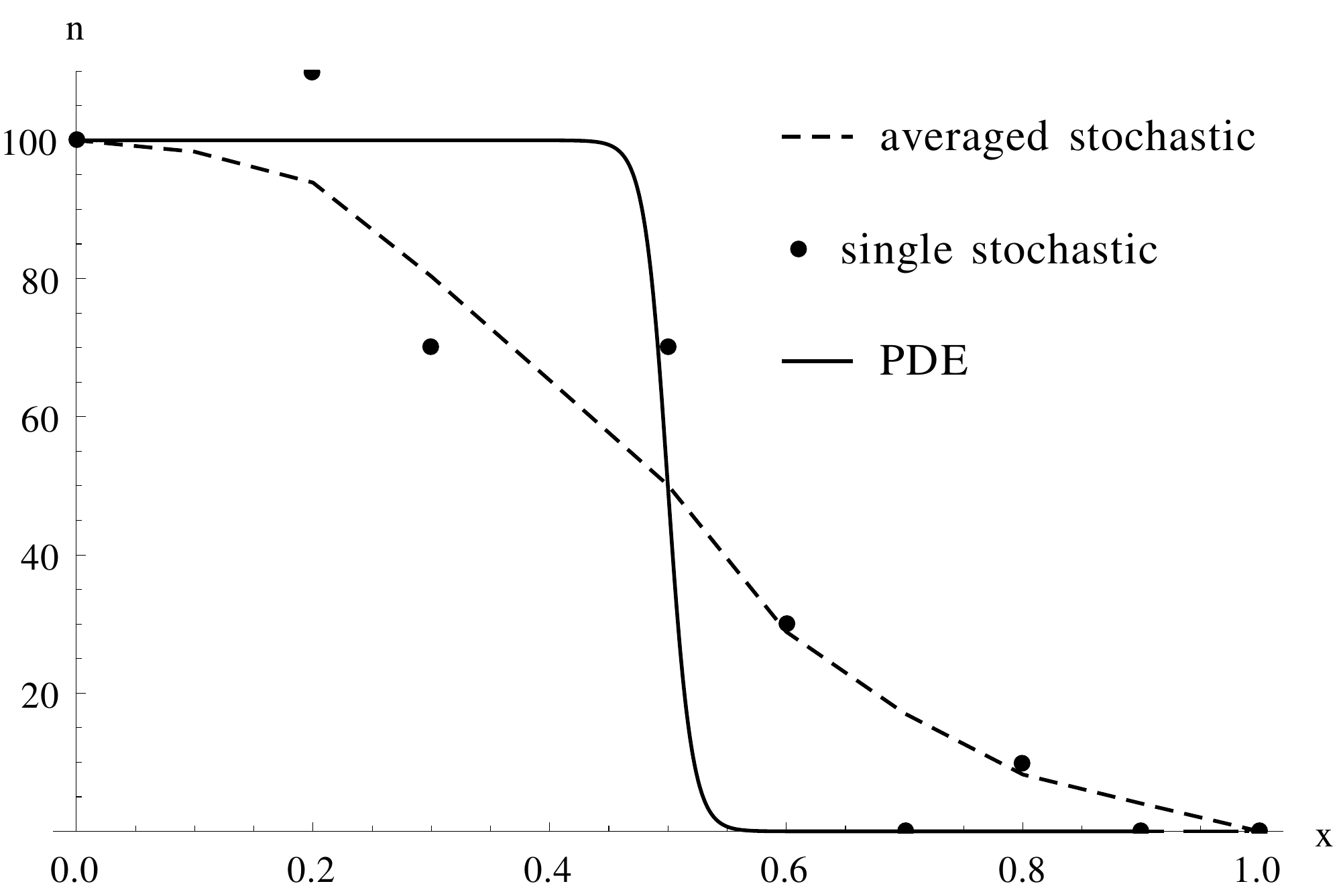}
}
\subfloat[Vessels $\rho(t,x)$ for h=0.1]
{
\includegraphics[width=0.48 \linewidth]{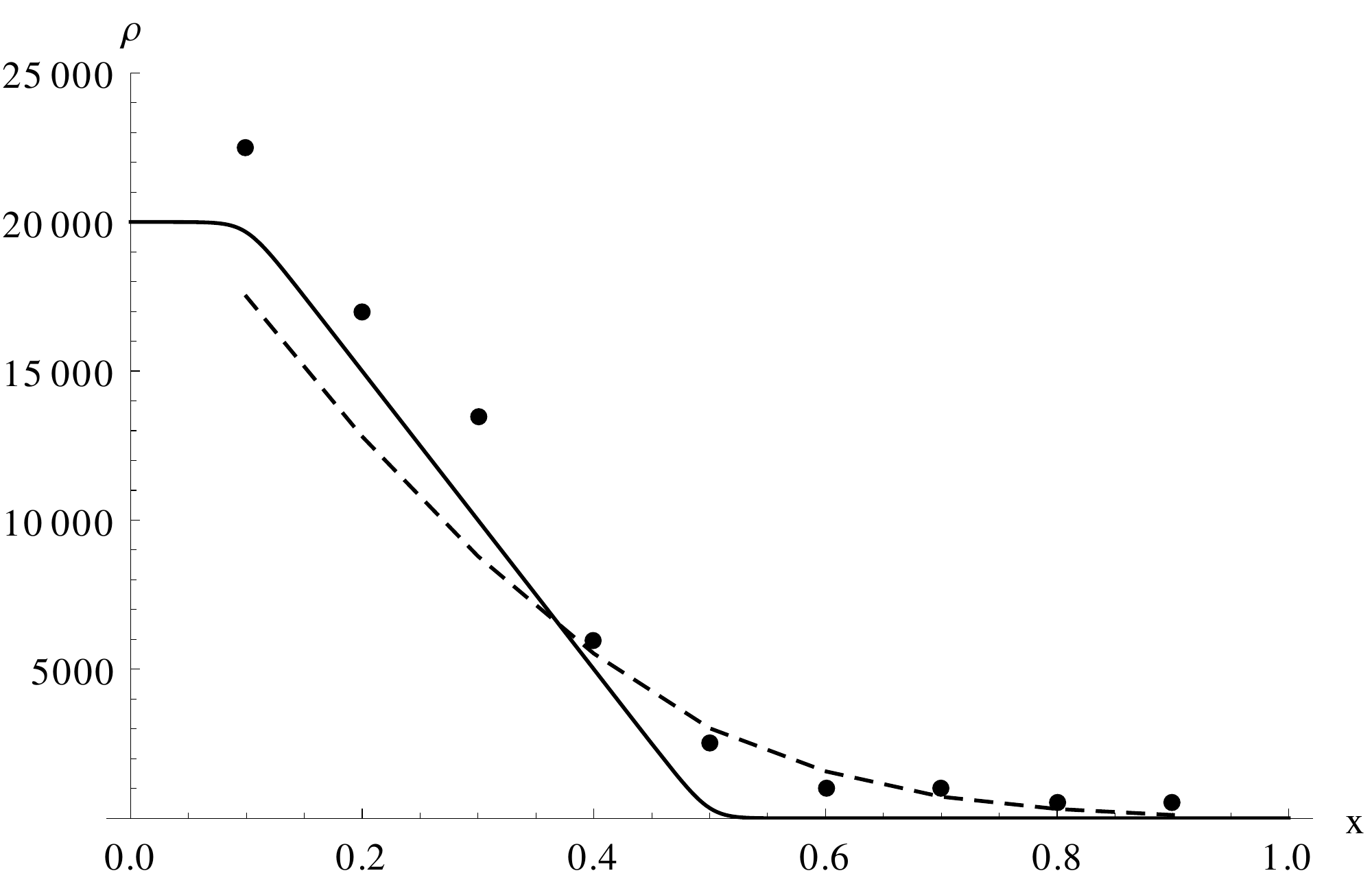}
}
\\
\subfloat[Tips $n(t,x)$ for h=0.05]
{
\includegraphics[width=0.48 \linewidth]{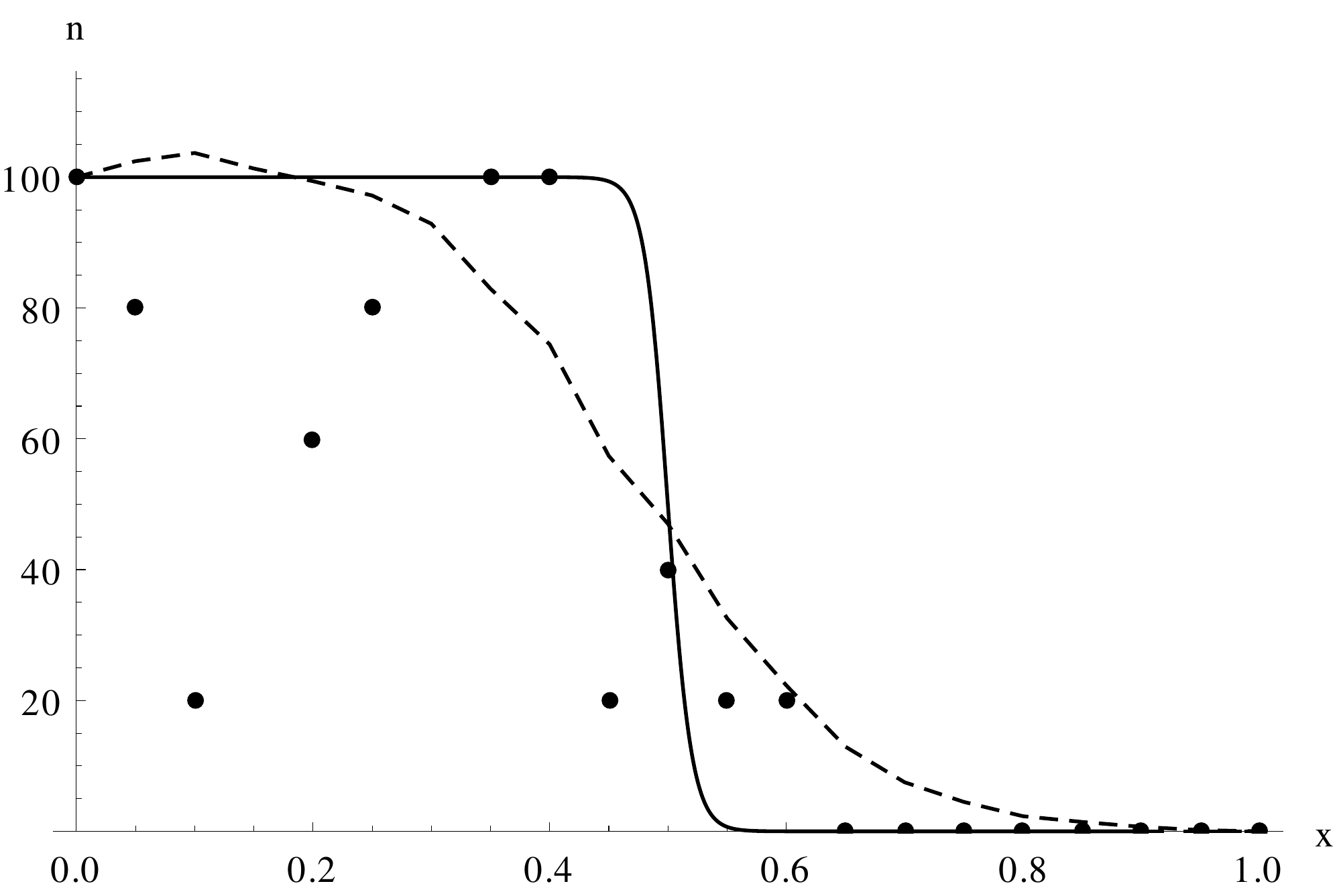}
}
\subfloat[Vessels $\rho(t,x)$ for h=0.05]
{
\includegraphics[width=0.48 \linewidth]{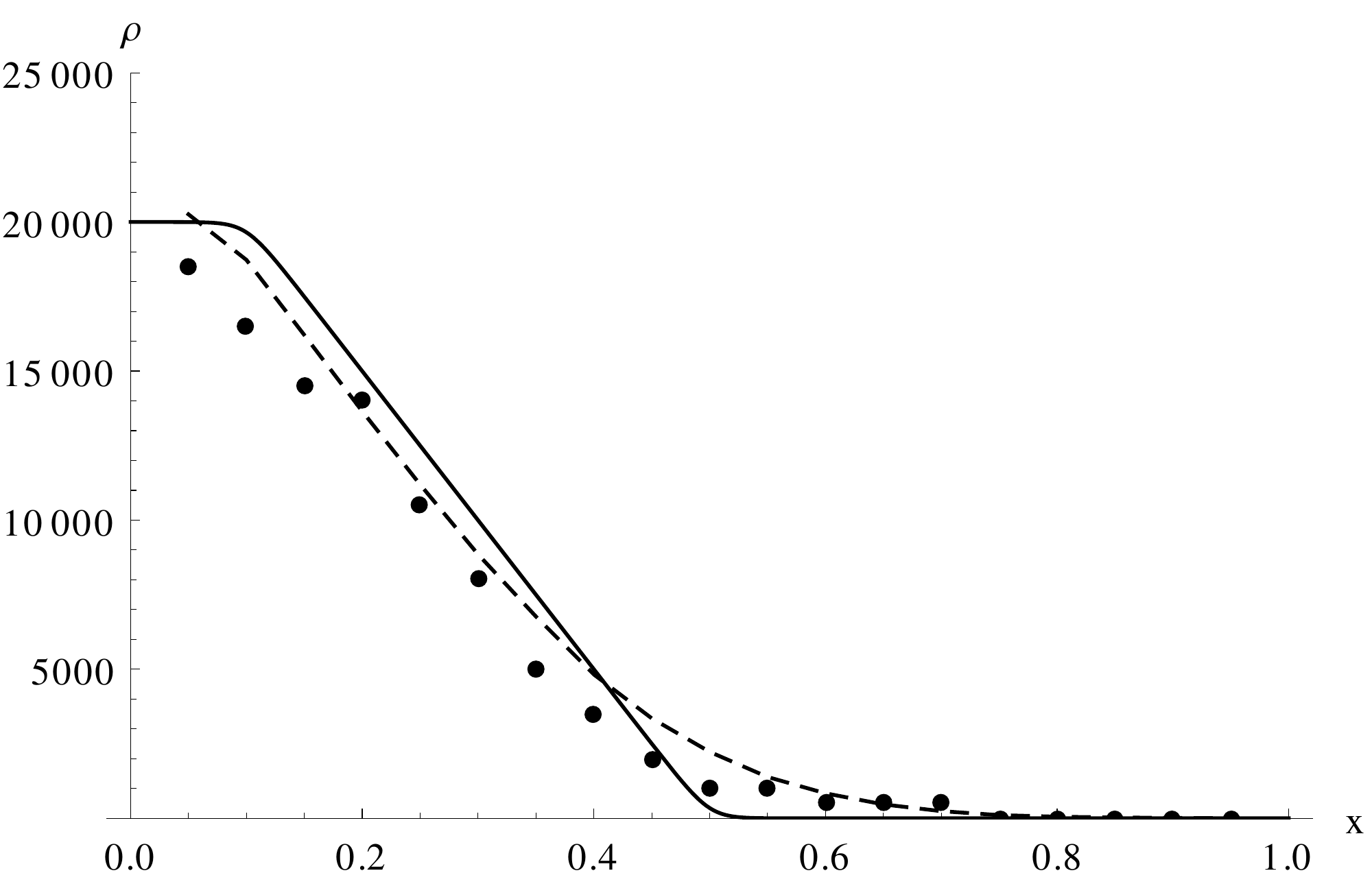}
}
\\
\subfloat[Tips $n(t,x)$ for h=0.02]
{
\includegraphics[width=0.48 \linewidth]{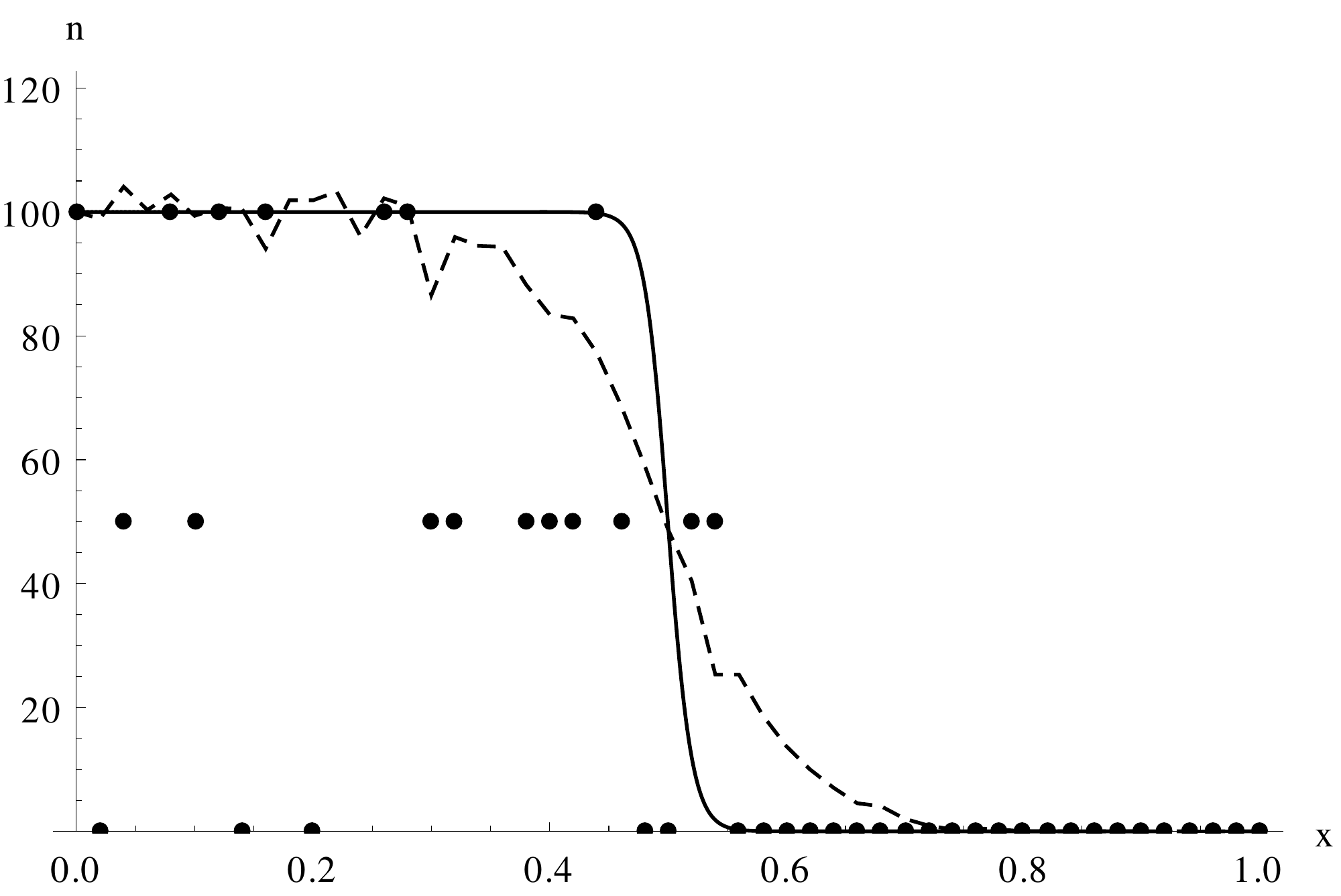}
}
\subfloat[Vessels $\rho(t,x)$ for h=0.02]
{
\includegraphics[width=0.48 \linewidth]{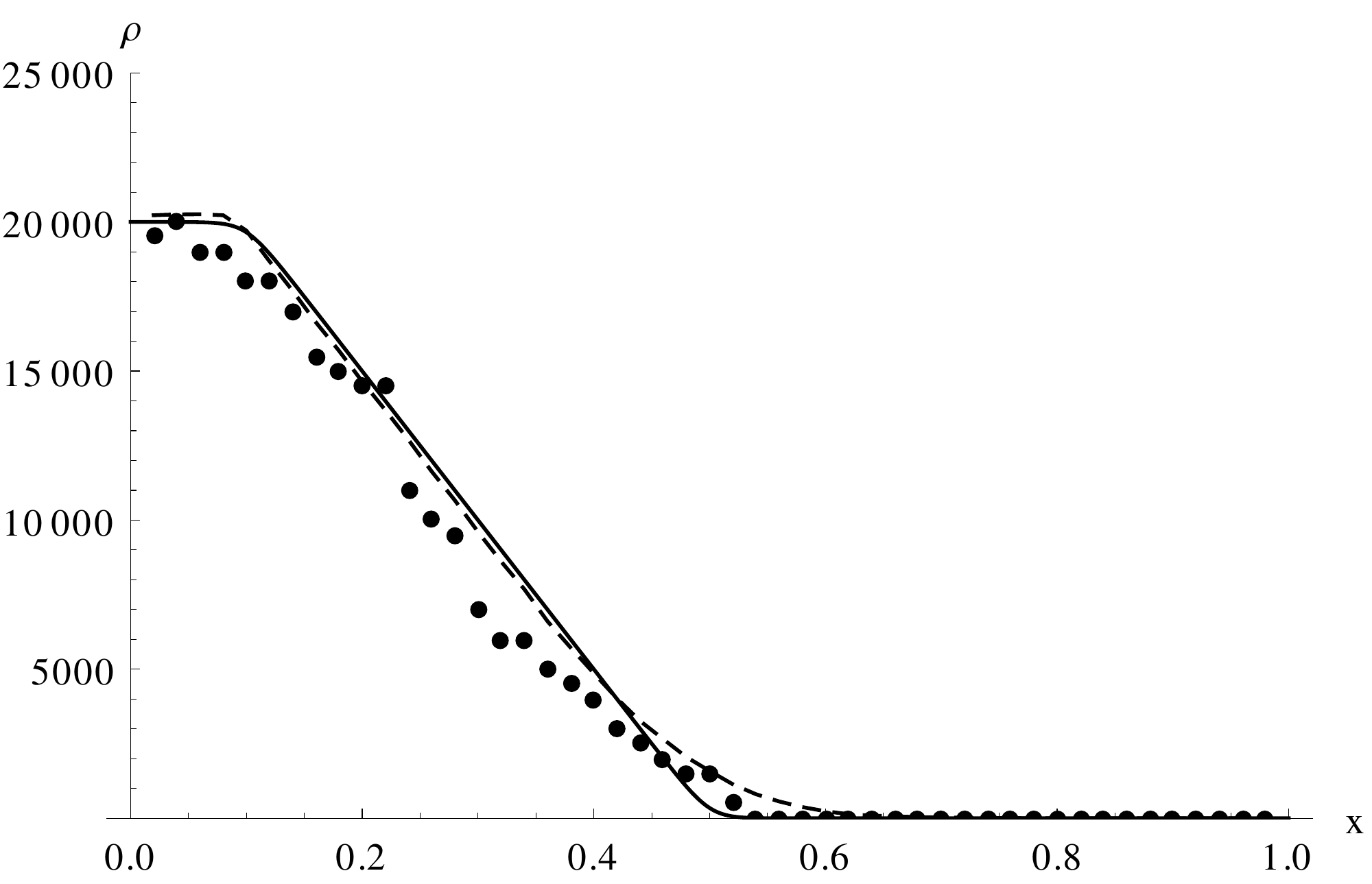}
}
\caption{\label{fig:vascularInvasionChemotaxisLatticeConstantDependence}
Tip cells migrating by chemotaxis, according to transition rate \eqref{eq:transitionRateChemotaxis} from a parent vessel to the left of the domain to the right. All plots shown at time $t=0.4$. The averaging of the stochastic model is over 1280 realisations.
} 
\end{figure}

A qualitative difference between the stochastic and PDE simulations in Figure \ref{fig:vascularInvasionChemotaxis} is that for the tip profile, the front of the wave in the stochastic model is less sharp. Due to the finite size of the lattice we have diffusive effects in the stochastic model, even though there is no explicit diffusion term. Whereas in the PDE model the wave front moves with constant speed, in the discrete stochastic model the movement of the front is composed of jumps by individual cells. This process cannot be sharp. However, we expect this effect will decrease as the lattice constant decreases. Indeed, we confirm this with the simulations shown in Figure \ref{fig:vascularInvasionChemotaxisLatticeConstantDependence}.
%

\subsection{Diffusion and chemotaxis}\label{sec:SimDiffusionAndChemotaxis}
\begin{figure}[h!]
\subfloat[Tips $n(t,x)$ at $t=0$]
{
\includegraphics[width=0.48 \linewidth]{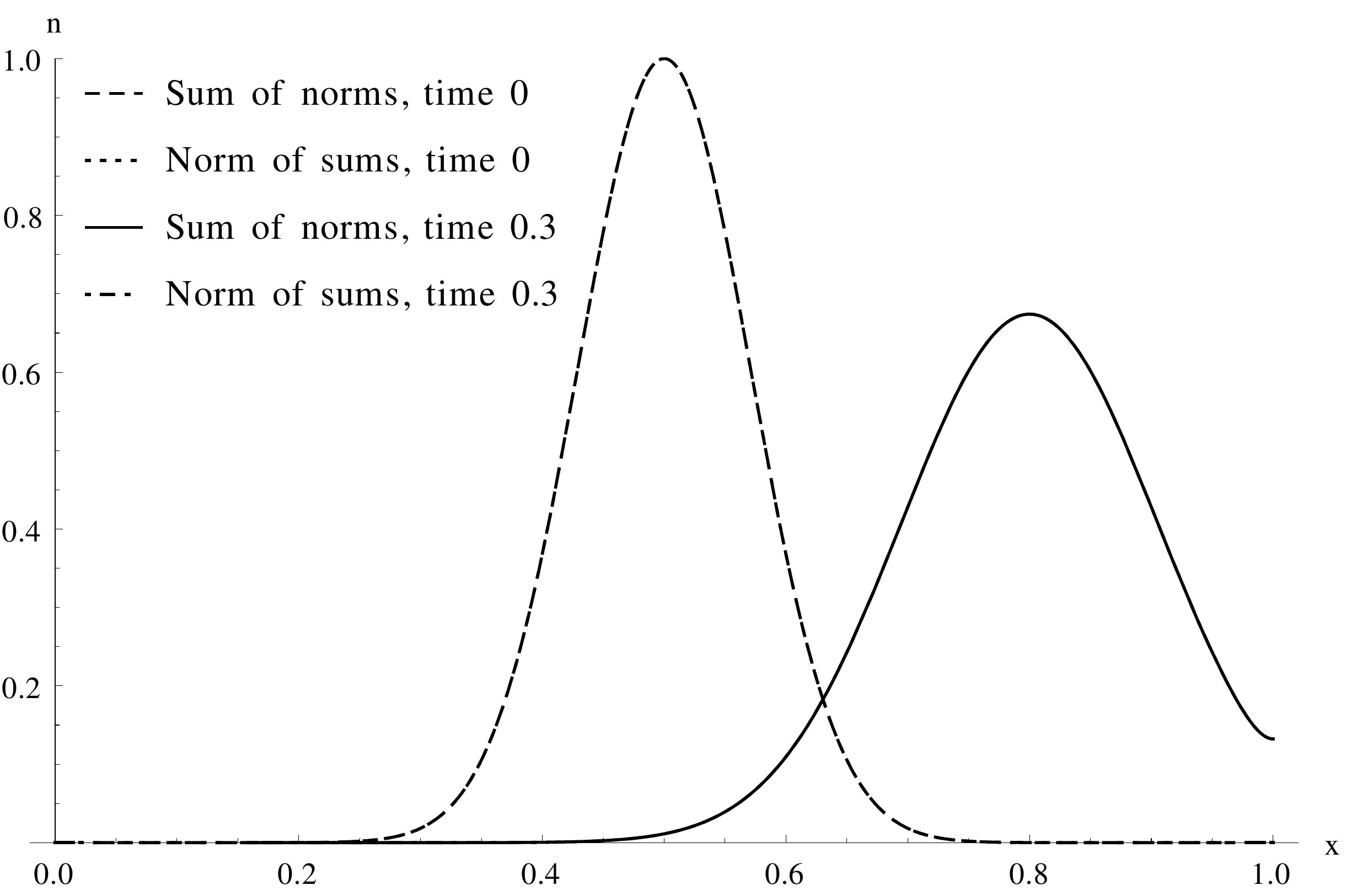}
\label{fig:vascularInvasionChemotaxisPlusDiffusionTips}
}
\subfloat[Vessels $\rho(t,x)$ at $t=0$]
{
\includegraphics[width=0.48 \linewidth]{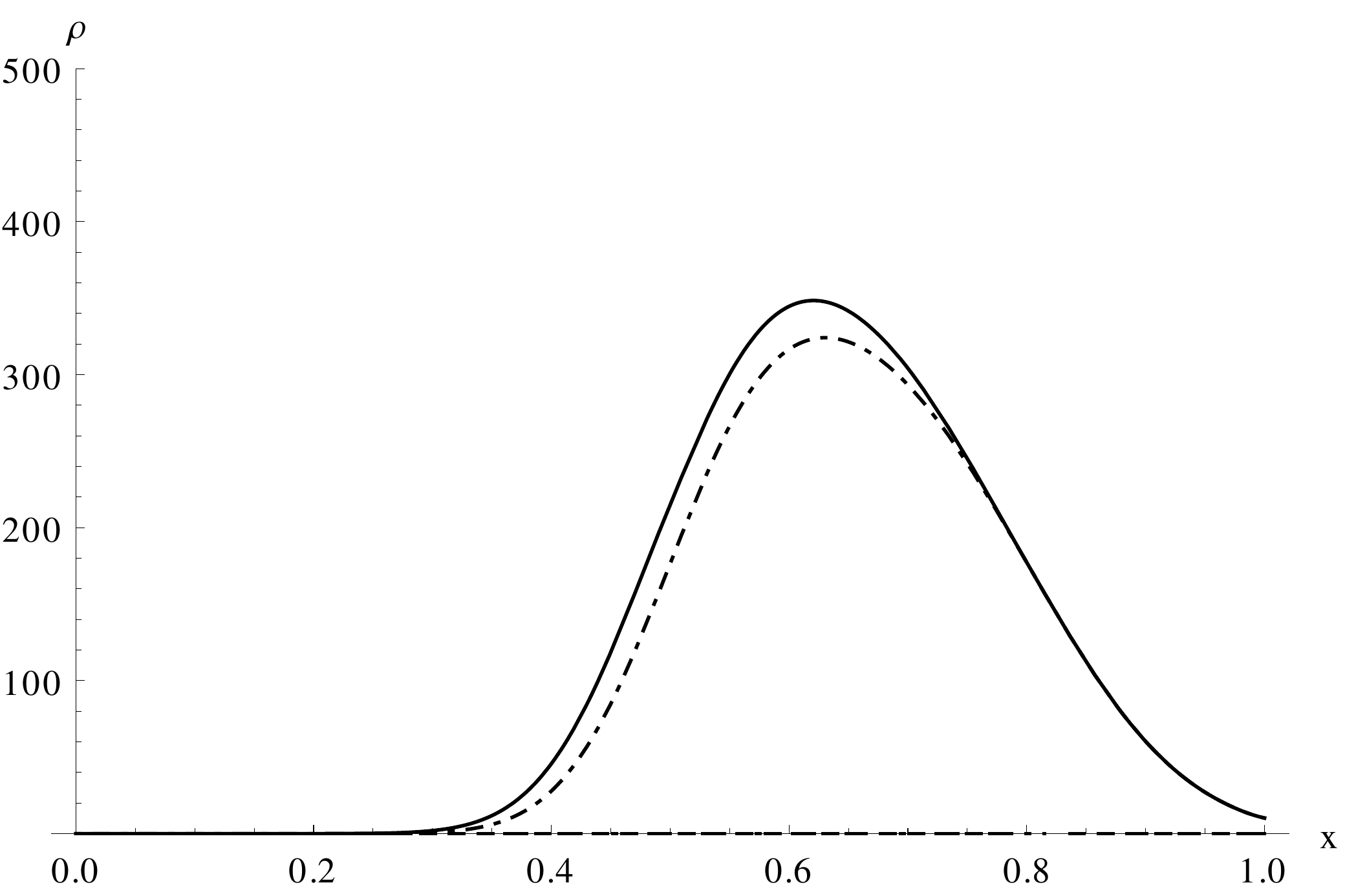}
\label{fig:vascularInvasionChemotaxisPlusDiffusionVessel}
}
\caption{\label{fig:vascularInvasionChemotaxisPlusDiffusion} 
Comparison of two PDE models combining chemotaxis and diffusion \eqref{eq:PDETotalMovementDifference1} and \eqref{eq:PDETotalMovementDifference2}, as described in section \ref{sec:MFcombinedMovement}. The tips in both cases behave in the same way, so the respective plots are identical. The vessel profile at time $t=0.3$ shows a slight difference for the two cases.
} 
\end{figure}

We now compare the different possibilities for combining random movement Case 2 with chemotaxis, as outlined in sections \ref{sec:TransitionRateTotalMovement} and \ref{sec:MFcombinedMovement}. We focus on the differences obtained from the PDEs \eqref{eq:PDETotalMovementDifference1} and \eqref{eq:PDETotalMovementDifference2}, ignoring the underlying stochastic models for now. The comparison between the stochastic and continuum models proceeds as per the last subsections. Recall that the difference between \eqref{eq:PDETotalMovementDifference1} and \eqref{eq:PDETotalMovementDifference2} is that the vessel densities evolve over time at rates which are proportional either to the sum of the norms of the diffusive and chemotactic fluxes in \eqref{eq:PDETotalMovementDifference1}, or to the norm of the sum of the fluxes in \eqref{eq:PDETotalMovementDifference2}. Thus, we expect to see differences when the diffusive and chemotactic fluxes are in opposing directions. In the following simulation, we fix 
\beq
\chi c(t,x) = x,\quad D=0.01.\nonumber
\eeq
We impose Neumann boundary conditions, $\delta_R=100$ and prescribe an initial distribution of the tip cells which is smooth and peaks at the middle domain, as shown in Figure~\subref*{fig:vascularInvasionChemotaxisPlusDiffusionTips} for time $t=0$. Since in both cases, the evolution of the tip cells is governed by the same PDE, the graphs in Figure \subref*{fig:vascularInvasionChemotaxisPlusDiffusionTips} are indistinguishable. The diffusive flux will always point away from the peak of the tip distribution, whereas the chemotactic flux always points to the right. Hence, the model based on the norm of the sum of the fluxes will produce fewer vessels in the domain to the left of the peak of tips, which is confirmed in Figure \subref*{fig:vascularInvasionChemotaxisPlusDiffusionVessel}.
%

\section{Simulations of the full angiogenesis model}\label{sec:simulationFullModel}

We will now perform simulations of a full stochastic model of angiogenesis, including terms representing tip cell movement, vessel production, sprouting, anastomosis and vessel regression. We summarise the transition rates discussed in section \ref{sec:stochasticChemotaxis}:
\begin{align}\label{eq:transitionRatesFullModel}
 \mathcal{T}_{N_k-1,N_l+1,R_k+\delta_R,R_l|N_k,N_l,R_k,R_l} &= \pospart{\stochDiff(N_k-N_l)} + \pospart{\stochChemo N_k (c_l-c_k)},\quad l=k\pm 1 \nonumber\\
 \mathcal{T}_{N_{k}+1|N_{k}} &= {\tilde a_0} R_k c_k + {\tilde a_1} N_k c_k, \nonumber \\
 \mathcal{T}_{N_{k}-1|N_{k}} &= {\tilde \beta_1} N_k R_k, \nonumber\\
\mathcal{T}_{R_{k}-1|R_{k}} &= {\tilde \gamma} R_k.
\end{align}
They are chosen in such a way that the corresponding continuum equations, together with the PDE governing the evolution of the concentration of the angiogenic factor, are given by
\begin{align}\label{eq:PDEFullModel}
\frac{\partial n}{\partial t} &= D\frac{\partial^2 n}{\partial x^2} - \chi\frac{\partial}{\partial x}\left(n\frac{\partial c}{\partial x}\right) + a_0\rho c + a_1 H(c-\hat{c})n c - \beta n\rho,\nonumber\\
\frac{\partial \rho}{\partial t} &= \frac{1}{\mu}\left|D\frac{\partial n}{\partial x}\right| +\frac{1}{\mu}\left|\chi n\frac{\partial c}{\partial x}\right|_1 -\gamma\rho, \nonumber\\
\frac{\partial c}{\partial t} &= D_c\frac{\partial^2 c}{\partial x^2} - \lambda c - a_1H(c-\hat{c})n c.
\end{align}
These continuum equations are almost identical to the continuum model of angiogenesis presented in \cite{byrne1995mathematical}, so we can directly compare the stochastic model to an established continuum model. The only difference is the appearance of the norm (either in this form or as the norm of the sum of the fluxes, as discussed in section \ref{sec:MFcombinedMovement}) in the evolution equation for the vessel densities in \eqref{eq:PDEFullModel}: the corresponding equation in \cite{byrne1995mathematical} or the related publications \cite{balding1985mathematical,chaplain1993model} had no norm. The appearance of the norm looks more natural and is certainly mathematically more consistent, as it ensures that the vessel densities cannot become negative. Furthermore, generalising the model with the norm to higher dimensions is straight-forward, whereas the model without norms makes no sense in higher dimensions (see appendix \ref{sec:app:higherDimensions}). We choose the same 
initial and boundary conditions for the PDEs as in \cite{byrne1995mathematical},
\begin{align}\label{eq:initialAndBCFullModel}
n(0,1) &= n_L, \quad \rho(0,1) = \frac{1}{\mu},\quad n(0,x)=\rho(0,x) = 0 \quad (0\leq x<1)\nonumber\\
c(0,0) &= 1, \quad c(0,x) = 0 \quad (0< x\leq1), \nonumber\\
n(t,1) &= n_L e^{-k t},\quad  \rho(t,1) = \frac{1}{\mu}\rho_{min}+\frac{1}{\mu}(1-\rho_{min})e^{-kt}, \nonumber\\
n(t,0) &= 0,\quad c(t,0) = 1, \quad c(t,1) = 0. 
\end{align}
The parameter values in non-dimensionalised units were given in \cite{byrne1995mathematical,mantzaris2004mathematical}\footnote{Not all parameters were listed in \cite{byrne1995mathematical}, so we take the remaining one from \cite{mantzaris2004mathematical}. Hence, the results we obtain are as in \cite{mantzaris2004mathematical}.} to be 
\begin{align}\label{eq:parameters}
a_0 &= 50\mu,& a_1 &= 10,&\beta &= 50\mu,&\gamma &= 0.25, \nonumber\\
\chi &= 0.4,&D&=10^{-3},& \hat{c} &= 0.2,&k &= 1.5, \nonumber\\
\lambda &= 1,&D_c &=1,& \rho_{min} &= 0.05,& n_L&=1,
\end{align}
where $\mu$ is the average cell length (see the discussion below).
The initial and boundary conditions have the following interpretation: on the boundaries of the domain, $x\in[0,1]$, we have a tumour at $x=0$ and a parent vessel at $x=1$. Note that the PDEs in \eqref{eq:PDEFullModel} do not depend on spatial derivatives of $\rho$, and hence the system \eqref{eq:initialAndBCFullModel} is overspecified and the boundary condition for $\rho$ in \eqref{eq:initialAndBCFullModel} is not required to solve the PDE. We interpret this boundary condition as a source term for new vessel cells. In practice, we solve the equations such that in the spatial discretisation of the PDE, we simply fix the boundary value corresponding to the left-most discretized value of $\rho$. This is consistent as the equation for $\rho$ in \eqref{eq:PDEFullModel} is an ODE at each point in space.

We remark that in \cite{byrne1995mathematical}, the modelling focus was on corneal assays with implanted tumour, where the vessels grow in a quasi two-dimensional environment. The actual model was formulated in one dimension, describing the projection onto a one-dimensional domain between tumour and parent vessel. Tip cell densities were measured in tips per unit area, and the vessel densities in length per 
unit area. In 
the present paper, we chose units as cell numbers per length for the one-dimensional model. As neither the model in \cite{byrne1995mathematical} nor the present paper account for the fact that cell sizes can vary, it is straightforward to translate between the two units: as the vessels form one-dimensional structures, the average length of a vessel segment of one cell is then simply $\mu$. Hence, to translate between our variable $\rho$ and the $\rho$ appearing in \cite{byrne1995mathematical}, we simply multiply by $\mu$. This explains the appearance of $\mu$ in the parameters in \eqref{eq:parameters}. Note that since the experimental setup describes angiogenesis in two spatial dimensions, the reduction to a one dimensional model also works in this simple form if tip movement is not diffusion dominated, as we have to assume that the contribution to movement is negligible for the dimension over which we integrate. However, two-dimensional diffusion is symmetric with respect to both dimensions. As in the experimental setup, migration is dominated by chemotaxis, we integrate over the direction perpendicular to the chemotactic gradient. In the simulation below we fix $\mu=0.1 h$. This is as we argued in earlier sections: the stochastic model is well defined when the cell size is smaller than the box size, so there are several cells per box. 

\begin{figure}[h!]
\subfloat[Tips $n(t,x)$ for $h=0.02$, $N^0=100$]
{
\includegraphics[width=0.48\linewidth]{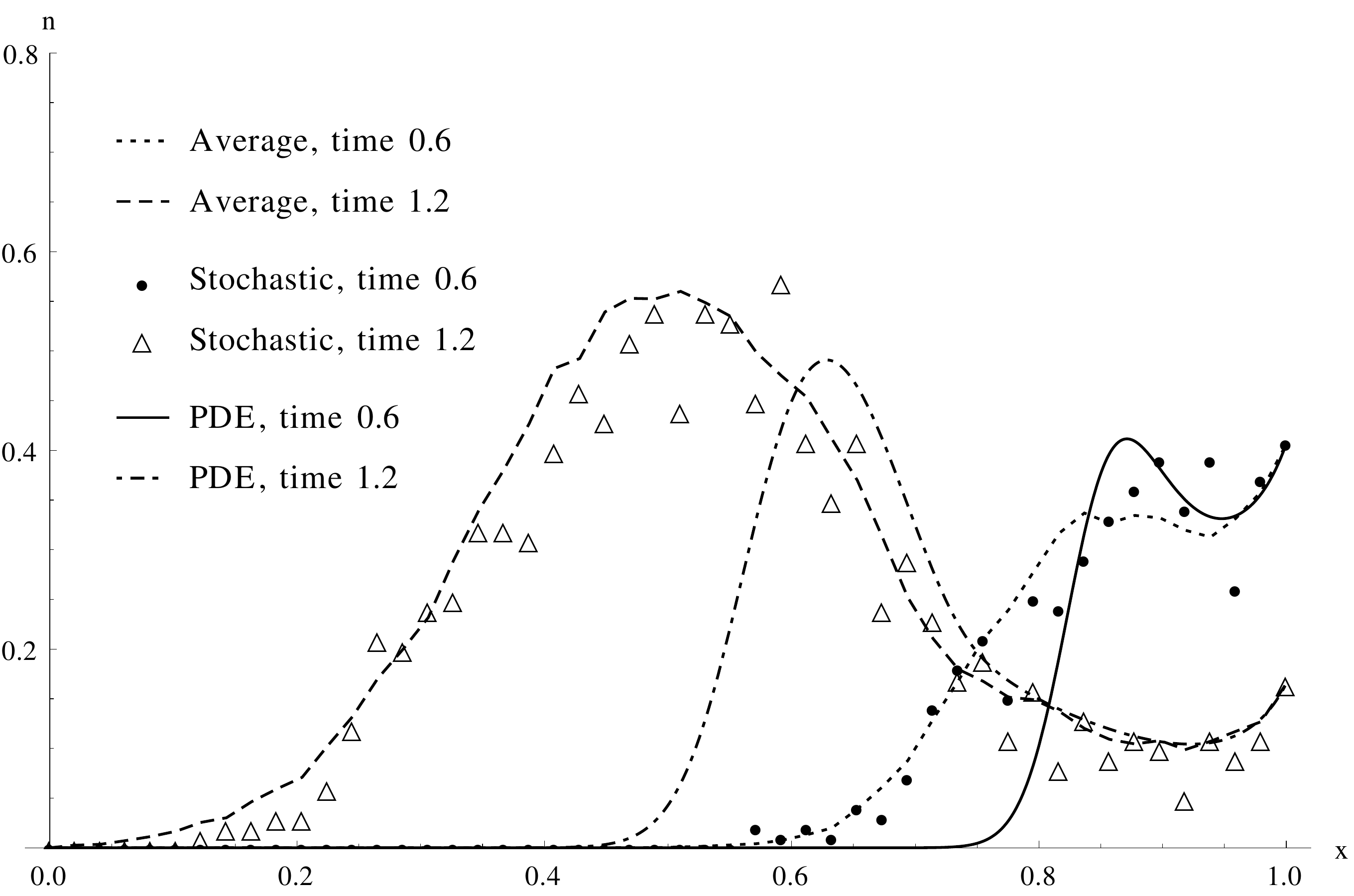}
\label{fig:simulationFullModelTips1}
}
\subfloat[Vessels $\rho(t,x)$ for $h=0.02$, $N^0=100$]
{
\includegraphics[width=0.48\linewidth]{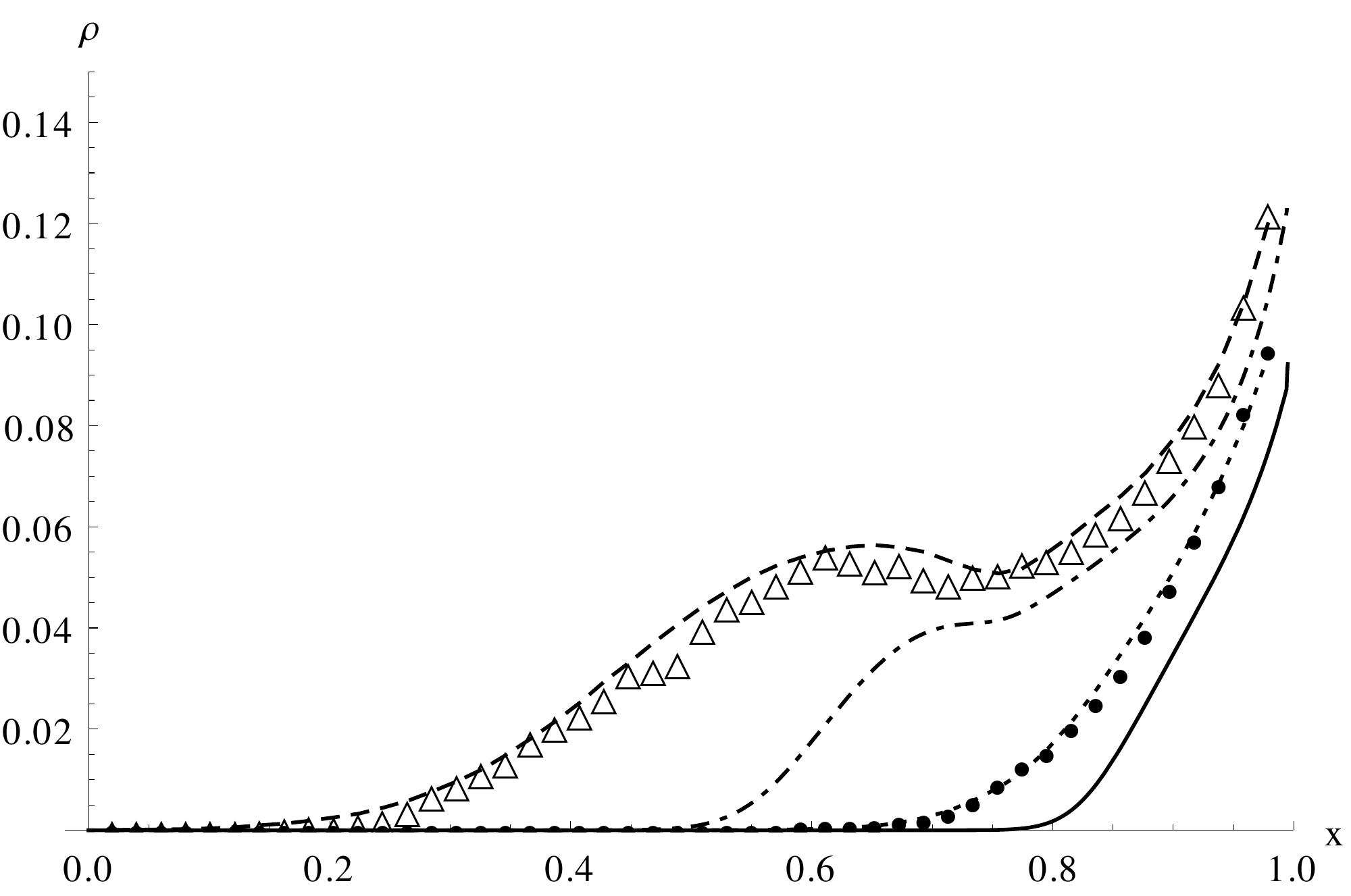}
\label{fig:simulationFullModelVessels1}
}
\\
\subfloat[Tips $n(t,x)$ for $h=0.005$, $N^0=25$]
{
\includegraphics[width=0.48\linewidth]{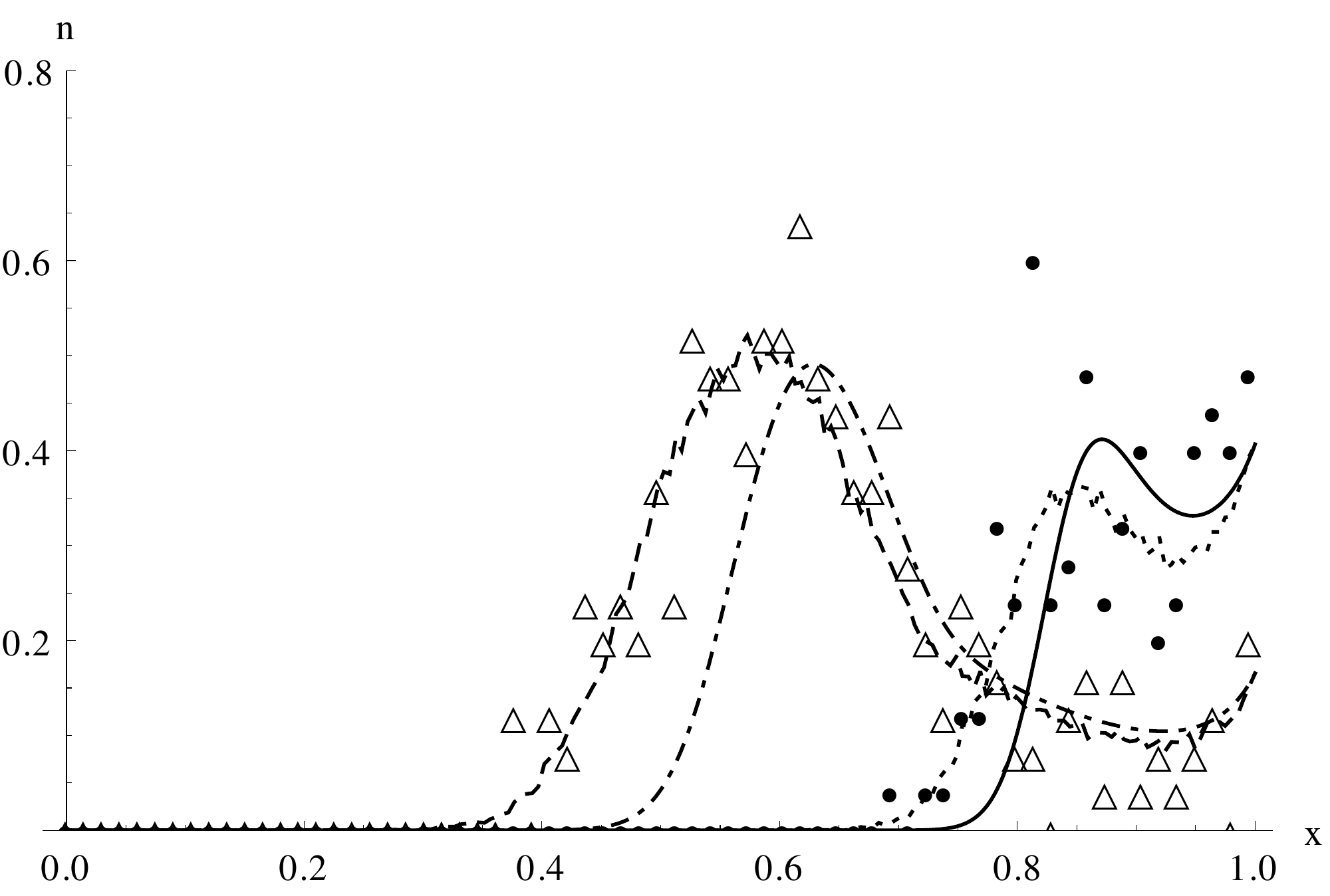}
\label{fig:simulationFullModelTips2}
}
\subfloat[Vessels $\rho(t,x)$ for $h=0.005$, $N^0=25$]
{
\includegraphics[width=0.48\linewidth]{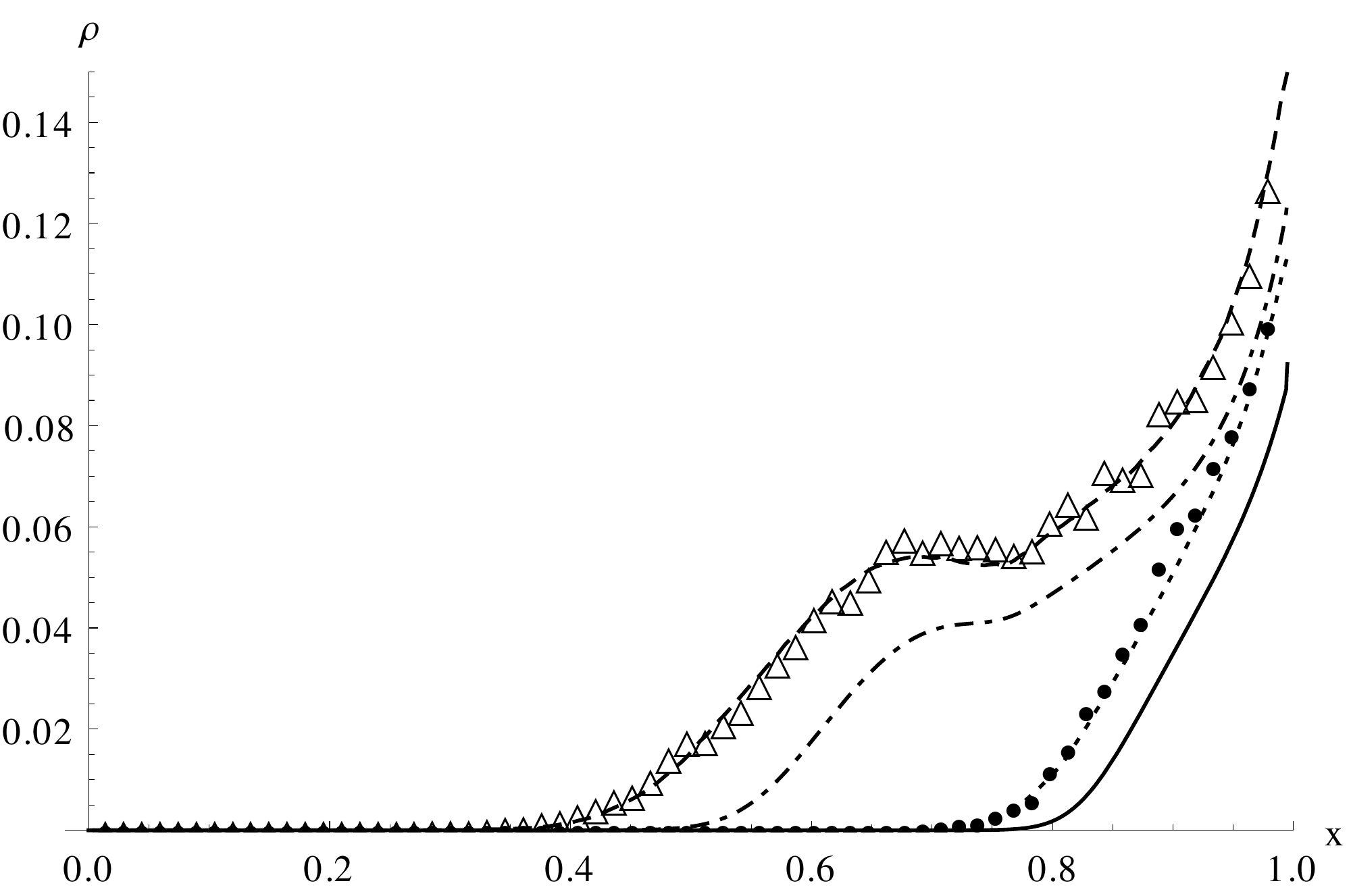}
\label{fig:simulationFullModelVessels2}
}\\
\subfloat[Tips $n(t,x)$ for $h=0.00125$, $N^0=50000$]
{
\includegraphics[width=0.48\linewidth]{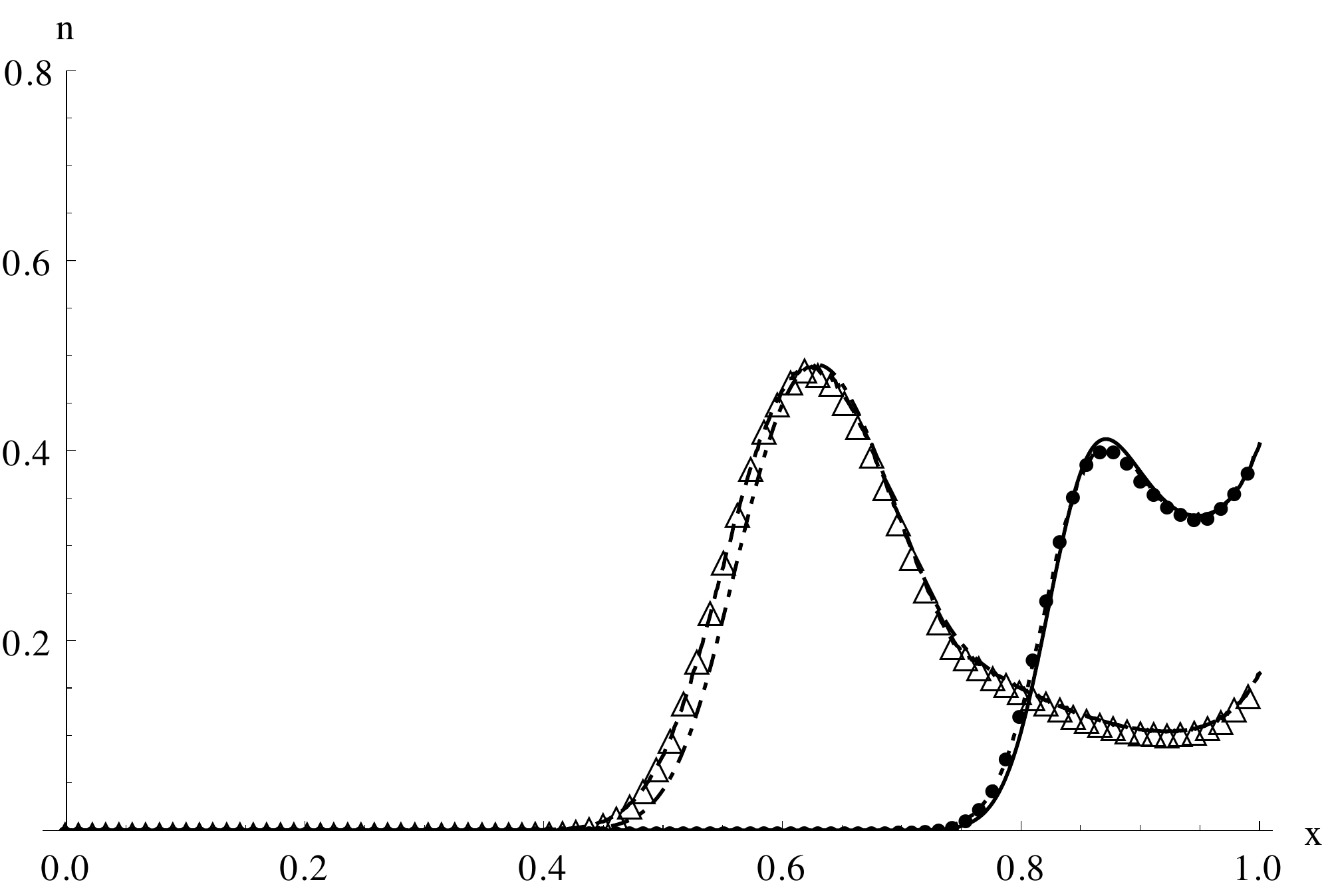}
\label{fig:simulationFullModelTips3}
}
\subfloat[Vessels $\rho(t,x)$ for $h=0.00125$, $N^0=50000$]
{
\includegraphics[width=0.48\linewidth]{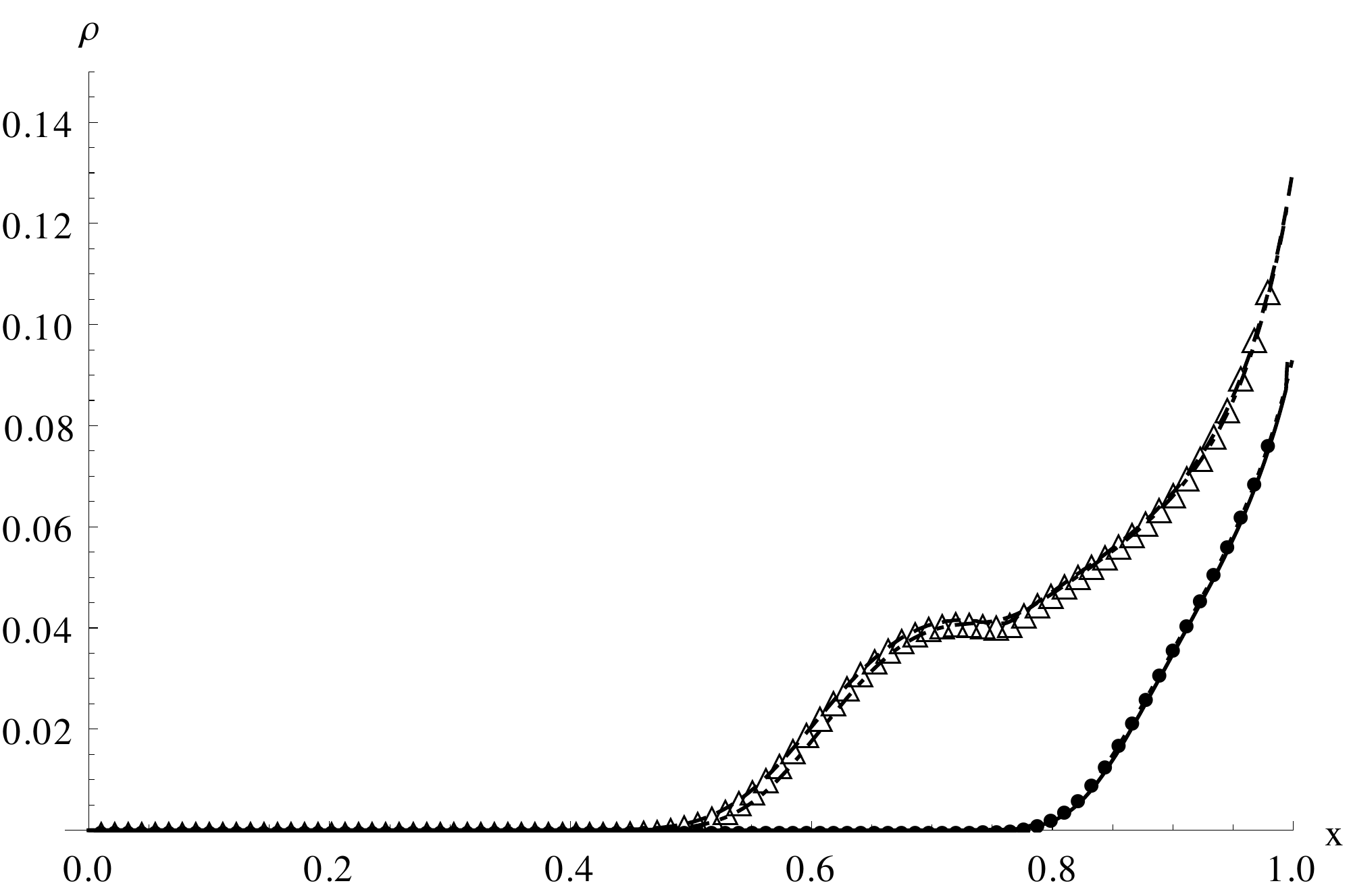}
\label{fig:simulationFullModelVessels3}
}
\caption{\label{fig:simulationFullModel} Results of the PDE \eqref{eq:PDEFullModel} and the stochastic model defined by \eqref{eq:transitionRatesFullModel}. The averaging of the stochastic model is over $100$ realisations. The plots show spatial tip and vessel distributions for different lattice constants and initial numbers of tip cells, at two different times. In \protect\subref{fig:simulationFullModelTips2}-\protect\subref{fig:simulationFullModelVessels3} we have omitted some points in the plot for the stochastic model for better visualisation. We used the units of \cite{byrne1995mathematical} for better comparison.}
\end{figure}

Figure \ref{fig:simulationFullModel} shows the results for the simulation of the PDE \eqref{eq:PDEFullModel} and the stochastic model defined by \eqref{eq:transitionRatesFullModel}. For better comparison with \cite{byrne1995mathematical,mantzaris2004mathematical}, we have rescaled $n$ and $\rho$ by $\frac{h}{N^0}$, where $N^0$ is the initial number of tip cells in the boundary box.
We see in Figures \subref*{fig:simulationFullModelTips1} and \subref*{fig:simulationFullModelVessels1} that when there are initially $N_0=100$ tips on the right boundary, and a lattice spacing of $h=0.02$, the qualitative shapes of the tip and vessel profiles obtained by averaging over $100$ realisations of the stochastic model look similar to those obtained from the PDE, but there are marked quantitative differences. Most importantly, we see that the wave front, for both tips and vessels, in the stochastic model moves considerably faster than in the PDE model. Recall that when we focused on chemotaxis only, as shown in Figure \ref{fig:vascularInvasionChemotaxis}, the wave front in the stochastic model was smeared out compared to the wave front in the PDE model, with some individual tip cells in the stochastic model moving faster than in the PDE model, where all tips were moving at the same speed. Here, in the full model of angiogenesis, those leading tip cells, accompanied by newly formed vessel cells, can sprout and hence increase the magnitude of the tip density at the wave front. 

In Figures \subref*{fig:simulationFullModelTips2} and \subref*{fig:simulationFullModelVessels2} we chose a smaller lattice constant $h=0.005$, with $N^0=25$ adjusted such that the density of the tip cells remains the same as in Figures \subref*{fig:simulationFullModelTips1} and \subref*{fig:simulationFullModelVessels1}. The difference between the PDE and the average of $100$ realisations of the stochastic model is still pronounced, but less so than in Figures \subref*{fig:simulationFullModelTips1} and \subref*{fig:simulationFullModelVessels1}. Individual realisations of the stochastic model are noisier for the tip cell movement shown in Figure \subref*{fig:simulationFullModelTips2} than that shown in Figure \subref*{fig:simulationFullModelTips1}, due to the lower number of tip cells per box.

In Figures \subref*{fig:simulationFullModelTips3} and \subref*{fig:simulationFullModelVessels3}, we decreased the lattice constant further to $h=0.00125$, but also increased the initial tip density, choosing $N^0=50000$. The results confirm that we obtain good agreement even between a single realisation of the stochastic model and the PDE. This is expected, as noise is suppressed for larger cell numbers, and discrete effects disappear in the limit as $h\to 0$. However, $N^0=50000$ is quite a large number of initial tip cells which might not be realistic in many modelling situations. The purpose of including Figures \subref*{fig:simulationFullModelTips3} and \subref*{fig:simulationFullModelVessels3} here is too confirm the mathematical consistency of our model.

We have also confirmed that we obtain similar results when making the alternative choice involving no norm in the vessel production term in \eqref{eq:PDEFullModel}, as discussed in section \ref{sec:TransitionRateTotalMovement} (results not shown). The reason is that the total flux of tip cells is dominated by the chemotactic flux for the parameters chosen here.

\section{Discussion}\label{sec:Discussion}

In this paper, we have developed a mesoscopic model of angiogenesis, where tip cell migration, production of new vessel cells, sprouting, anastomosis and vessel regression are modelled as stochastic events that affect individual cells. The model is defined on a fixed lattice, and is mesoscopic in the sense that each box of the lattice can accommodate several cells. By investigating how the model behaves when we change the lattice constant, we were able to take the continuum limit of the mean field equations of the stochastic model. We have studied situations under which the continuum model provides a good approximation to the discrete stochastic model. In particular, we have shown that in many situations, even if the movement of individual tip cells is highly stochastic, the resulting vasculature has low noise and is very similar to that predicted by the continuum model. 
Comparison of the continuum model derived from the stochastic model with existing continuum models of angiogenesis reveals that we cannot use a conventional approach to model random movement of tip cells; rather, we need to employ a novel transition rate in the stochastic model in order to recover the standard snail-trail model. One might argue that this novel transition rate, which differs from conventional ones by assuming that tip cells can only migrate in directions of lower tip cell concentrations, is not a particularly realistic way to take crowding and finite cell size effects into account. Our model could be easily modified to include more realistic terms for crowding, that affect cell movement terms and proliferation rates. However, these terms would not lead to the well-known snail-trail equations on the deterministic continuum scale. Likewise, there are many ways in which the stochastic and PDE models presented here could be extended. Other modes of tip movement (for example, haptotaxis \cite{chaplain2000mathematical}), or more realistic dependencies of the chemotaxis term (see for instance \cite{painter2002volume,mantzaris2004mathematical,hillen2009user}) or the sprouting term on the chemoattractant could be incorporated easily. Such considerations are postponed for future work. As snail-trail models of angiogenesis such as \cite{balding1985mathematical,byrne1995mathematical} were motivated by similar models of fungal growth \cite{edelstein1982propagation}, and we can reproduce those angiogenesis models in the deterministc continuum limit of our stochastic model, it would also be interesting to apply our modelling framework to fungal growth.

Another avenue which we are currently investigating involves extending our model into a stochastic/ continuum hybrid model. This can be motivated by the requirement to simulate large domains of tissue, where it is not computationally feasible to keep track of all individual cells, but in certain regions stochastic effects are important. Our modelling approach is suitable for constructing such hybrid models: regions in space with low variance can be treated as continuous while those with high variance can be treated as stochastic, and we can directly relate the stochastic model to the continuum one. Hence, we can quantify the error arising when using continuum models, and only switch to a continuum model when this error is small.

Finally, for more biological realism, we would need to couple the stochastic model developed here to a model which includes the source of the angiogenic factor, which could be, for instance, tumour cells or macrophages in a wound, itself as a stochastic entity, and study the interaction of the dynamics of the AF source and the dynamics of the vasculature.
\appendix


\section{Detailed derivation of mean field equations}\label{sec:app:derivationMeanField}

The transition rates discussed in section \ref{sec:stochasticModel} fall into two categories: some that describe tip cell movement and the associated production of vessel cells; others represent local reaction terms which change the cell content inside a single box. We will now analyse the general structure of the mean field equations for these two cases.
\subsection{Movement of tip cells}

The movement terms discussed in section \ref{sec:stochasticModelMovement} were modelled by transition rates of the form $\mathcal{T}_{N_{k}-1,N_{l}+1,R_{k}+\delta_R,R_{l}|N_{k},N_{l},R_{k},R_{l}}$, $l=k\pm1$.
We now restrict ourselves to the interior of the domain, $k=2,\dots,k_{max}-1$,
and refer to appendix \ref{sec:app:BCs} for the treatment of the boundary.
The term in the master equation \eqref{eq:generalMasterEquationAngiogenesis} describing movement was 
 $\sum_{k,l\in \nene{k}}(E_{N_k}^{+1}E_{N_l}^{-1}E_{R_k}^{-\delta_R}-1)\mathcal{T}_{N_{k}-1,N_{l}+1,R_{k}+\delta_R,R_{l}|N_{k},N_{l},R_{k},R_{l}}P$.
We see that the contribution to the mean field equations \eqref{eq:generalMFequationAllTerms} from moving tip cells is given by
\begin{align}\label{eq:meanFieldTipsGeneral}
\mathcal{E}^{N_k}_M &= \sum_{\{N_{j}\},\{R_{j}\}}\sum_{m,l\in
\nene{m}}N_k(E_{N_m}^{+1}E_{N_l}^{-1}E_{R_m}^{-\delta_R}-1)\mathcal{T}_{N_{m}-1,N_{l}+1,R_{m}+\delta_R,R_{l}|N_{m},N_{l},R_{m},R_{l}}P\nonumber\\
 &= \sum_{\{N_{j}\},\{R_{j}\}}\sum_{l\in\nene{k}}N_k\left((E_{N_k}^{+1}E_{N_l}^{-1}E_{R_k}^{-\delta_R}-1)\mathcal{T}_{N_{k}-1,N_{l}+1,R_{k}+\delta_R,R_{l}|N_{k},N_{l},R_{k},R_{l}}\right.\nonumber\\
&\quad+\left.(E_{N_l}^{+1}E_{N_k}^{-1}E_{R_l}^{-\delta_R}-1)\mathcal{T}_{N_{l}-1,N_{k}+1,R_{l}+\delta_R,R_{k}|N_{l},N_{k},R_{l},R_{k}}\right) P\nonumber\\
  &= \sum_{\{N_{j}\},\{R_{j}\}}\sum_{l\in\nene{k}}N_k\left(\mathcal{T}_{N_{k},N_{l},R_{k},R_{l}|N_{k}+1,N_{l}-1,R_{k}-\delta_R,R_{l}}P(N_k+1,N_l-1,R_k-\delta_R,R_l)\right.\nonumber\\
  &\quad-\mathcal{T}_{N_{k}-1,N_{l}+1,R_{k}+\delta_R,R_{l}|N_{k},N_{l},R_{k},R_{l}}P(N_k,N_l,R_k,R_l)\nonumber\\
  &\quad+\mathcal{T}_{N_{l},N_{k},R_{l},R_{k}|N_{l}+1,N_{k}-1,R_{l}-\delta_R,R_{k}} P(N_k-1,N_l+1,R_k,R_l-\delta_R)\nonumber\\
  &\quad-\left.\mathcal{T}_{N_{l}-1,N_{k}+1,R_{l}+\delta_R,R_{k}|N_{l},N_{k},R_{l},R_{k}} P(N_k,N_l,R_k,R_l)\right)\nonumber\\
  &= \sum_{\{N_{j}\},\{R_{j}\}}\sum_{l\in\nene{k}}\left((N_k-1)\mathcal{T}_{N_{k}-1,N_{l}+1,R_{k}+\delta_R,R_{l}|N_{k},N_{l},R_{k},R_{l}}P(N_k,N_l,R_k,R_l)\right.\nonumber\\
  &\quad-N_k\mathcal{T}_{N_{k}-1,N_{l}+1,R_{k}+\delta_R,R_{l}|N_{k},N_{l},R_{k},R_{l}}P(N_k,N_l,R_k,R_l)\nonumber\\
  &\quad+(N_k+1)\mathcal{T}_{N_{l}-1,N_{k}+1,R_{l}+\delta_R,R_{k}|N_{l},N_{k},R_{l},R_{k}} P(N_k,N_l,R_k,R_l)\nonumber\\
  &\quad-N_k\left.\mathcal{T}_{N_{l}-1,N_{k}+1,R_{l}+\delta_R,R_{k}|N_{l},N_{k},R_{l},R_{k}} P(N_k,N_l,R_k,R_l)\right)\nonumber\\
 &= \sum_{\{N_{j}\},\{R_{j}\}}\sum_{l\in \nene{k}}(\mathcal{T}_{N_{l}-1,N_{k}+1,R_{l}+\delta_R,R_k |N_{l},N_{k},R_{l},R_{k}}\nonumber\\
 &-\mathcal{T}_{N_{k}-1,N_{l}+1,R_{k}+\delta_R,R_{l} |N_{k},N_{l},R_{k},R_{l}})P.\nonumber\\
\end{align}
Here, $\nene{m}=m\pm1$ denotes nearest neighbouring boxes of box $m$. In the first line, we simply substitute in the movement part of the master equation  \eqref{eq:generalMasterEquationAngiogenesis}. Then we note that only those parts of the sum which are nearest neighbours of $k$ contribute: otherwise we apply the shift operators to $\mathcal{T}_{N_{m}-1,N_{l}+1,R_{m}+\delta_R,R_{l}|N_{m},N_{l},R_{m},R_{l}}P$ and then shift the summation index of the sum over the state space, and recover the original sum, so the combination $(E_{N_m}^{+1}E_{N_l}^{-1}E_{R_m}^{-\delta_R}-1)$ will yield zero. For the third equality we apply the shift operators, making only the shifted arguments in $P$ explicit. $P$ without arguments denotes, for brevity, a dependence on the unshifted cell numbers. Then, the summation indices will be shifted in such a way that all probability densities $P$ will depend on the unshifted arguments.

Likewise, we can derive a general form for the mean field equations for the vessels:
\begin{align}\label{eq:meanFieldVesselsGeneral}
\mathcal{E}^{R_k}_M &= \sum_{\{N_{j}\},\{R_{j}\}}\sum_{m,l\in
\nene{m}}R_k(E_{N_m}^{+1}E_{N_l}^{-1}E_{R_m}^{-\delta_R}-1)\mathcal{T}_{N_{m}-1,N_{l}+1,R_{m}+\delta_R,R_{l}|N_{m},N_{l},R_{m},R_{l}}P\nonumber\\
 &= \sum_{\{N_{j}\},\{R_{j}\}}\sum_{l\in\nene{k}}R_k\left((E_{N_k}^{+1}E_{N_l}^{-1}E_{R_k}^{-\delta_R}-1)\mathcal{T}_{N_{k}-1,N_{l}+1,R_{k}+\delta_R,R_{l}|N_{k},N_{l},R_{k},R_{l}}\right)P\nonumber\\
 &= \sum_{\{N_{j}\},\{R_{j}\}}\sum_{l\in\nene{k}}R_k\left(\mathcal{T}_{N_{k},N_{l},R_{k},R_{l}|N_{k}+1,N_{l}-1,R_{k}-\delta_R,R_{l}}P(N_k+1,N_l-1,R_k-\delta_R,R_l)\right.\nonumber\\
&\quad-\left. \mathcal{T}_{N_{k}-1,N_{l}+1,R_{k}+\delta_R,R_{l}|N_{k},N_{l},R_{k},R_{l}}P(N_k,N_l,R_k,R_l)\right)\nonumber\\
 &= \sum_{\{N_{j}\},\{R_{j}\}}\sum_{l\in \nene{k}}\delta_R\mathcal{T}_{N_{k}-1,N_{l}+1,R_{k}+\delta_R,R_{l} |N_{k},N_{l},R_{k},R_{l}}P.\nonumber\\
\end{align}
The derivation is similar to the derivation of \eqref{eq:meanFieldTipsGeneral}, but the shift of summation index here is by $\delta_R$ rather than by $\pm1$. We see that the qualitative difference between \eqref{eq:meanFieldVesselsGeneral} and the mean field equation for the tips, \eqref{eq:meanFieldTipsGeneral}, is that to the rate of change of mean of vessels in box $k$ only those transition rates contribute which account for tips jumping out of box $k$. In contrast, the mean number of tips in box $k$ changes with the difference of transition rates describing incoming and outgoing tips from box $k$. This difference is clearly understood by the fact that we model vessel cells to be static, so there is no loss of vessel cells in any box due to movement. There will only be a loss of vessel cells by other mechanisms such as regression, which we shall now discuss. 

\subsection{Local source and sink terms}
We now discuss the general structure of the master equation which takes into account sprouting, i.e. the production of a new tip cell. This involves a transition rate of the form
\beq
\mathcal{T}_{N_{k}+1|N_{k}}.\nonumber
\eeq
The term in the master equation \eqref{eq:generalMasterEquationAngiogenesis} describing only tip birth takes the form $\sum_{k}(E_{N_k}^{-1}-1)\mathcal{T}_{N_{k}+1|N_{k}}P$.
From this we are led to the contribution to the mean field equation \eqref{eq:generalMFequationAllTerms} 
\begin{align}\label{eq:MFSourceGeneral}
\mathcal{E}^{N_k}_S &= \sum_{\{N_{j}\},\{R_{j}\}} N_k(E_{N_k}^{-1}-1)\mathcal{T}_{N_{k}+1|N_{k}}P.\nonumber\\
&=\sum_{\{N_{j}\},\{R_{j}\}} N_k\left(\mathcal{T}_{N_{k}|N_{k}-1}P(N_k-1)-\mathcal{T}_{N_{k}+1|N_{k}}P(N_k)\right),\nonumber\\
&=\sum_{\{N_{j}\},\{R_{j}\}} \left((N_k+1)\mathcal{T}_{N_{k}|N_{k}-1}P(N_k)-N_k\mathcal{T}_{N_{k}+1|N_{k}}P(N_k)\right),\nonumber\\
&=\sum_{\{N_{j}\},\{R_{j}\}} \mathcal{T}_{N_{k}+1|N_{k}}P,\nonumber\\
\mathcal{E}^{R_k}_S &= 0.
\end{align}
The derivation again makes use of the application of the shift operators and subsequent shift of summation indices, as for the derivation of equations \eqref{eq:meanFieldTipsGeneral},\eqref{eq:meanFieldVesselsGeneral}. Similar expressions can be derived for contributions of anastomosis and vessel regression to the mean field equations.

\section{Angiogenesis model in higher dimensions}\label{sec:app:higherDimensions}

We will now comment on the generalisation of our model to higher spatial dimensions $d$. Consider the transition rate describing tip movement from box $k$ to box $l$, with the structure 
\beq
\mathcal{T}_{N_k-1,N_l+1,R_k+\delta_R,R_l|N_k,N_l,R_k,R_l}.\nonumber
\eeq
In one dimension, we had $l=k\pm 1$. In higher dimensions, we can think of $k,l$ as multiindices, i.e. 
\beq
k = \{k_1,k_2,\dots,k_d\}.
\eeq
Here, $k_j = 1,\dots,k_{max,j}$.

In this way, the above transition rate requires no change at all. We need only define the nearest neighbours of $k$. On a regular grid, we define the nearest neighbours $l$ to be any of 
\begin{align}
l&=\{k_1\pm 1,k_2,\dots,k_d\}\nonumber\\
l&=\{k_1,k_2 \pm 1,\dots,k_d\}\nonumber\\
\hdots\nonumber\\
l&=\{k_1,k_2,\dots,k_d\pm 1\}.\nonumber
\end{align}
With this definition, we can calculate the mean field equations, take the continuum limit, and obtain the continuum equations for the different cases of movement given in equations \eqref{eq:PDEDiffusionStandard}, \eqref{eq:PDEDiffusionDifference} and \eqref{eq:PDEchemotaxis}, with the Laplace and Nabla operators now defined in $d$ dimensions. The norms are  $L_1$ norms, due to the choice of our grid. The relation between the discrete and continuous variables is now given by 
\begin{align}
 n(t,x_1,\dots,x_d) &= \frac{N_k}{h^d},
\end{align}
with $x_j=k_j h$.

Concerning the generalisation of the local interaction terms, which were used in the model to describe sprouting, anastomosis and regression, we remark that one needs to scale the parameters such that these rates scale like $\mathcal{T} = h^d f(\frac{N_k}{h^d},\frac{R_k}{h^d},\frac{C_k}{h^d})$, where  $f$ is a function of the intensive variables only, which means these variables do not scale with system size.

\section{Stochastic treatment of the angiogenic factor}\label{sec:app:stochasticChemical}
\begin{figure}[h!]
\begin{center}
\includegraphics[width=0.48 \linewidth]{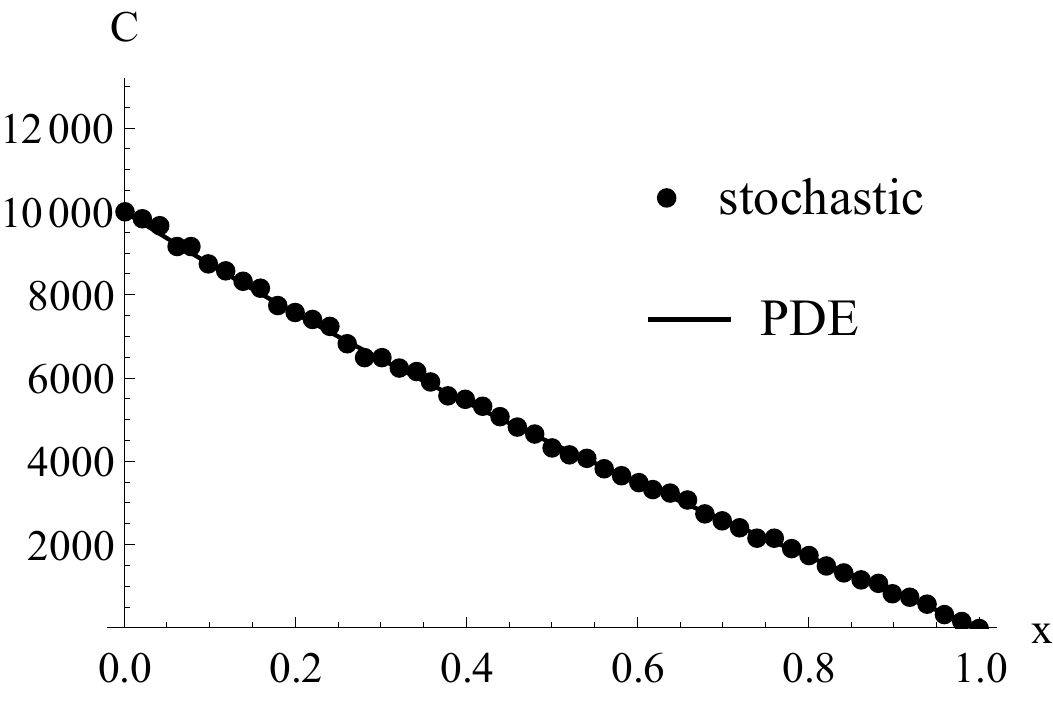}
\end{center}
\caption{\label{fig:stochasticChemical} The steady state profile of the angiogenic factor simulated from the PDE \eqref{eq:PDEappendix} and the stochastic model defined by \eqref{eq:transitionRateAFMovement} and \eqref{eq:transitionRateAFReaction}, but no vessel cells present. Dirichlet boundary conditions where imposed such that $C_1=10000$ and $C_{50}=0$.
}
\end{figure}

It is straightforward in our model to treat the AF stochastically. For this purpose, we introduce a new variable $C_k$, $k=1,\dots,k_{max}$, which is related to the continuous, deterministic $c$ used in this paper by
\beq
\mean{C_k} = h c(x).
\eeq
As before, $x=kh$ in one spatial dimension. Then, a state in the stochastic model is specified by $N_k, R_k$ and $C_k$. We can now introduce additional transition rates in the stochastic model describing the movement of molecules:
\beq\label{eq:transitionRateAFMovement}
 \mathcal{T}_{C_k-1,C_l+1|C_k,C_l} = D_c^hC_k. 
\eeq
As for the cell motility terms, we have $D_c^h = \frac{D_c}{h^2}$. 
Furthermore, we introduce a transition rate describing the consumption and degradation of the angiogenic factor:
\beq\label{eq:transitionRateAFReaction}
 \mathcal{T}_{C_k-1|C_k} = \lambda C_k + \frac{a_1}{h}H(C_k-\hat{C})N_kC_k.
\eeq
In the continuum limit, we obtain
\begin{align}\label{eq:PDEappendix}
 \frac{\partial c}{\partial t} &= D_c\Delta c -\lambda c - a_1H(c-\hat{c})n c,
\end{align}
reproducing \eqref{eq:PDEAFgeneral}, \eqref{eq:AFSourceTerm}. Other reaction terms can be implemented in a similar way. 
Let us now estimate the size of the stochastic effects associated with the angiogenic factor. As stochastic effects are expected to be stronger when smaller numbers of molecules are involved, we are conservative in underestimating the number of molecules. In the corneal assay modelled in section \ref{sec:simulationFullModel}, the concentration of the angiogenic factor (here, acidic fibroblast growth factor) was taken to be $10^{10}M$ \cite{byrne1995mathematical}, and the distance between tumour and initial vessel, i.e. the size of our modelling domain, was $3mm$. There are various estimates of the thickness of the cornea, and a conservative choice is given by $100 \mu m$ \cite{henriksson2009dimensions}. Then a box in a $2D$ model with a discretisation of 50 sites in both dimensions would represent a volume of $100\mu m \left(\frac{3mm}{50}\right)^2$, which is again a conservative choice for the $1D$ model where we integrate over one dimension. Thus, the total number of molecules of the angiogenic factor in one box is of the order
\beq
C_k\approx 10^{-10}M 100\mu m \left(\frac{3mm}{50}\right)^2 \approx 6\times 10^{23}10^{-10}\frac{1}{10^{-3}m^3} \frac{9}{2500}10^{2-6-6}m^3\approx 10^{4}.
\eeq
If we solve \eqref{eq:PDEappendix} and the stochastic model defined by \eqref{eq:transitionRateAFMovement} and \eqref{eq:transitionRateAFReaction} for the case considered in section \ref{sec:simulationFullModel}, i.e. Dirichlet boundary conditions, with a source on the left boundary, $C_1=10000$, and a sink with $C_{50}=0$ on the left, and a lattice of $50$ sites,  we obtain the steady state profile presented in Figure \ref{fig:stochasticChemical}.
With $C_1=10000$ and $C_{50}=0$ we typically have $C_k\leq 10000$, giving again a conservative estimate of the real stochastic fluctuations. We see in Figure \ref{fig:stochasticChemical} that the stochastic fluctuations are relatively small throughout the domain, justifying the use of the PDE for the angiogenic factor throughout the domain. We note also that it is justified to use the same spatial discretisation to solve the PDE as was used to simulate the stochastic model, as variations between neighbouring lattice sites are small.

\section{Numerical solution of the model}\label{sec:app:Numerics}

\subsection{Stochastic model}
To simulate a realisation of the stochastic model, we use the Gillespie algorithm \cite{gillespie1976general,gillespie1977exact}, see also the review \cite{erban2007practical}. This means that if we let $\alpha_m$, $m=1,\dots,m_{tot}$ denote all non-zero transition rates in the model, enumerated by integers $m$, then $\alpha_0=\sum_m\alpha_m$ denotes the total rate. The time to the next event is exponentially distributed, so we draw a uniformly distributed random number $r_1\in[0,1]$ such that the time to the next event will be
\beq
\tau=\frac{1}{\alpha_0}\log{\frac{1}{r_1}}.
\eeq
We then draw a second uniformly distributed random number $r_2\in[0,1]$ which determines which event will take place. In particular, if 
\beq
\frac{\sum_m^{s-1}\alpha_m}{\alpha_0}\leq r_2 <\frac{\sum_m^{s}\alpha_m}{\alpha_0},
\eeq
with $0<s\leq m_{tot}$, then the event corresponding to rate $\alpha_s$ will occur. A principal assumption concerning the AF is that it diffuses and reacts on a timescale much faster than that of events affecting the cells, which are movement, sprouting, anastomosis or regression. Hence, between consecutive Gillespie events we solve the PDE describing the evolution of the concentration of the AF $c$ using a finite difference approximation, with a forward Euler method.

\subsection{Solution of the PDEs}

The PDEs appearing in this paper were solved with Mathematica 9 using the method of lines, and in most cases we could rely on the built-in solver to choose the correct time integration method. We have compared our results with  those obtained using several, alternative ODE integration methods such as explicit and implicit 4th order Runge-Kutta, and also to a Matlab implementation using the pdepe function, and a C++ implementation of a finite difference scheme with explicit Euler integration. In the reduced models of tip cell movement and vessel production shown in section \ref{sec:simulationsMovement}, the tips evolve according to pure diffusion or pure chemotaxis with linear gradient. In these cases, the PDEs admit explicit solutions by Fourier expansion or direct integration, respectively. For the case of pure chemotaxis as well as the simulations of the full model in section \ref{sec:simulationFullModel}, the explicit Euler method is unstable and cannot be used.

\section{Boundary conditions in the stochastic model}\label{sec:app:BCs}

We will now briefly discuss different choices of boundary conditions in the stochastic model. We only need to focus on the part of the model involving the transition rate describing tip cell migration, $\mathcal{T}_{N_k-1,N_l+1,R_k+\delta_R,R_l|N_k,N_l,R_k,R_l}$, as the other transition rates are local, depending on a single box.

To facilitate comparison of the stochastic and PDE models, it is useful to choose slightly different conventions from the main part of this paper and center the boxes at $x=(k-1)h$, so that box $k$ extends from $x\in[k-\frac{3h}{2},k-\frac{h}{2}]$. Then the domain extends from $x\in [-\frac{h}{2},L+\frac{h}{2}]$, if we let $k=1,\dots,k_{max}$, and $(k_{max}+1)h = L$. This has the advantage that, when choosing Dirichlet boundary conditions, we can directly relate the contents of boxes $k=1, k_{max}$ to corresponding values of the PDE model at $x=0,L$. As we assume $h\ll L$ this redefinition of the domain does not affect the solution of the PDE.

\paragraph{Dirichlet boundary conditions:}
We implement the boundary conditions by specifying the tip numbers in boxes $k=1$ and $k=k_{max}$, so
\begin{align}
N_1(t)&=f_L(t),\nonumber\\
N_{k_{max}}(t)&=f_R(t).
\end{align}
Here, we assume that $f_L$ and $f_R$ are slowly varying functions of time, compared to the timescale of events occurring in the stochastic model. This means we can hold $N_1$ and $N_{k_{max}}$ fixed between successive stochastic events. Another option is to implement Dirichlet boundary conditions in terms of stochastic reactions. Rather than fixing $N_1$ and $N_{k_{max}}$ deterministically, they could undergo birth and death processes so that only on average would we get $N_1=f_L(t)$, $N_{k_{max}}=f_R(t)$. For simplicity, in this paper we view $N_1$ and $N_{k_{max}}$ as deterministic.

\paragraph{Neumann boundary conditions:}
For simplicity we restrict our study to zero flux boundary conditions. A simple implementation is such that for box $k=1$, there is only one outgoing transition rate, 
\beq\label{eq:boundaryTransitionRate}
\mathcal{T}_{N_1-1,N_2+1,R_1+\delta_R,R_2|N_1,N_2,R_1,R_2},
\eeq
and similarly for box $k=k_{max}$. In this way, there is automatically no flux through the boundary. However, the number of vessel cells produced in the boundary boxes will be considerably smaller than the number produced in boxes $k=2,\dots,k_{max}-1$, as there will be less net movement. One way to overcome this is to add a reflection transition rate 
\beq
\mathcal{T}_{N_1,R_1+\delta_R|N_1,R_1},
\eeq
and similarly for $k=k_{max}$. This has the interpretation that when a cell attempts to move to the left, it is reflected at the impenetrable boundary, and, as a result, remains in the same box from where it starts. It will still leave some vessel cells behind in this process. Such a term would only make sense for random movement Case 1, as outlined in section \ref{sec:stochasticModelMovement}, since Case 2, as well as chemotaxis, depend on the difference in cell numbers or concentrations in neighbouring boxes. Here, the incoming and outgoing boxes are identical.

An alternative implementation of zero-flux boundary conditions is similar to that commonly used for boundary conditions in finite difference schemes of PDEs. One fixes the number of tips in boxes $1$ and $k_{max}$, respectively, as 
\begin{align}
N_1&=N_2\nonumber\\
N_{k_{max}}&=N_{k_{max}-1}.\nonumber
\end{align}
For simplicity, we implemented Neumann no-flux boundary conditions using the first method \eqref{eq:boundaryTransitionRate}.


\begin{acknowledgements}
This publication was based on work supported in part by Award No KUK-C1-013-04, made by King Abdullah University of Science and Technology (KAUST).
TA and PG gratefully acknowledge the Spanish Ministry for Science and Innovation (MICINN) for funding under grant MTM2011-29342 and Generalitat de Catalunya for funding under grant 2009SGR345.
PKM was partially supported by the National Cancer Institute, National Institutes of Health grant U54CA143970.
\end{acknowledgements}

\bibliographystyle{spmpsci}      
\bibliography{angiogenesis}   

%
%

\end{document}